\documentclass[useAMS,usenatbib,fleqn]{mn2e}

\usepackage{times}

\usepackage{amsmath}
\usepackage{upgreek}
\usepackage{graphicx}

\title[ALMA and SOFIA water masers]{The Physics of Water Masers observable with ALMA and SOFIA: Model Predictions for Evolved Stars}
\author[M. D. Gray, A. Baudry, A. M. S. Richards, E. M. L. Humphreys, A.M. Sobolev and J. A. Yates]
{M. D. Gray$^{1}$, A. Baudry$^{2,3}$,
A. M. S. Richards$^{1}$, E.M.L. Humphreys$^{4}$,\newauthor A.M. Sobolev$^{5}$ and J. A. Yates$^{6}$\\
$^{1}$Jodrell Bank Centre for Astrophysics, School of Physics and Astronomy, University of Manchester,
M13 9PL, UK\\
$^{2}$Universit\'{e} de Bordeaux, LAB, UMR 5804, F-33270 Floirac, France\\
$^{3}$CNRS, LAB, UMR 5804, F-33270 Floirac, France\\
$^{4}$ESO Karl-Schwarzschild-Str. 2, 85748 Garching, Germany\\
$^{5}$Ural Federal University, Ekaterinberg, Russia\\
$^{6}$Dept. of Physics and Astronomy, University College London, WC1E 6BT, UK}
\begin{document}

\date{}

\pagerange{\pageref{firstpage}--\pageref{lastpage}} \pubyear{2002}

\maketitle

\label{firstpage}

\begin{abstract}
We present the results of models that were designed to study all possible water maser
transitions in the frequency range 0-1.91\,THz, with particular emphasis on maser transitions that may be
generated in evolved-star envelopes and observed with the \textit{ALMA} and \textit{SOFIA}
telescopes. We used tens of thousands of radiative transfer models of both spin species of H$_2$O,
spanning a considerable parameter space in number density, kinetic
temperature and dust temperature. Results, in the form of maser optical depths, have
been summarized in a master table, Table~\ref{hugetab}.
Maser transitions identified in these models were grouped according to
loci of inverted regions in the density/kinetic temperature
plane, a property clearly related to the dominant mode of pumping. A more detailed
study of the effect of dust temperature on maser optical depth enabled us to
divide the maser transitions into three groups: those with both collisional and
radiative pumping schemes (22,96,209,321,325,395,941 and 1486\,GHz), a much
larger set that are predominantly radiatively pumped, and another large group
with a predominantly collisional pump.
The effect of accelerative and decelerative velocity shifts of up to 5\,km\,s$^{-1}$
was found to be generally modest, with the primary effect of reducing computed
maser optical depths. More subtle asymmetric effects, dependent on line
overlap, include maximum gains offset from zero shift by $>$1\,km\,s$^{-1}$,
but these effects were predominantly found under conditions of weak amplification.
These models will allow astronomers to use multi-transition water maser observations 
to constrain physical conditions down to the size of individual masing clouds (size
of a few astronomical units).
\end{abstract}


\section{Introduction}
\label{intro}

Water masers from the $6_{16}-5_{23}$ transition of ortho-H$_2$O (o-H$_2$O) at
22.23508\,GHz are abundant in star-forming regions, in the
extended atmospheres of evolved stars and in some external
galaxies (`megamasers'). The original detection towards the Orion
star-forming region \citep{1969Natur.221..626C} was followed up by observations
that included the red supergiant star VY~CMa \citep{1969Sci...165..180M}. Modern
surveys of star-forming and evolved star sources have been carried out, or are in progress. Probably
the majority of work has concentrated on masers from star-forming regions, including
both high-mass and low-mass sources.

Masers at 22\,GHz, from any source type,  are not the main focus of the present work, but 
22\,GHz is so far the only water maser frequency at which large-scale surveys have been
carried out, giving us some idea of the Galactic distribution of sources. With the
implicit assumption that the distribution of water masers at other frequencies
broadly follows the 22-GHz distribution, we therefore initially consider 22-GHz
surveys for all Galactic source types.

Selective surveys of water masers in high-mass young stellar objects (HMYSOs) 
\citep{2009A&A...507..795U,2011MNRAS.418.1689U}
typically observe
towards sources that satisfy certain infra-red colour criteria based, for example,
on the Red MSX Source survey, or 4.5\,$\umu$m emission detected by the \textit{GLIMPSE} instrument
aboard the Spitzer satellite 
\citep{2013ApJ...769...20Y}. Unlike 6.7-GHz methanol masers that are believed
to be associated only with the formation of high-mass stars
\citep{2008A&A...485..729X}, 22-GHz water masers are also
associated with low-mass protostellar objects, and surveys of objects of this
type include work by
\citet{2001ApJ...559L.143F} and
\citet{2003ApJS..144...71F} (respective errata in
\citet{2007ApJ...659L..81F,2007ApJS..171..349F})
that showed that water masers are associated primarily
with low-mass protostars of class~0 with some in class~1. No masers were found
towards class~2 protostars or pre-stellar cores. An association with 100-AU-scale
jets suggests that the water maser emission is coupled to shock waves, as in other
source types. \citet{2013ApJ...773...70H} and \citet{1996ApJ...456..250K} showed
that the conditions for shocks to excite water masers can occur in star-forming regions,
notably (but not exclusively) due to the impact of protostellar jets, discussed further in
Section~\ref{ss:theo}. A caveat regarding such an interpretation is that the collimation
and precession of jets is linked to variability of masers, so long-term monitoring
of sources is preferable to observations at a single epoch. Other excitation schemes
involve the boundaries of wide-angle outflows
\citep{1984ApJ...280..618W,1994ApJ...427..914M} and discs
\citep{1997A&A...327..758F,2003ApJ...586..306G}.

Detection of water masers towards only 5 of 99 observed low-mass
protostars in Orion, selected on the basis on infra-red colours, suggests that
water masers probably appear rarely in such objects
\citep{2013ApJS..209...25K}. Water masers were found towards 9 per cent of
the intermediate-mass objects observed by
\citet{2011ApJS..196...21B}, so the detection rate probably rises with
protostellar mass. Water masers associated with HMYSOs are highly variable:
a 20-yr study of 43 such objects by \citet{2007A&A...476..373F} includes
velocity-time-flux plots that may be used to compare short-duration features
and more stable spectral components. Observations with the \textit{ATCA} (Australia Telescope Compact Array) towards
dust clumps emitting strongly at 1.2\,mm \citep{2011MNRAS.416..178B} revealed
a much better correlation between 22-GHz masers in MYSOs and these clumps than is
typically found for masers and infra-red colours. The 1.2-mm clumps
associated with H$_2$O masers are of order 1\,pc across.

Unbiased surveys of water masers naturally include sources of both
protostellar and evolved-star types. Small-scale interferometric pathfinder
surveys of the Galactic centre \citep{2011MNRAS.410.1283C} and other
regions of the Southern Galactic plane \citep{2010MNRAS.407.2599C} with the \textit{ATCA}
show that variability on timescales of order 1\,yr is common, that the
spatial density of water masers exceeds that of methanol and OH, even
in regions where the latter two species are known to be common, and
that the positions of water maser sources are significantly more
stable over time than their spectra. The spectral ranges of water maser
sources were already known to exceed those typical of methanol and
OH \citep{2007A&A...476..373F}. In the sample observed by \citet{2010MNRAS.407.2599C},
nine objects (28 per cent) exhibited extreme velocity spectral components, defined
as separated from the systemic velocity by $>$30\,km\,s$^{-1}$. These components,
associated with outflows,
are dominated by blue-shifted emission.
The H$_2$O Southern Galactic Plane Survey (HOPS) \citep{2011MNRAS.416.1764W} surveyed 100 square degrees of the Southern Galactic
plane with a broad-band spectrometer that included the 22-GHz water maser line
amongst many other transitions down to an rms noise level of typically 1-2\,Jy.
The survey results suggest that 800-1500 22-GHz maser sources exist in the Milky Way down to
this noise level. A scale-height of only 0.5$\degr$, similar to that for 6.7-GHz methanol
masers, suggests that most HOPS detections are associated with high-mass star
formation. This view has been confirmed by \textit{ATCA} positional associations
\citep{2014MNRAS.442.2240W} that identify, of the sources associated
with otherwise known astrophysical objects, 69 per cent with star formation, 19 per cent
with evolved stars and 12 per cent unknown.

It has long been known that individual 22-GHz water maser features are among the smallest
and brightest known. Early global VLBI (Very Long Baseline Interferometry)
experiments \citep{1972SvA....16..379B,1973R&QE...16..613B} determined
that features as small as 300\,microarcsec (4.5\,AU) existed in the W49 star-forming
region, with a brightness temperature of $T_b$ = 10$^{15}$\,K. A recent VLBI experiment
with a space-based antenna (\textit{RadioAstron}) found structure on a scale of only
10\,microarcsec ($\sim$1\,R$_{\sun}$) in Cep~A\footnote{vestnik.laspace.ru/eng 2014 vol.3, p4}, with 
a brightness temperature of at
least 1.5$\times$10$^{14}$\,K; a flare in Orion~KL observed with \textit{HALCA} (Highly
Advanced Laboratory for Communications and Astronomy) \citep{1999PASJ...51..333O}
had a brightness temperature of 10$^{17}$\,K.

A single-dish survey of 401 evolved stars, mostly of Mira, OH/IR and semi-regular
variable types, was used to calculate useful statistics for both the 22-GHz
H$_2$O line, and the SiO 43-GHz masers from the vibrational states $v=1$ and $2$
\citep{2014AJ....147...22K}.
Results from such a single-phase snapshot of many sources are broadly consistent
with those derived from long-term monitoring of a small number of sources over
at least several stellar periods, an example of the latter type of observation
being \citet{2005A&A...437..127L}. The snapshot revealed that the 22-GHz masers
are weaker (in photon luminosity, peak and
frequency-integrated intensity) than the SiO masers at most stellar phases, but that the water masers
become relatively more powerful as stars become more massive, and have greater
mass-loss rates - that is as stellar properties move from the Miras to the OH/IR stars
\citep{1986A&A...167..129E}. More information about
the effect of the circumstellar velocity field on the masers can be gleaned from
the velocity extent and red/blue dominance in the spectra. In Mira variables,
\citet{2014AJ....147...22K} find a very similar spectral extent for SiO and H$_2$O 22\,GHz masers
(respectively 13.7 and 12.9\,km\,s$^{-1}$). However, the H$_2$O emission becomes significantly 
broader in OH/IR stars (25.5\,km\,s$^{-1}$ against 13.3). In both types, the SiO velocities
are consistent with an average expansion velocity of $\leq$7\,km\,s$^{-1}$ in the maser zone,
as required by observational constraints \citep{2007ApJ...671.2068R}. 
Individual 22-GHz H$_2$O maser sources show a wide range of values for the expansion
velocity, with spectral extents exceeding 40\,km\,s$^{-1}$ for supergiants and some
OH/IR stars, whilst Miras have values between 4 and $\sim$40\,km\,s$^{-1}$.
For example, VLBI observations 
of RT~Vir \citep{2003ApJ...590..460I} found an approximate velocity of expansion of
8\,km\,s$^{-1}$ in the water maser zone.
Infra-Red Astronomy Satellite (\textit{IRAS}) colour-colour diagrams in \citet{2014AJ....147...22K} indicate that
H$_2$O masers form at an earlier evolutionary stage than SiO masers, but the H$_2$O masers
may also persist into the PPN (proto-planetary nebula) stage, probably in a new form dominated by
asymmetric outflow. Development of water (and SiO) maser emission in the transition from
AGB (Asymptotic Giant Branch) to post-AGB stars is surveyed in more detail in \cite{2014ApJS..211...15Y}.

The influence of pulsation shocks on 22-GHz H$_2$O masers is not clear. 
Model predictions (for example \citealt{2008MNRAS.391.1994I}, \citealt{2011MNRAS.418..114I}) are of weakening shocks with increasing
radius.
\citet{1992ApJ...394..221E} modelled the different appearance of maser beaming from
quiescently expanding clouds or shocked slabs, and \citet{2010evn..confE...5R} showed that
the symptoms of shocks are only observed in a minority of clouds, mostly in the 
thinner-shelled CSEs in the study. This is consistent with the predictions of
\citet{2008MNRAS.391.1994I}. However, in this context, one would expect to see evidence
for shocks close to the inner rim of the water maser shell, but
\citet{2003ApJ...590..460I} observed a shock-accelerated feature further out in the
shell around RT~Vir, and VY~CMa also shows evidence that suggests shocks far from
the star (Section~\ref{ss:vycma}). It is clear that shocks alone cannot explain water
maser variability, since monitoring shows that features throughout the entire 22-GHz
shells (perhaps 15/150\,AU thick around AGB/RSG stars) vary in concert
within weeks to years - much faster than even a highly supersonic shock.

The 22-GHz transition remains by far the most widely studied, but maser
action has been confirmed in several more lines
at higher frequencies. Twelve (possibly thirteen) additional maser transitions were 
tabulated in \citet{2007IAUS..242..471H}, and a similar list of eleven maser transitions
appears in Chapter~2 of \citet{mybook}.
A newer search for detected masing lines in the literature shows that at least 18
transitions have now been identified as masers in addition to 22\,GHz (see entries
with Y in the final column of Table~\ref{almatab} and the text of
Section~\ref{ss:alma}). More maser transitions will almost certainly
be identified in future, and it is worth noting that several high frequency lines
that were detected with Herschel at low spectral resolution \citep{2014MNRAS.437..532M}
are predicted to be masers in this work, and are listed in Table~\ref{hugetab}.
Most of the known high-frequency masers in Table~\ref{almatab}, like 22-GHz, have been detected towards both star-forming regions and evolved 
stars. However, a sub-set of maser transitions within the first excited
state of the vibrational bending mode ($v_2=1$), and two $v_2=2$ maser transitions, have been detected
exclusively towards evolved stars (see final column of Table~\ref{almatab} and footnotes).

Almost all of the high-frequency water masers have rest frequencies above 100\,GHz,
making them difficult to observe from low-altitude sites. They have been much less
useful scientifically than the 22-GHz transition because, prior to the advent
of \textit{ALMA}, there were few interferometric observations of any of the high-frequency
lines, and none to compare with the milli-arcsecond resolution typically achieved
with continental VLBI at 22\,GHz. \textit{ALMA} offers the exciting possibility of extending
routine interferometric observations with angular resolution of typically tens of
milli-arcseconds to most of the known mm-wave and sub-mm water maser transitions.
\textit{SOFIA} is a single-dish instrument, but provides a new window that may allow us to
detect new water maser transitions with frequences greater than 1\,THz.
 
\subsection{Water masers in AGB and post-AGB stars}

Work on high-frequency water masers prior to 2007, including most of the initial
detections, has been reviewed by \citet{2007IAUS..242..471H}. An exception is the
interferometric work carried out with the Submillimeter Array (\textit{SMA}) on the 658-GHz transition \citep{2007IAUS..242..481H},
a transition in the $v_2=1$ bending mode of o-H$_2$O, where the lower energy level of
the transition lies at 2297\,K above the ground state. In a sample of four Mira variables,
the spectral width of the 658-GHz line was found to be similar to that of the SiO $v=1,J=5-4$
line at 215\,GHz. This similarity, and comparable energies above ground 
for the lower maser levels, suggest that these o-H$_2$O and SiO
transitions originate from a similar region: the SiO zone that lies much closer
to the star ($<$ 5 stellar radii) than is typical for 22-GHz water masers.
The 22-GHz maser shell, resolved by interferometry, typically extends from about
5 to 20 stellar radii, over which distance the expansion velocity approximately doubles,
exceeding the escape velocity in the process (for example, \citealt{2012A&A...546A..16R}).

The Mira R~Leo and the semi-regular variable W~Hya were observed in eight sub-millimetre
water transitions with \textit{APEX} (Atacama Pathfinder EXperiment) \citep{2008A&A...477..185M}. The very high excitation line
at 354.8\,GHz (17$_{4,13}$-16$_{7,10}$) was not detected towards either source, despite predictions
of inversion by \citet{1991ApJ...368..215N} and \citet{1997MNRAS.285..303Y} (respectively NM91 and YFG97). However, both sources exhibited
strong masers at 437.3\,GHz, and weaker emission at 443.0\,GHz, that are not predicted
by NM91. Photon luminosities of the detected lines, and the velocity extents of the spectra,
were consistent with the detected lines arising from approximately the same region of the
outflowing atmosphere. In R~Leo, near contemporary 22-GHz observations showed that this line
was anomalously weak, whilst the 437-GHz maser was much stronger than the others. These
observations also resulted in the first detection of a maser at 474.7\,GHz - a line predicted
by NM91, but only at the higher temperature of 1000\,K, and by YFG97.
Herschel HIFI (Heterodyne Instrument for the Far Infrared) observations \citep{2012A&A...537A.144J} detected 658-GHz masers towards several
AGB stars with new maser detections at 620.7\,GHz towards IRC+10011, W~Hya and IK~Tau and possible
masers at 970.3\,GHz towards the latter two stars. 

A type of post-AGB star, known as a water-fountain source, has evolved to a state
in which a fast, highly asymmetric post-AGB outflow is interacting with the older
spherically symmetric shell from the AGB phase. Water masers form in the interaction
zone of the fast and slow outflows. \citet{2014A&A...562L...9T} studied seven water
fountain sources with \textit{APEX}, detecting 321-GHz H$_2$O masers towards three of them, but detecting
no emission at 325\,GHz. The breadth of the 321-GHz spectrum ($>$100\,km\,s$^{-1}$) links these
masers to the fast wind, rather than the AGB material that has an expansion velocity 
of only $\sim$20\,km\,s$^{-1}$. In two of the sources, the authors co-locate the 321-GHz
and 22-GHz masers, but invoke a more complicated model for the third source, where the
spectral widths of the maser lines are substantially different.

\subsection{Water masers in red supergiants}
\label{ss:vycma}

Although the known water-masing red supergiants (RSGs) are at ditances $>$800 pc, they have bright 
22-GHz masers and searches for higher transitions have been fruitful,
especially in VY CMa, for example \citet{1989ApJ...341L..91M},
\citet{1990ApJ...350L..41M}, and see Table~\ref{almatab}. High frequency lines have also
been detected from other RSGs, for example 321- and 325-GHz emission \citep{1995MNRAS.273..529Y}, 
658-GHz \citep{1995ApJ...450L..67M} from NML~Cyg and VX~Sgr, plus
183-GHz emission also from S~Per and $\umu$~Cep \citep{1998A&A...334.1016G}.

VY~CMa has the brightest water masers and the largest CSE, and has
been studied in the most detail. Specific properties of VY~CMa are 
summarized in \citet{1969Sci...165..180M}. This star was one of the
four bright sources that were the subject of the first interferometric observations
of 22-GHz water masers \citep{1970ApJ...160L..63B}, and it is also a source of several
mm-wave and sub-mm water masers. VY~CMa was among the stars observed by \citet{2007IAUS..242..481H}, who
showed that its 658-GHz emitting region is unresolved at scales of $\sim$1\farcs

VY~CMa was also a target of the multi-frequency \textit{APEX} observations by \citet{2008A&A...477..185M}. Detections
were achieved in seven of the nine transitions observed, and all had photon luminosities between
6 $\times$ 10$^{44}$ and 5 $\times$ 10$^{45}$\,s$^{-1}$, some three orders of magnitude more powerful
than typical values for AGB stars. The photon luminosity of the 22-GHz maser in VY~CMa is similar
to the level of its sub-mm lines. A distinct difference between VY~CMa and AGB stars is in the 
width and shape of its maser spectral profiles: each profile is remarkably individual in VY~CMa,
suggesting different regions of origin, whilst the profiles from AGB stars have similar widths
for a given object. The VY~CMa spectra are also typically very broad ($>$50\,km\,s$^{-1}$) compared
to a few km\,s$^{-1}$ in AGB stars. The 437-GHz line, not predicted by NM91 or YFG97, was also
detected towards VY~CMa.

\textit{SMA} observations of VY~CMa \citep{2013ApJS..209...38K} included six water transitions
in the frequency range 293-337\,GHz. The interferometer did not resolve the emission
in any of these transitions (synthesized beam of approximately 0\farcs9). The expected
ground vibrational state masers at 321 and 325\,GHz were accompanied by a possible
weak maser in the ($0,1,0$) state at 293.7\,GHz.

By far the most detailed information about a subset of sub-mm water maser lines
(at 321, 325 and 658\,GHz) comes from \textit{ALMA} observations \citep{2014A&A...572L...9R}.
For the first time, the continuum emission and masers were, simultaneously, well resolved, establishing
the almost certain site of the star to coincide with the centre of the water maser
expansion, some 0\farcs4 to the North-West of the brightest continuum region. With
spatial resolution as good as 60\,milliarcsec at 658\,GHz, the spatial relationship
of maser features at the three sub-mm frequencies, and at 22\,GHz, could be established.
While the maser features share a common sky-plane area up to $\sim$1\,arcsec ($\sim$850\,AU)
in diameter,
they rather avoid each other at smaller scales, consistent with significantly
different pumping regimes. Linear distributions of masers with a
velocity gradient are common. Although the 658-GHz masers are mainly concentrated
in the central 0\farcs1, as expected for a transition with its lower level so far above
ground, some 658-GHz masers are found much further from the star, and this behaviour
is unexplained. It was suggested that the farthest-out 658-GHz masers could be excited
by shocks in the stellar wind, for example from the collision of fast-flowing material
with more slowly moving dense clumps.

The 620.7-GHz transition was detected as a maser for the first time towards VY~CMa
by \citet{2010A&A...521L..51H} with the HIFI instrument aboard \textit{HERSCHEL}. Their
experiment was polarization sensitive, and found negligible polarization near the
peak of the line, but rising values to $\sim$6 per cent in the line wings, in a
manner that is consistent with the theory by \citet{1981ApJ...243L..75G}. The
same instrument also detected the 970.3-GHz transition as a maser \citep{2011ASPC..445..317D}.
SPIRE (Spectral and Photometric Imaging Receiver) and PACS (Photoconductor Array Camera and Spectrometer) observations 
of VY~CMa detected a number of emission lines of water
that may be masers \citep{2014MNRAS.437..532M} at 1158.3,1172.5,1278.3,1296.4,1308.0,1322.1,1435.0
and 1440.8\,GHz. These transitions are all predicted to be strongly inverted in our
Table~\ref{hugetab}. The limited resolution of the \textit{HERSCHEL} observations makes it difficult
to tell whether there are masers in these transitions. Numerical fits that accompany the
observations agree with our predictions of inversion in 7 out of 9 cases, the exceptions
being 1435.0 and 1542.0\,GHz. Similar HIFI observations of VY~CMa
by \citet{2013A&A...559A..93A} find emission in 7 additional transitions at 968.0,970.3,1000.9,
1153.1,1205.8,1718.7 and 1753.9\,GHz that are in our table of maser predictions,
Table~\ref{hugetab}. The 968-GHz transition is particularly identified as a maser, though the
velocity resolution of about 1\,km\,s$^{-1}$ is arguably not good enough to confirm this.
One additional transition at 1194.8\,GHz is claimed by \citet{2013A&A...559A..93A} as a
possible maser, but is not predicted to be so in this work.

\subsection{Water masers in star-forming regions}

As for evolved stars, most work on water masers other than 22\,GHz, prior
to 2007, is summarised in \citet{2007IAUS..242..471H}. The range of transitions found
as masers in star-forming regions was long
thought to be restricted to the vibrational ground state, but is now known to include
masers from the bending mode, $v_2=1$. It is also notable that 183.3-GHz masers
cover an extended region in Orion IRc2 (as covered by a mosaic of \textit{IRAM} 30-m observations), and
these appear to be generated under conditions that are too cold and rarefied to be
considered in the usual shock-style pumping schemes \citep{1994ApJ...432L..59C}. A
similar conclusion was reached for 325.2-GHz masers from the same
source \citep{1999ApJ...520L.131C}. The 183-GHz line is one of very few masers to be
observed towards low-mass protostellar objects: for example, 3 emission spots were detected towards
Serpens~SMM1, a class~0 low-mass protostar envelope with a mass of only 2.7\,M$_{\sun}$
\citep{2009ApJ...706L..22V}. This detection supports the view that 183-GHz masers
can be pumped in cooler, more rarified, conditions than those at 22\,GHz.

The first imaging observations of the 321-GHz, 10$_{2,9}$-9$_{3,6}$ transition of o-H$_2$O
with the \textit{SMA} \citep{2007ApJ...658L..55P} revealed that, in Ceph~A, the distribution of
these masers lies along the jet of a disc-outflow system, while 22-GHz masers in this
source are associated with the disc.
\textit{SMA} observations were also made in the 183-GHz line of p-H$_2$O towards Serpens~SMM1. 
As for the 321-GHz maser in Cepheus~A, the 183-GHz emission
appears associated with a protostellar jet, and is spatially distinct from 22-GHz features.

The advent of \textit{ALMA} observations has led to significantly more detailed
information in some transitions, particularly towards the relatively close high-mass star-forming
region Orion~KL. \citet{2012ApJ...757L...1H} (and see also the erratum
in \citealt{2014ApJ...797L..35H}) detected the $v_2=1$ line at 
232.7-GHz towards Orion~KL; previously this maser transition was known only
in evolved star envelopes. The 232.7-GHz emission was clearly associated with
Source~I, but the maser distribution was not resolved. The 232.7-GHz spectrum was more
similar to that of 22-GHz water masers than to that of SiO masers. 325-GHz
masers in Source~I were imaged with the \textit{SMA} by \citet{2012IAUS..287..184N}. These
masers trace a collimated outflow, with rotation, and probably sample similar pumping conditions
to 22-GHz masers. Emission at 321\,GHz towards Orion Source~I may be of maser origin
and appears to sample similar conditions to 43-GHz SiO masers \citep{2014ApJ...782L..28H}.

The \textit{HERSCHEL} spacecraft has much poorer spatial resolution than \textit{ALMA}, but it
has the advantage of observing free from atmospheric absorption that becomes
increasingly problematic at frequencies $>$500\,GHz. The 620.7-GHz (5$_{3,2}$-4$_{4,1}$) transition was
detected as a maser by \citet{2013ApJ...769...48N} towards Orion~KL, Orion~S and
W49, with clear variability over 2\,yr in the last source. The 620.7-GHz line appears to
be pumped under physical conditions that are a subset of those that pump 22-GHz masers.
The same \textit{HERSCHEL} instrument (HIFI) was used by \cite{2014A&A...567A..31J} to study
polarization and variability properties of the 620.7-GHz masers in Orion~KL. No
significant polarization was found ($<$2 per cent), but whilst some spectral features at 22-GHz show
strong polarization (up to 75 per cent), the one closest in velocity to the 620.7-GHz line
exhibits much weaker polarization of order 10 per cent. 

\subsection{Extragalactic water masers}
\label{ss:extragal}

Extragalactic water megamasers are a well-known phenomenon, and have been employed,
via their arrangement in discs with Keplerian rotation, as tools to measure geometrical
distances to NGC4258 and other galaxies (\citealt{2010ApJ...718..657B} and references therein).
High-frequency water masers are also observed towards some megamaser sources, based
on active galactic nucleus (AGN) excitation. 183 and 439-GHz lines were detected towards NGC3079 by
\citet{2005ApJ...634L.133H}. Emission at 183\,GHz without accompanying masers at
22-GHz has been observed towards Arp~220 \citep{2006ApJ...646L..49C}.
More recent \textit{ALMA} observations detected 321-GHz
masers in the Circinus galaxy \citep{2013ApJ...768L..38H}, but these 
extragalactic sources will not be considered in detail here.

\subsection{Water maser theory and modelling}
\label{ss:theo}

The pumping mechanism of the 22-GHz water maser has long been understood
in general terms, with a model going back as far as the work of \citet{1973A&A....26..297D}.
Put simply, efficient radiative coupling of the `backbone' levels (those that
have the smallest allowed value of $K_a$ for a given value of $J$) preserves
near-Boltzmann populations in these levels, while significant departures from Boltzmann
predictions appear in the populations of non-backbone levels.
The upper level of the maser, $J_{K_a,K_c}=6_{1,6}$, is
a backbone level, but the lower level, $5_{2,3}$, can be drained efficiently into
a lower backbone level, $4_{1,4}$, by collisions and spontaneous emission. A similar
explanation can immediately be applied to other maser transitions where the
upper level is a member of the backbone, namely 183, 325 and 380\,GHz.

Detailed explanations of the 22-GHz maser pump in terms of a dissociative J-type
shock were introduced by \citet{1989ApJ...346..983E}, where masers form in the
cooling post-shock gas following the chemical re-formation of water. Extension
to non-dissociative MHD shocks was carried out by \citet{1996ApJ...456..250K}, and
these models can produce water at a typically higher temperature in the maser
zone than J-shocks. This appears to make
C-type shocks the more likely site of masers from transitions with lower levels $\ga$1000\,K.
More recent work \citep{2013ApJ...773...70H} has revisited
J-shocks as the maser formation zone, establishing likely boundaries in density
and shock velocity for the formation of saturating 22-GHz masers in both C-type
and J-type shocks. It is important to note that the aspect ratios of masers
formed in shocks generally, and J-shocks in particular, are large \citep{2013ApJ...773...70H},
that is they are typically 30-100 times more extended parallel to the shock
than perpendicular to it, and are therefore strongly beamed parallel to the
shock front.

Models involving many levels and transitions of water, including most of the
known water masers, have been used to test understanding of the basic
pumping scheme, and to make predictions of additional masers. The best known
works of this
type are probably NM91 and YFG97. The former used escape probability methods,
whilst the latter used accelerated lambda iteration (ALI) to compute molecular
energy level populations, as does the present work. However, both NM91 and YFG97
used molecular data that has since been superceded, and neither
allowed for far-infrared line overlap. NM91 showed that the
collision plus spontaneous emission pumping scheme, as usually applied to 22-GHz
masers, actually applies also to the great majority of the then known higher frequency
water masers. YFG97 showed that some transitions, for example those
at 437, 439 and 471\,GHz are best pumped by a strong infra-red continuum 
radiation field, for example from dust local to the maser zone.

Studies of extended (15''$\times$15'') 183-GHz maser emission towards Sgr~B2(M)
\citep{2006ApJ...642..940C}
included both LVG (Large Velocity Gradient) and more exact radiative transfer models to obtain the physical
conditions. To explain this extended emission requires a very different
regime from that considered by most models, including NM91 and YFG97. The
dominant pumping process for the 183-GHz masers in this source is radiative,
driven by FIR photons, and covers a density and temperature regime that is
only at $T_K\sim 30$\,K and $n_{H_2}\sim 10^5 - 10^6$\,cm$^{-3}$. The LVG models
demonstrate the switch to the conventional collision-based pumping scheme
at higher temperatures (400-500\,K). The 183-GHz maser in O-rich evolved star
envelopes was modelled by \citet{1999ApJ...525..845G}, using LVG and exact methods.
A significant radiative component to the pumping mechanism was found for stars
of higher mass-loss rates, proceeding through radiative excitation of the first
excited state of the vibrational bending mode.

A model of the water masers in a pulsating AGB star \citep{2001A&A...379..501H}
was one of the first to try and establish the spatial relationship of the 
principal mm-wave and sub-mm masers, and the possibility of their co-existence
with each other and with 22-GHz masers. This model represented the maser clumps
as LVG blocks with physical conditions drawn from a pulsating atmosphere model
of a Mira variable \citep{1988ApJ...329..299B}. The pulsating atmosphere provides
radial shock-waves, attenuating with increasing radius. By $\sim$5\,R$_*$, maser 
emission from many 22-GHz clouds appears consistent with a smoother, radial outflow,
although wind clumping and changes in mass loss rate may lead to shocks at
larger radii. \citet{2001A&A...379..501H} classified the water masers into three groups:
the first, with 321\,GHz as the prototype, lie close to the star (radial extent
2-5\,R$_*$), sampling similar conditions to SiO masers, and emit tangentially with
emission dominated by a few bright spots. Group 2, exmplified by 22- and 325-GHz masers
have a moderate extent, with most emission in the range 2-10\,R$_*$, peaking in the
zone where the motion of the gas is changing from radial pulsation towards
a steady outflow, but we note that 22-GHz masers are found closer
to the star in the model than in observations.
A third group, with 183\,GHz as the prototype, have an even
larger extent (2-20\,R$_*$) with mostly radial emission provided by a very
large number of comparatively weak spots. \citet{2001A&A...379..501H} did not
include any vibrationally excited states of water, and so were unable to make
predictions about the 658-GHz line. An early attempt to model the $(0,1,0)$
vibrational state \citep{1977PASJ...29..669D} did not find this transition to
be inverted, but recent work \citep{2015MNRAS.449.2875N} reproduces the
maser, using the same molecular data as the present work.

Recent ALI modelling of the 22-GHz maser with up-to-date molecular data
\citep{2013JPhCS.461a2009N} reinforces the usual view that a substantial
radiation field in the far- and mid-infra-red is harmful to the 22-GHz
inversion.

Polarization in H$_2$O maser transitions other than 22\,GHz has been
considered by \citet{2013A&A...551A..15P} in the regime where the magnetic
field defines a good quantization axis, but where the Zeeman
rate is much smaller than the inhomogeneous line width. However, we do not
model the Zeeman structure of H$_2$O in the present work.

\subsection{Origins of clumping}
\label{ss:lumps}

In circumstellar envelopes, 22-GHz H$_2$O masers are observed to arise from
clumps of gas with a size that correlates with the stellar radius \citep{2012A&A...546A..16R}.
In the smaller types (for example Miras and semi-regular variables) the cloud
size is apparantly restricted enough to make most masers unsaturated. The
H$_2$O maser zone encompasses the critical range of radii where the radial
velocity passes through the gravitational escape velocity, and periodic
pulsations develop into a smooth wind. In fact, we observe this wind as
a two-phase medium, with warm, dense clumps, suitable for pumping water
masers, enclosed within a more rarefied phase that supports
OH main-line maser emission \citep{1999MNRAS.306..954R}. The density
contrast between the phases has been calculated to be of order $10-100$
\citep{2011A&A...525A..56R}, with the filling factor of the dense clumps
likely to be no more than 0.01.

Consideration of typical main-line pumping
conditions for OH masers (for example \citealt{1994A&A...282..213F}) strongly 
suggests that the more rarefied phase
is also cooler than the H$_2$O clouds. If this is so, the 
denser phase is over-pressured with respect
to the more rarefied phase, and must be unstable unless otherwise constrained,
for example by locally strong magnetic fields. If the clumps are indeed
pressure unstable, they should disperse over the order of a dynamical
time of $t\sim L/c_s$, where $L$ is the cloud scale and $c_s$, the sound
speed in the clump. Expanding this formula, the expected dispersal
time is
\begin{equation}
\tau = 1.98 [\bar{\mu}/(\gamma_{7/2}T_{1000})]^{1/2} L_{AU}\;\;\mathrm{yr},
\label{t_disp}
\end{equation}
where $\bar{\mu}$ is the mean molecular mass in H$_2$ units, $\gamma_{7/2}$
is the ratio of specific heat capacities relative to that for a rigid diatomic 
molecule, and $T_{1000}$ is the temperature in units of 1000\,K. It appears that
an additional constraint mechanism is required, since individual maser clumps
survive for decades (comparable to the crossing time of
the 22-GHz maser shell, \citealt{2012A&A...546A..16R}) even though their maser
brightnesses may fluctuate on timescales of
5\,yr or less.

Theoretically, we can understand a bifurcation of the medium arising in the
SiO zone via thermal and magneto-thermal instabilities 
\citep{1989A&A...209..305C,1994ApJ...433..303C,1993ApJ...418..263N,1998MNRAS.295..970G}. 
Regions of gas that
are well cooled by radiation from rovibrational transitions of CO and H$_2$O
go on to become dense clouds that support SiO and H$_2$O maser emission, and probably also
have a greater abundance of these molecules than their surroundings.
The possible collision of the rarefied and
dense components of the gas at larger radii (the water-maser zone) may cause additional
shocks in some stars, plausibly resulting in the observed linear features
in VY~CMa \citep{2014A&A...572L...9R}. An additional source of shock heating
is certainly attractive, since the radial stellar pulsation shocks 
become much attenuated as they progress through the water maser zone
\citep{1988ApJ...329..299B,2001A&A...379..501H}. Both shock mechanisms however
lead to variability of the water masers on typical timescales of
months to years.

Although an explanation in terms of instabilities, with an associated physical
scale corresponding to an optimum growth rate, may at first appear to be at odds
with the observed scaling with star size \citep{2012A&A...546A..16R}, the
observation and clumping theory may still be compatible. Recent work
(Gray et al. in preparation) on instabilities in a spherical geometry generates
perturbations in terms of spherical harmonics that have a natural radial scaling
as a multiple of the stellar radius.

\section{Observational Possibilities for Water Masers}
\label{s:theo}

There is a long history observing the 22-GHz maser line with many instruments, including
VLBI; other telescopes have been mentioned in Section~\ref{intro}.
At higher frequencies, single dishes and arrays such as \textit{IRAM} and \textit{SMA}
have provided first detections of a number of other bright water masers but in most cases 
are unable to resolve them alone. VLBI instruments such as, for example, the \textit{KVN}
(Korean VLBI Network)
and \textit{GMVA} (Global mm-VLBI)
could be employed to study the 67- and 96-GHz lines if new receivers
become available. However, water masers will be an ideal target for \textit{ALMA},
which can already observe high-frequency cicumstellar
H$_2$O masers interferometrically. There are no current instruments covering
most of the spectral regions between 0.1 and 1.2\,THz that are not covered by \textit{ALMA}
bands. One maser transition that falls outside the \textit{ALMA} bands is at 380\,GHz,
detected by the Kuiper Airborne Observatory \citep{1980IAUS...87...21P}. At frequencies
above the upper limit of \textit{ALMA}, the emphasis is more on detection of
 hitherto unobservable H$_2$O maser
lines, rather than on high-resolution imaging, and \textit{SOFIA} has two useful observing bands
at frequencies above 1\,THz. The availability of a new generation of instruments, but
particularly \textit{ALMA} and \textit{SOFIA}, make it timely to produce fresh models
of the inversions in water transitions at frequencies from 0 to 1.91\,THz.
The capabilities of \textit{ALMA} and \textit{SOFIA}, with particular
attention to maser observations, are considered below.

\subsection{\textit{ALMA} Observations}
\label{ss:alma}

\textit{ALMA} \footnote{almascience.org} will eventually observe in all the atmospheric windows between
31.3 and 950\,GHz, specifically:
31-45, or perhaps 35-52,\,GHz (Band~1), 65-90\,GHz (Band~2), 84-116\,GHz (Band~3), 125-163\,GHz (Band~4), 
163-211\,GHz (Band~5),
211-275\,GHz (Band~6), 275-373\,GHz (Band~7), 385-500\,GHz (Band~8), 602-720\,GHz (Band~9)
and 787-950\,GHz (Band~10). 
Bands 1, 2 and 5 are not yet (2016) available. Taking 30-80\,mas resolution as an example,
and with at least 36 12-m antennas, we achieve, after 
5 minutes observing on-source, a 4-sigma brightness
temperature sensitivity of around 5$\times$10$^4$\,K or better in all
bands, except for a few masers lying in the worst regions for
atmospheric transmission where 10-15 minutes are required.  This assumes a
velocity resolution of ~0.1\,km\,s$^{-1}$ in a maximum velocity span of ~150\,km\,s$^{-1}$
per spectral window (spw). At $\lambda=3$\,mm (Band 3) this velocity resolution requires
the finest normally-available frequency
resolution of 32.278\,kHz in dual polarization (taking on-line Hanning
smoothing into account).  The velocity resolution at longer
wavelengths is broader in proportion to the wavelength, or narrower
resolution is similarly possible at shorter wavelengths, down to 0.01\,km\,s$^{-1}$
in Band 10 (0.3\,mm wavelength), a spectral resolving power of
3$\times$10$^7$.  The correlator was designed to reach 3.8\,kHz resolution 
in single polarization in a limited total bandwidth, but spectral resolution
may in practice be limited by other factors.

\textit{ALMA} can observe multiple spw in different configurations in four
2-GHz basebands, subject to placement rules, allowing several
spectral lines to be observed simultaneously (within a band-dependent maximum span)
and for the stellar continuum to be detected,
facilitating alignment of observations taken at different times.  

In Table~\ref{almatab},
we list the known water maser lines from both water spin species, locating
them within an \textit{ALMA} band where possible. We exclude transitions with frequencies
below 65\,GHz that may fall into the currently unavailable band~1 and the 380.2-GHz
line between bands~7 and 8, and detected from the \textit{KAO} \citep{1980IAUS...87...21P}. Also
included in Table~\ref{almatab} are a number of predicted maser lines from NM91 and
YFG97 that should be reasonably clear of atmospheric extinction. We note that the
transitions at 645.77,645.91,863.84 and 863.86\,GHz are not predicted to be
strong masers in the present work.
\begin{table}
\caption{Known and predicted H$_2$O masers observable by \textit{ALMA}}
\begin{tabular}{@{}lccccl}
\hline
$\nu$ & Transition & Spin & A-value & Band & Known\\
GHz   & $J'_{K'_a,K'_c}-J_{K_a,K_c}$ & o/p & Hz &  & \\
\hline
67.804  & $(4_{1,4}-3_{2,1})^*$   & o & 1.950(-7) & 2 & N      \\
96.261  & $(4_{4,0}-5_{3,3})^*$   & p & 4.76(-7) & 3 & Y(M89) \\
183.31  & $3_{1,3}-2_{2,0}$       & p & 3.65(-6) & 5 & Y(C90) \\ 
232.69  & $(5_{5,0}-6_{4,3})^*$   & o & 4.76(-6) & 6 & Y(M89) \\
268.15  & $(6_{5,2}-7_{4,3})^{**}$ & o & 1.53(-5) & 6 & Y(T10)\\
293.66  & $(6_{6,1}-7_{5,2})^*$   & o & 7.22(-6) & 7 & Y(M06)\\
297.44  & $(6_{6,0}-7_{5,3})^*$   & p & 7.51(-6) & 7 & Y$^a$(K13)\\
321.23  & $10_{2,9}-9_{3,6}$      & o & 6.16(-6) & 7 & Y(M90a)\\
325.15  & $5_{1,5}-4_{2,2}$       & p & 1.17(-5) & 7 & Y(M90b)\\
331.12  & $(3_{2,1}-4_{1,4})^{**}$ & o & 3.38(-5) & 7 & Y$^a$(K13) \\
336.23  & $(5_{2,3}-6_{1,6})^*$   & o & 1.08(-5) & 7 & Y(F93)\\
354.81  & $17_{4,13}-16_{7,10}$   & o & 9.21(-6) & 7 & Y(F93) \\
390.13  & $10_{3,7}-11_{2,10}$    & p & 8.51(-6) & 8 & N \\
437.34  & $7_{5,3}-6_{6,0}$       & p & 2.15(-5) & 8 & Y(M93) \\
439.15  & $6_{4,3}-5_{5,0}$       & o & 2.82(-5) & 8 & Y(M93) \\
443.02  & $7_{5,2}-6_{6,1}$       & o & 2.23(-5) & 8 & Y(M08) \\
470.89  & $6_{4,2}-5_{5,1}$       & p & 3.48(-5) & 8 & Y(M93) \\
474.69  & $5_{3,3}-4_{4,0}$       & p & 4.82(-5) & 8 & Y(M08) \\
488.49  & $6_{2,4}-7_{1,7}$       & p & 1.38(-5) & 8 & N \\
620.70  & $5_{3,2}-4_{4,1}$       & o & 1.11(-4) & 9 & Y(H10) \\
645.77  & $9_{7,3}-8_{8,0}$       & p & 4.62(-5) & 9 & N \\
645.91  & $9_{7,2}-8_{8,1}$       & o & 4.62(-5) & 9 & N \\
658.01  & $(1_{1,0}-1_{0,1})^*$   & o & 5.57(-3) & 9 & Y(M95) \\
863.84  & $10_{8,3}-9_{9,0}$      & o & 9.46(-5) & 10 & N \\
863.86  & $10_{8,2}-9_{9,1}$      & p & 9.46(-5) & 10 & N \\
906.21  & $9_{2,8}-8_{3,5}$       & p & 2.22(-4) & 10 & N \\
\hline
\end{tabular}
Note that transitions surrounded by $()^*$, $()^{**}$ are respectively in the $v_2=1,2$ or
$(0,1,0),(0,2,0)$ excited states (the first and second excited states of the bending
mode). All other transitions are in the vibrational ground state.\\
Transitions in the $(0,2,0)$ vibrational state were observed by \citet{2010ApJ...720L.102T} (T10).\\
Superscript $^{**}$ indicates a $(0,2,0)$ line present in the \citet{2013ApJS..209...38K} (K13).
Supescript 'a' indicates a line not identified as a water line but present in the spectral survey by Kaminski et al.
(2013)(K13).
Codes for other first detections are given in the final column of the table as follows:\\
M89 - \citet{1989ApJ...341L..91M};C90 - \citet{1990A&A...231L..15C}\\
M06 - \citet{2006A&A...454L.107M};M90a - \citet{1990ApJ...350L..41M}\\
M90b - \citet{1990ApJ...363L..27M};F93 - \citet{1993LNP...412...65F}\\
M93 - \citet{1993ApJ...416L..37M};M08 - \citet{2008A&A...477..185M}\\
H10 - \citet{2010A&A...521L..51H};M95 - \citet{1995ApJ...450L..67M}\\
\label{almatab}
\end{table}

\subsection{\textit{SOFIA} Observations}
\label{ss:sofia}

\textit{SOFIA} is an airborne infra-red telescope.
Details of the project, and the observers' handbook, may
be obtained from the observatory web page\footnote{www.sofia.usra.edu/Science/index.html}.
At typical operational altitudes, the telescope is above
99 per cent of the atmospheric water vapour, and atmospheric transmission is typically
above 80 per cent. However, there are are still several frequency bands where the
atmospheric opacity is severely limiting. The 2.7-m main mirror (effective area
corresponds to a 2.5-m instrument) gives a 16\arcsec \,beam at a wavelength of 160\,$\umu$m.

For H$_2$O maser observations, the most useful instrument on board {\it SOFIA}
is the heterodyne spectrometer, {\it GREAT} (German REceiver for Astronomy at Terahertz frequencies). This has, in principle, five spectral
windows in the frequency range 1.2-5\,THz. However, as of Cycle~3, only the lowest frequency pair,
L1 and L2, were suitable for maser observations. Owing to atmospheric opacity,
the L1 window is split into two sub-bands from 1.262 to 1.396 and from 1.432 to
1.523\,THz. The L2 window is 1.800-1.910\,THz. {\it GREAT} feeds a backend system
that features two possible Fourier transform spectrometers. The higher and lower resolution
instruments offer respective frequency resolutions of 44 and 212\,kHz, which are
adequate to resolve maser and thermal lines.

Unlike {\it ALMA}, {\it SOFIA} does not have the spatial resolution necessary to
resolve water masers in circumstellar envelopes and star-forming regions. However,
it does have the capacity to make new detections, and we list below likely target
lines in Table~\ref{sofiatab}. The transitions have been selected from NM91 and
YFG97, and are therefore all in the vibrational ground state.
\begin{table}
\caption{Predicted H$_2$O maser lines observable by {\it SOFIA}}
\begin{tabular}{@{}lcccl}
\hline
$\nu$ & Transition & Spin & A-value & Band \\
GHz   & $J'_{K'_a,K'_c}-J_{K_a,K_c}$ & o/p & Hz &  \\
\hline
1269.3  & $13_{3,11}-12_{4,8}$& p & 5.57(-4) & L1$_{lo}$ \\
1278.5  & $7_{4,3}-5_{5,2}$   & o & 1.55(-3) & L1$_{lo}$ \\
1295.6  & $8_{2,7}-7_{3,4}$   & o & 1.07(-3) & L1$_{lo}$ \\
1308.5  & $8_{4,5}-9_{1,8}$   & o & 2.60(-4) & L1$_{lo}$ \\
1321.8  & $6_{2,5}-5_{3,2}$   & o & 2.33(-3) & L1$_{lo}$ \\
1345.0  & $7_{4,4}-8_{1,7}$   & p & 2.55(-4) & L1$_{lo}$ \\
1435.8  & $9_{4,6}-10_{1,9}$  & p & 3.33(-4) & L1$_{hi}$ \\
1440.3  & $7_{2,6}-6_{3,3}$   & p & 2.30(-3) & L1$_{hi}$ \\
1885.0  & $8_{4,5}-7_{5,2}$   & o & 7.40(-3) & L2       \\
1901.9  & $12_{3,10}-11_{4,7}$& o & 2.75(-3) & L2       \\
\hline
\end{tabular}
\label{sofiatab}
\end{table}

\section{The Model}
\label{s:model}

There are perhaps two viewpoints that one may take when constructing a computational
model of maser action, or for that matter, of any radiation transfer problem. The
first is to select a particular source, and attempt to model its geometry, dynamics
and other physical properties as accurately as possible, while accepting that the
resulting model may have a very limited predictive power when applied to other sources.
The second viewpoint, and this is the one adopted in the present work, is to run a
very large number of fairly simple models, sampling a large parameter space in temperature,
density and velocity fields, while accepting that none of these models apply very
well to any individual source.

The computational results discussed below were produced by the code {\sc mmolg}, an
ALI radiation transfer code with a slab geometry. This code is indeed a direct development
of that used by YFG97, and is based on an original code by \citet{1985JCoPh..59...56S}.
The major improvement of {\sc mmolg} over the code used by YFG97 is the incorporation
of local (from thermal and microturbulent widths) and non-local 
(from bulk velocity gradients) line overlap. The general overlap theory for
{\sc mmolg} is based on \citet{1992LNP...401..431S}, and is also written down
in \citet{mybook}. The line overlap operates within each water spin-species, but does
not couple them radiatively, so that o-H$_2$O and p-H$_2$O were modelled independently.
{\sc mmolg} also benefits from an improved dust model, and more modern molecular
data (see below). An increased number of levels and transitions makes line overlap
more probable: in a typical example from p-H$_2$O, a total of 7341 radiative transitions
were compressed into 6503 overlapping groups or blends, with a maximum overlap
order (largest number of lines in a single group) of 10. The lowest frequency transition
affected by overlap was 1428\,GHz, so overlap does not affect the propagation of any
masers visible to \textit{ALMA}, but may,
under rather extreme conditions, influence some target transitions for \textit{SOFIA}.

The slab geometry used in the current work is
probably not ideal, but is not unreasonable given the linear shock-based observational
features that are frequently identified in the water-maser zones of evolved
stars. Two cloud scales were used to represent, respectively, maser features
in AGB stars and supergiants. For the former model, the total thickness of the
water-containing slab was 4.5$\times$10$^{13}$\,cm (approximately 3\,AU), whilst in the latter
case 2.25$\times$10$^{14}$\,cm (15\,AU) was used. In both cases, this water-containing
slab was divided into 80 logarithmically-spaced layers, except for certain test models
where a different number of layers was used to check invariance. In addition to the
water-bearing slab discussed above, a further 5 layers were added to the side of
the slab further from the observer. These layers have an exponentially decreasing
abundance of water, and an exponentially increasing continuum opacity for the purpose
of enforcing an optically thick boundary condition on the slab. 

Silicate dust was modelled by means of the optical efficiencies for absorption
and scattering by spherical grains computed by \citet{1984ApJ...285...89D}.  These data are tabulated
by grain radius and wavelength with 81 radius blocks between $r=0.001$\,$\umu$m and $r=10$\,$\umu$m.
Each block is subdivided into 241 wavelength entries between $\lambda =1000$\,$\umu$m and 
$\lambda =0.001$\,$\umu$m. The dust in the present model used the full range of radii
and wavelengths available in the data. The size-spectrum of grains was taken to be a
power law with an index of $-3.5$. The density of the grain material was set to
3300\,kg\,m$^{-3}$, and the mass fraction of dust was 0.01. Dust absorption coefficients
were obtained from the tabulated data and the size spectrum, whilst emission coefficients
were calculated from Kirchoff's law, on the assumption of thermal emission at the
dust temperature, $T_d$, one of the physical variables of the slab (see Section~\ref{ss:cloud} below).
There was no independent stellar radiation field (inconsistent with the optically thick
boundary condition), nor interstellar field incident on the optically thin boundary.

With the dust parameters introduced above, the dust is typically optically thin at
wavelengths of order 1\,mm and longer; it therefore has 
little direct effect at the maser wavelengths. At important wavelengths for radiative pumping, such as
the $\nu_2$ band (6.27\,$\umu$m) and the $\nu_1$ band (2.73\,$\umu$m), the dust is generally
optically thick, with optical depths in the typical range 10-1000.

A severe defect of the slab geometry with respect to inverted transitions is that the
maser depth parallel to the slabs is, in principle, infinite. We avoid this problem by
adopting the same \textit{ansatz} that was used in YFG97.

\subsection{Molecular Data}
\label{ss:molecdat}

Energy levels, statistical weights, electric dipole transition frequencies, Einstein A-coefficients and
collisional rate coefficients were obtained as monolithic data files in the `RADEX' format
from the Leiden Database for Molecular Spectroscopy\footnote{http://home.strw.leidenuniv.nl/~moldata/H2O.html} \citep{2005A&A...432..369S}. 
Within these files, one for each spin species of H$_2$O,
additional comment lines credit the energy level data to \citet{2001JPCRD..30..735T}
and the Einstein A-values to \citet{2006MNRAS.368.1087B}.
Collisional data is divided between two collisional partners: H$_2$ molecules and free electrons \citep{2008A&A...492..257F}.
The collisional rate coefficients are supplied for 11 kinetic temperatures: 200,400,800,1200,1600,2000,2500,3000,3500,4000
and 5000\,K. The free-electron collisional data was not used in the present work, since no electron abundance or
ionization fraction was specified. In data tables associated with \citet{2011MNRAS.418..114I}, we note that the
electron density is unlikely to exceed 10$^{-5}$ of the total number density in the SiO maser zone, and
$n_e<2$\,cm$^{-3}$ at a distance comparable to the outermost H$_2$O maser regions \citep{1998MNRAS.297.1151S}.
Collision data for both spin species of H$_2$O and molecular hydrogen
do not specify the spin species of H$_2$ used; in fact they are an average of ortho- and
para-H$_2$ contributions, assuming the thermal abundance ratio of 3 o-H$_2$ to 1 p-H$_2$. This
is considered adequate for temperatures above 200\,K (the lowest tabulated
by \citealt{2008A&A...492..257F}). The effect of changing sets of rate coefficients in numerical calculations
has been considered by \citet{2013A&A...553A..70D}, who concluded that, in models of H$_2$O masers, the
likely uncertainty introduced into maser optical depths is a factor of 2 in the worst cases.

The collisional rate coefficients that are the most uncertain are the vibrationally inelastic
sets, but this uncertainty is mitigated by the fact that the collisional de-excitation rates
at moderate number densities around 10$^9$\,cm$^{-3}$ are very small: 0.0013, 0.0023 and 0.0008\,Hz
respectively for $(0,1,0)-(0,0,0)$, $(0,2,0)-(0,1,0)$ and $(0,2,0)-(0,0,0)$ at $T_K=$300\,K from data
in \citet{2008A&A...492..257F}. These rates would be larger at higher kinetic temperatures of course,
but even at 3000\,K, the largest value considered in the present work, they should be compared to typical
radiative decay rates of 0.2-25\,Hz.

Energy levels (411 for o-H$_2$O and 413 for p-H$_2$O) are ordered strictly by energy within each file, and
therefore include levels from several vibrational states: in addition to the vibrational ground state,
there are levels from the first two excited states of the bending mode, and from the first excited
state of each of the stretching modes. There are a number of notational difficulties relating
to the RADEX data files, at least in the versions used in the present work. These manifest themselves
when attempting to identify transitions from the model with equivalents from the JPL
database\footnote{http://spec.jpl.nasa.gov/ftp/pub/catalog/catform.html}. Details are deferred
to Appendix~\ref{app_filehack}.

We did not consider 
quasi-resonant energy transfer in H$_2$/H$_2$O collisions, as described in
\citet{2014AstL...40..425N}. This effect arises from transfer of rotational energy
in H$_2$ molecules to H$_2$O when the rotational populations of the hydrogen
molecules do not follow the Boltzmann distribution. Significant effects on at least
some H$_2$O inversions are predicted, particularly at high density
($n_{H_2} > 10^9$\,cm$^{-3}$). 

\subsection{Water spin species}
\label{ss:spinspec}

Water exists in two nuclear spin species: o-H$_2$O with a nuclear spin of 1, and
p-H$_2$O (spin 0). It is consisdered most unlikely that nuclear spin conversion between
these species can occur in the gas-phase interstellar medium
\citep{2012PhRvA..85a2521C}. However, the presence of other molecules or complexes, as could be
found, for example, on a grain surface, could make interconversion possible on
a timescale of minutes \citep{2011JPCA..115.9682S}.
In thermodynamic equilibrium, and for kinetic
temperatures above $\sim$50\,K, the abundance ratio for the water spin species should
take the value $n_o/n_p=3/1$, where $n_o$, $n_p$ refer respectively to the abundances
of o-H$_2$O and p-H$_2$O. By treating o-H$_2$O and p-H$_2$O separately, the present
work avoids setting any specific abundance ratio. However, the discussion above may
help when attempting to match the results described below to observed intensity
ratios between maser lines of o-H$_2$O and p-H$_2$O.

\subsection{Cloud Structure}
\label{ss:cloud}

The slabs used in the {\sc mmolg} code are considered uniform in all physical variables
over the main part of the slab, usually comprising 80 depth points. The exception
is the bulk velocity, which may vary with depth. In the 5 boundary slabs, the
bulk velocity is always constant, but the overall density is exponentially
increased (with a constant mass fraction of dust) to increase the continuum
opacity and enforce an optically thick boundary, whilst an exponential decrease
in the water abundance removes any line contribution to the radiative transfer.
The optically thick boundary condition allows the radiative diffusion approximation
to be used to solve the radiative transfer equation in the final slabs. The effect
of changing this condition to a symmteric type was investigated in YFG97 and found to
produce little difference in the final answers whilst significantly increasing the 
computation time. We therefore adopted the optically thick boundary as in YFG97.
 
The physical variables used in the models are the kinetic and dust temperatures,
respectively $T_K$ and $T_d$, the number density of H$_2$, $n_{\mathrm{H}_2}$, the
fractional abundance of H$_2$O, $f_{\mathrm{H_2O}}$, the bulk velocity, $v$, and the
microturbulent velocity, $v_T$. The kinetic and dust temperatures are independently
varied between jobs, so the model does not compute a self-consistent $T_d$ from
a combination of the interactions with the radiation field and collisions with gas molecules.
The fractional abundance of H$_2$O refers to either o-H$_2$O or p-H$_2$O, depending 
on the set of molecular data used in the particular job. The microturbulent
velocity is used simply as a temperature-independent line-broadening parameter, and
is added in quadrature to the thermal line width derived from $T_K$.

All physical variables have been changed between jobs, even if only for a 
small number of tests. In Table~\ref{phys_var} we show, for each variable,
a standard, or typical, value, and, for those quantities that were varied
routinely, a range of variation. When a range is given, the `standard' value
is equal to the median of the range for temperatures, and 
ten to the power of the median of the
common logarithm for densities. The number of steps is given for all
variables that were routinely changed between jobs; multiplying all the
step numbers together yields the number of jobs run for each slab thickness
and spin species.
\begin{table}
\caption{Values and ranges of physical variables}
\begin{tabular}{@{}lccc}
\hline
Variable & Standard & Range & Steps \\
\hline
$T_K$ & 1750\,K & 300-3000\,K & 30 \\
$T_d$ & 1025\,K & 50-2000\,K  & 14 \\
$n_{\mathrm{H}_2}$ & 10$^9$\,cm$^{-3}$ & 10$^7$ - 10$^{11}$\,cm$^{-3}$ & 13 \\
$f_{\mathrm{H_2O}}$ & 3$\times$10$^{-5}$ & tests only & - \\
$v$  & 0.0\,km\,s$^{-1}$ & -5.0 - +5.0\,km\,s$^{-1}$ & 9 \\
$v_T$ & 1.0\,km\,s$^{-1}$ & tests only & - \\
\hline
\end{tabular}
\label{phys_var}
\end{table}

Maser depths (negative optical depths) are calculated along the direction
perpendicular to the model slabs, and therefore give the smallest possible value.
Maser depths for modest angles away from this diection may be calculated by
dividing by the cosine of the off-axis angle.

\section{Results of Computations}
\label{s:comput}

Results were drawn from a total of more than 30000 jobs for each spin species
of H$_2$O in each of the models (supergiant or AGB slab scale). There was therefore
a vast amount of raw data, and some rather severe measures were necessary to reduce
this to a volume that is presentable in a paper of this type. Since this work
concentrates on maser lines, we discard data for all transitions that do not
show a minimum maser depth (negative optical depth), in the frequency bin
at zero velocity, somewhere
in the parameter space. We specify different minima for each model (see below)
since the masers in supergiants are most likely strong and saturated, whilst
those in AGB stars are more often weak and unsaturated. The number of transitions
considered here is therefore much smaller than the 320 lines of o-H$_2$O 
and the 311 lines of p-H$_2$O that
lie longward of the short-wavelength limit of \textit{SOFIA} at 156.96\,$\umu$m
(corresponding to 1910\,GHz).

Jobs were run until they achieved a convergence accuracy of $\epsilon=10^{-4}$.
Very few jobs failed to achieve this level, and most of the failures succeeded
on a re-run in which more ALI iterations were allowed. A standard run allowed
a maximum of 250 ALI iterations.

\subsection{VY~CMa and supergiants}
\label{ss_vycma_results}

In this model, we expect the common masers (for example 22\,GHz, 321\,GHz, 325\,GHz
and 658\,GHz) to be saturated, at least based on the
example of 22-GHz masers in S~Per \citep{2011A&A...525A..56R}. We have therefore considered
two possible cut-off values for the maser depth: a value of 3 to obtain information about
all the maser transitions that are likely to be detectable, and a larger
cut-off of 10 for this model in maser depth (negative optical depth) to define a
strong saturated maser. The selected depths correspond to respective amplification factors of
$e^3 \simeq 20$ and $e^{10} \simeq 22000$. A total of 55 o-H$_2$O transitions, and 48 p-H$_2$O transitions
were found at the cut-off of $\tau=3$, falling to 33 o-H$_2$O transitions and 30
 p-H$_2$O transitions
when $\tau=10$ was set. All transitions satisfying $\tau > 3$
are listed in Table~\ref{hugetab} with a marker in column 9 that indicates
whether the transition is likely to be observable by \textit{ALMA}, 
\textit{SOFIA}, or neither of these
telescopes. This initial selection was based on a dust temperature of $T_d=$50\,K, and
a velocity shift of zero, so radiatively pumped transitions and masers requiring
certain line-overlap effects for pumping are initially excluded.

The maser transitions recorded in Table~\ref{hugetab}
include examples of pure rotational transitions, both within the ground vibrational state and
excited vibrational states. All the included vibrational states play host to masers in rotational
transitions. There are also rovibrational maser transitions, most commonly between an upper
level in an excited vibrational state and a lower level in the ground state. However, there
are also examples of rovibrational masers between excited vibrational states.

\begin{figure*}
  \includegraphics[width=150mm,angle=0]{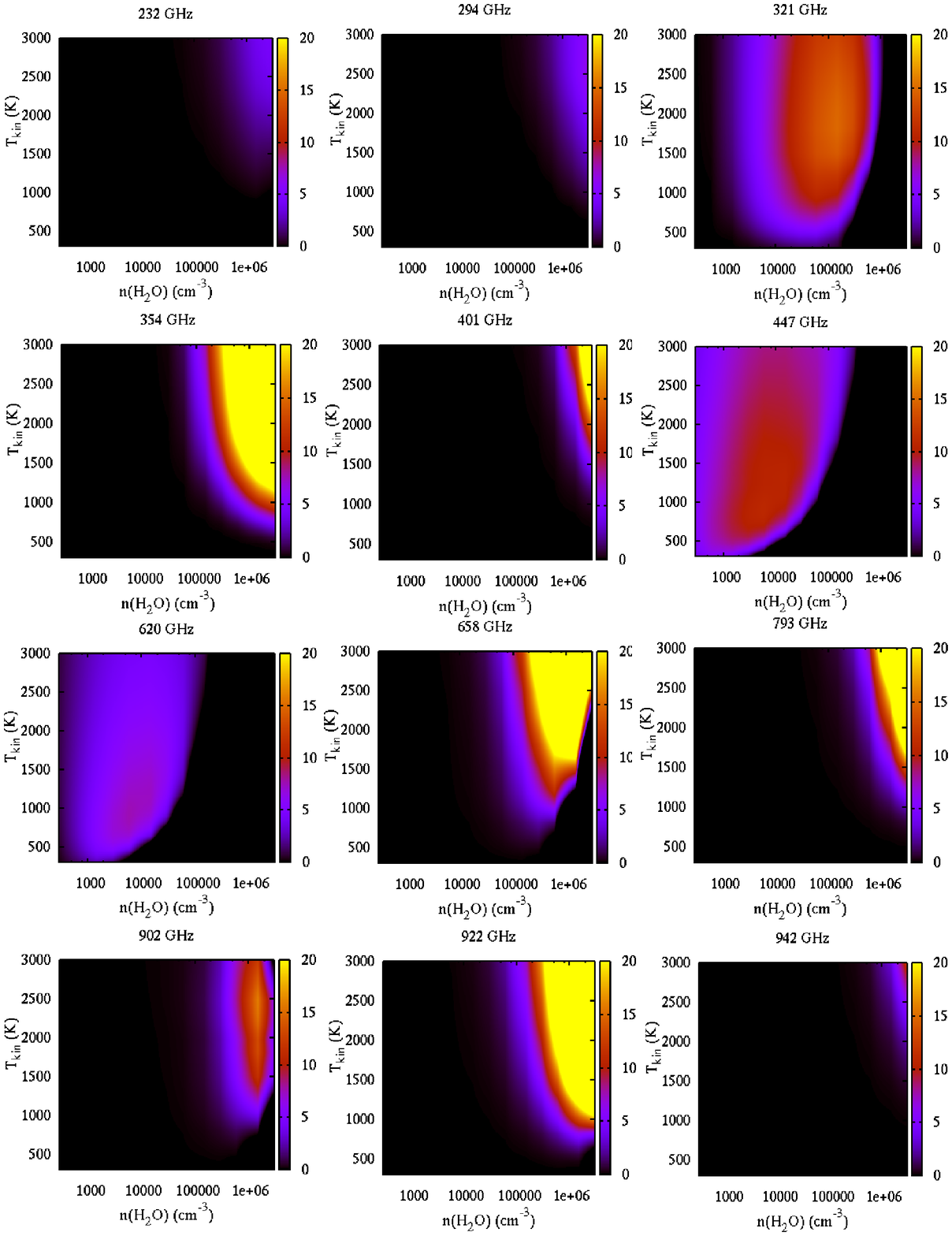}
  \caption{Maser depths (negative optical depths) for the lines of
o-H$_2$O visible to \textit{ALMA} in bands 3-10 as functions of
kinetic temperature and o-H$_2$O number density. The maser depths are
represented by a colour table that is the same for all transitions. These
plots represent a sample of the parameter space of the model
at a dust temperature of 50\,K and zero velocity shift.
Plots of the transitions at 826 and 848\,GHz have been
omitted: they resemble 232\,GHz, but are even more extremely concentrated
to the top-right-hand corner of the plane. Note that black indicates only that the
transition is not inverted; the level of absorption is not displayed. As the water
spin-species are treated independently, conversion of the $x$-axis to n(H$_2$) may
be effected simply by dividing by the fractional o-H$_2$O abundance of 3$\times$10$^{-5}$.}
\label{oh2o_alma_v0_td50}
\end{figure*}

In Figure~\ref{oh2o_alma_v0_td50} we plot the maser depth of 
the lines of o-H$_2$O that are visible to \textit{ALMA}
as a function of the number density of o-H$_2$O and of kinetic temperature. The colour scale
adopted shows black for all regions where a transition is not inverted. These plots are
all for a dust temperature of $T_d =$50\,K, and zero velocity shift through the slab. The
physical conditions used for Figure~\ref{oh2o_alma_v0_td50} also include the standard turbulent velocity
magnitude of 1\,km\,s$^{-1}$. 

The maser transitions plotted in Fig.~\ref{oh2o_alma_v0_td50} may be
divided into two families on the basis of the region occupied by the
positive maser depths in the $n_{\mathrm{H_2O}}$ versus $T_K$ plane. 
One family, comprising
321,447,620,658 and 902\,GHz, have a peak maser depth within the region
plotted, whilst the remainder, and the omitted 826 and 848-GHz lines have
depths that are still rising at the maximum density and $T_K$ covered
by the model. 

These families of transitions are not simply a convenient classification based
on the appearance of plots: they have a physical basis that reflects the
excitation of the transitions involved. The first family all have an upper
energy level less than 2600\,K above ground, whilst the second family have 
upper state energies corresponding to at least 3100\,K, and rising to 7165\,K
in the case of 395\,GHz. The first family are therefore mostly pure rotational
transitions, sited either within the ground vibrational state, or within
the $v_2=1$ excited state, whilst the second family include rotational
transitions in more highly
excited vibrational states and a number of rovibrational transitions.
At $T_d=$50\,K, this is exactly what we expect: upper energy levels are populated
almost entirely by collisions, so $T_K$ must be comparable to the upper-state
energy, and high densities are favoured, increasing the collision rate until
a critical density is reached that thermalises the populations and destroys
any inversion.

Transitions for which the bulk of the inverted region is captured by the figure,
for example 321, 447, 620 and 902\,GHz, show a much steeper decay of the maser
depth on the high-density side than on low-density side. This is evidence of a
well-defined critical density, above which the energy levels in the transition
become thermalized. However, it is apparent that the critical density depends
quite strongly on the kinetic temperature, particularly when this is near the
lower limit of the inverted range. The logarithmic density scale in the plots
compresses this variation: the critical density at 321\,GHz, for example, drops
by a factor of 5 as the kinetic temperature falls from 1500 to 500\,K.

\begin{figure*}
  \includegraphics[width=150mm,angle=0]{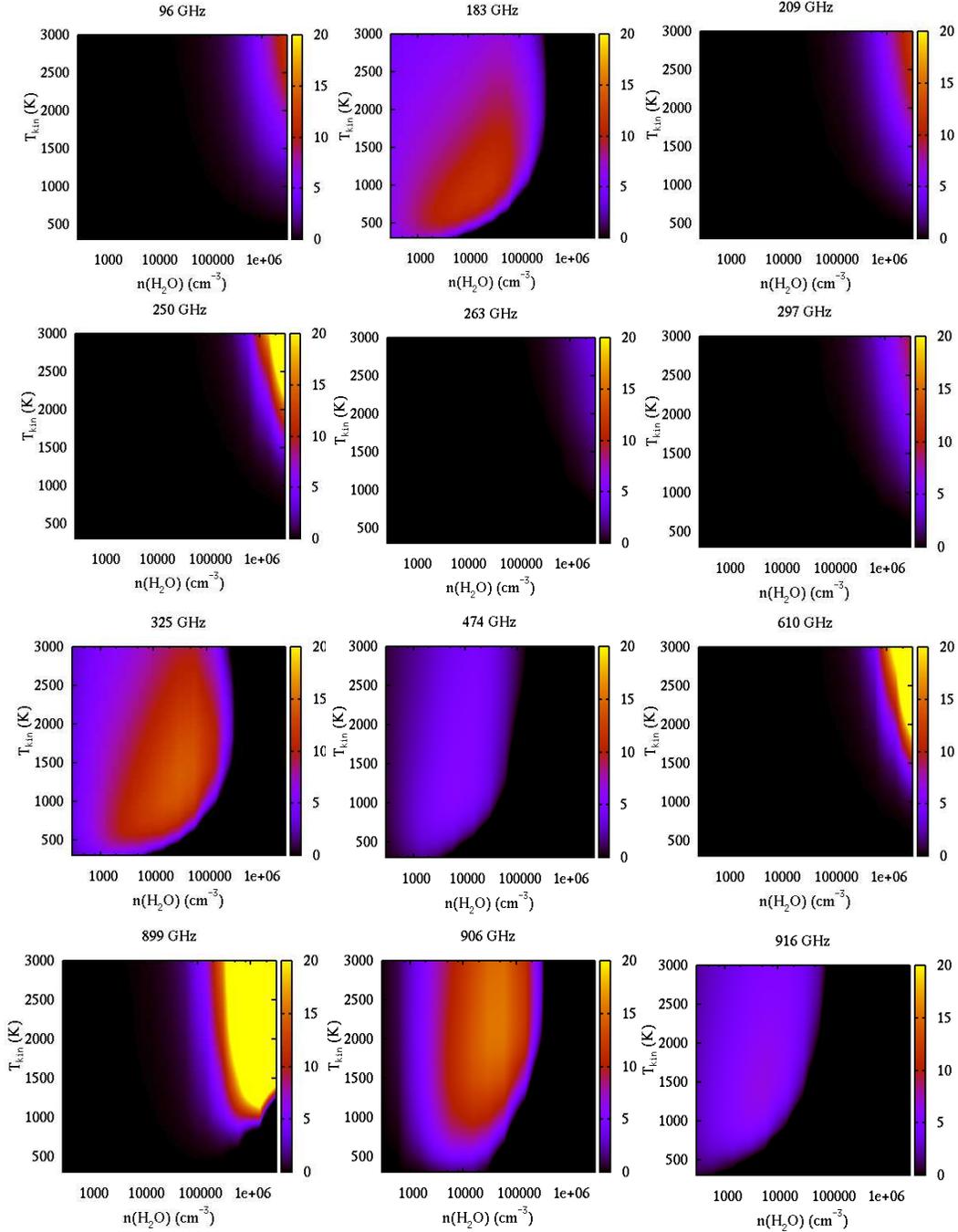}
  \caption{As for Fig.~\ref{oh2o_alma_v0_td50}, but for
p-H$_2$O. Plots of the transitions at 137, 488 and 832\,GHz have been
omitted: they resemble 262\,GHz, but are even more extremely concentrated
to the top-right-hand corner of the plane. As the water
spin-species are treated independently, conversion of the $x$-axis to n(H$_2$) may
be effected simply by dividing by the fractional p-H$_2$O abundance of 3$\times$10$^{-5}$.}
\label{ph2o_alma_v0_td50}
\end{figure*}

In Fig.~\ref{ph2o_alma_v0_td50}, we plot the p-H$_2$O counterpart to
Fig.~\ref{oh2o_alma_v0_td50}. Again, some transitions (137,488 and 832\,GHz) have
been omitted from the plot: they resemble the panel for 262\,GHz, but
are even more concentrated towards the high density and temperature regime; the
most extreme transition in this respect is 488\,GHz.
There is also a similar division into low- and high-excitation families
of lines, with the low-excitation family comprising the 183, 325, 474, 899, 906
and 916\,GHz transitions and the high-excitation family, 96, 137, 209, 250,
263, 297, 488, 610 and 832\,GHz.

\begin{figure*}
  \includegraphics[width=150mm,angle=0]{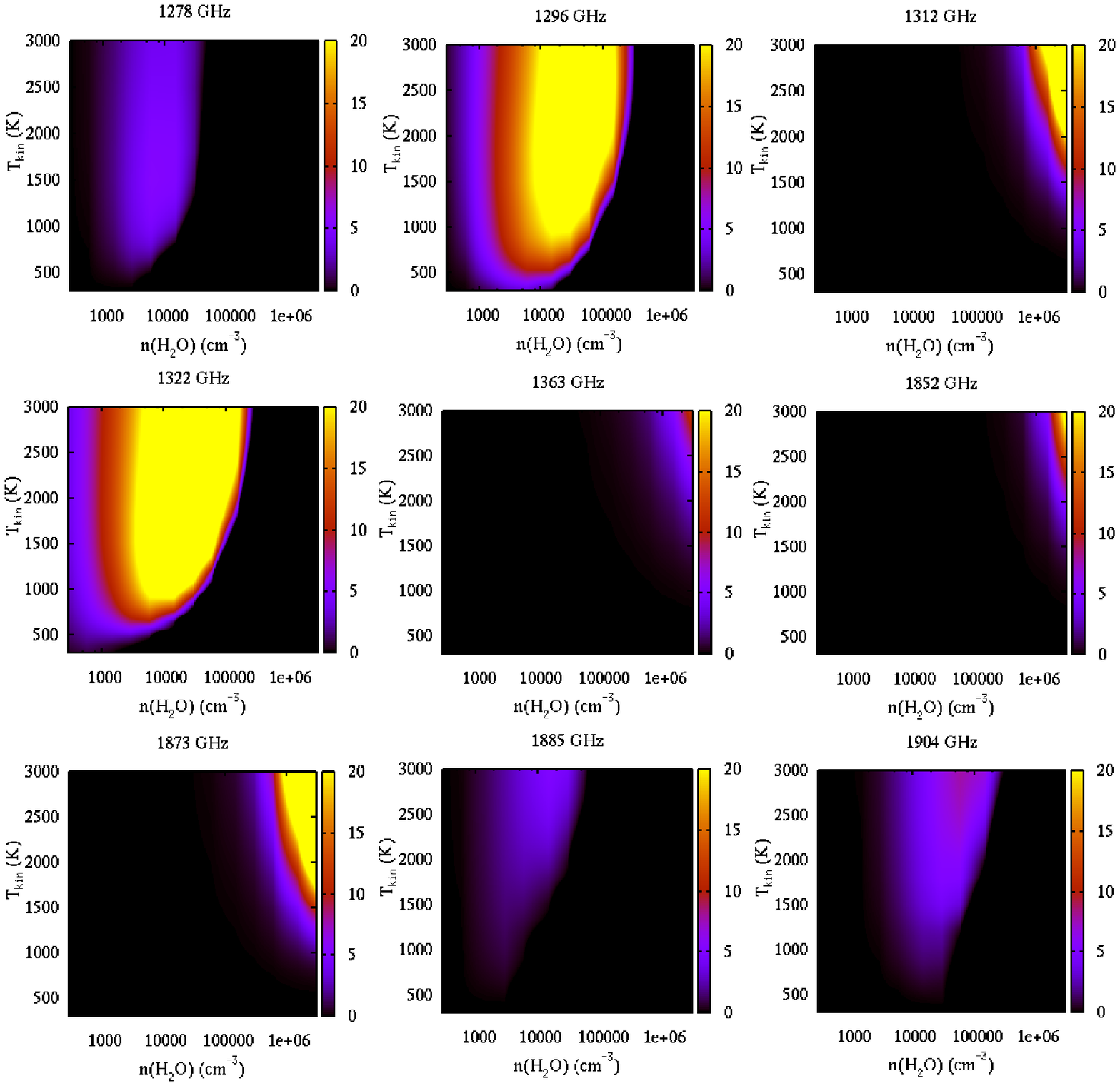}
  \caption{As for Fig.~\ref{oh2o_alma_v0_td50}, but for
o-H$_2$O transitions accessible to  \textit{SOFIA}. Plots of the transitions at 1459,1478,1491 and 1816\,GHz have been
omitted: they resemble 1363\,GHz plot, but the inverted region is more extremely concentrated
to the top-right-hand corner of the plane.}
\label{oh2o_sofia_v0_td50}
\end{figure*}

\begin{figure*}
  \includegraphics[width=150mm,angle=0]{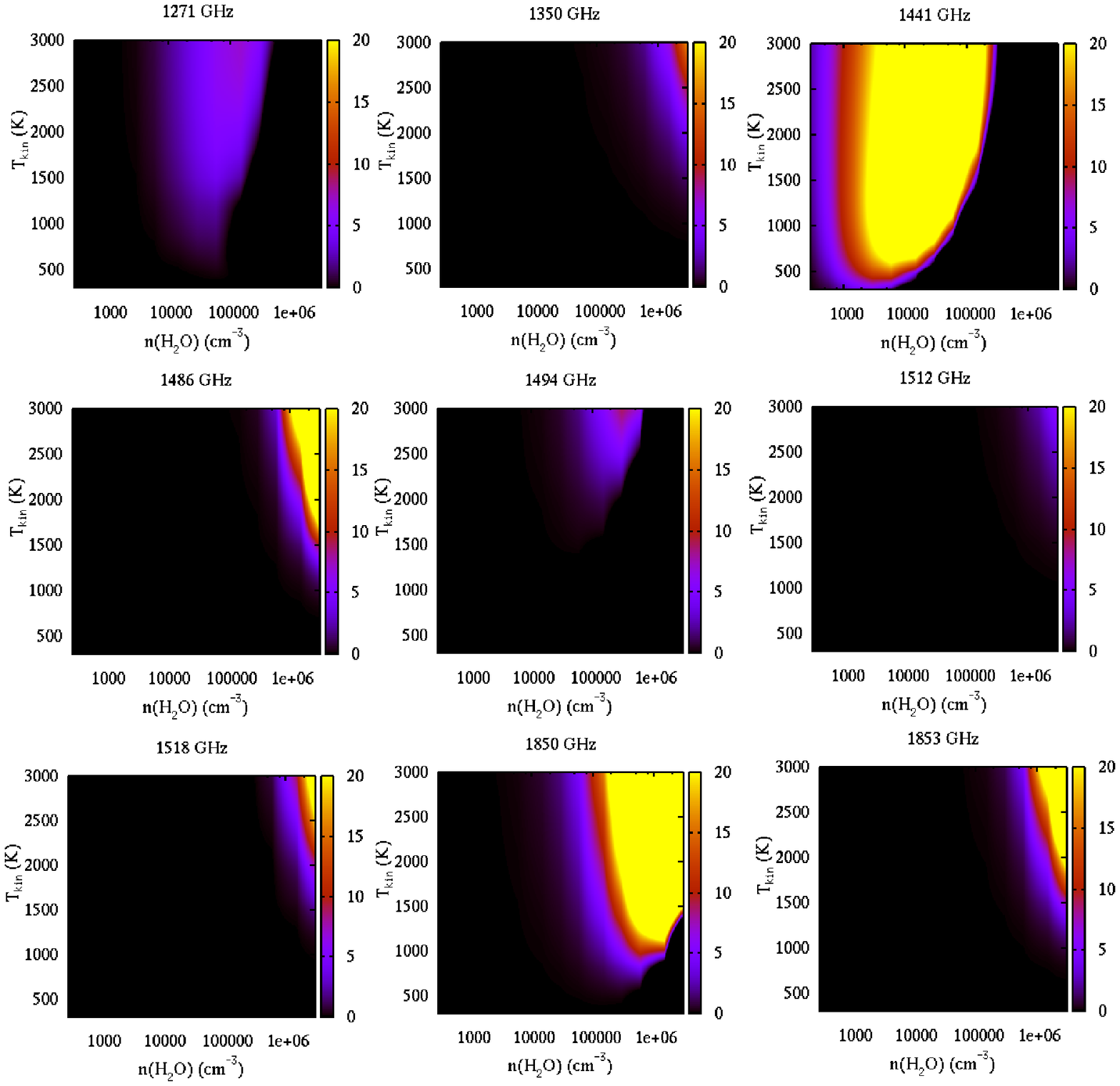}
  \caption{As for Fig.~\ref{oh2o_alma_v0_td50}, but for
p-H$_2$O transitions accessible to  \textit{SOFIA}.}
\label{ph2o_sofia_v0_td50}
\end{figure*}

Fig.~\ref{oh2o_sofia_v0_td50} and Fig.~\ref{ph2o_sofia_v0_td50} are the respective
analogues of Fig.~\ref{oh2o_alma_v0_td50} and Fig.~\ref{ph2o_alma_v0_td50} in the
range of frequencies accessible to \textit{SOFIA}.
The set of transitions visible to \textit{SOFIA} contains a larger fraction of
rovibrational transitions, and rotational transitions from vibrationally excited
states, than the \textit{ALMA} set, and there are consequently more lines without
laboratory-measured frequencies and large frequency uncertainties,
see Table~\ref{hugetab}.

\begin{figure*}
  \includegraphics[width=150mm,angle=0]{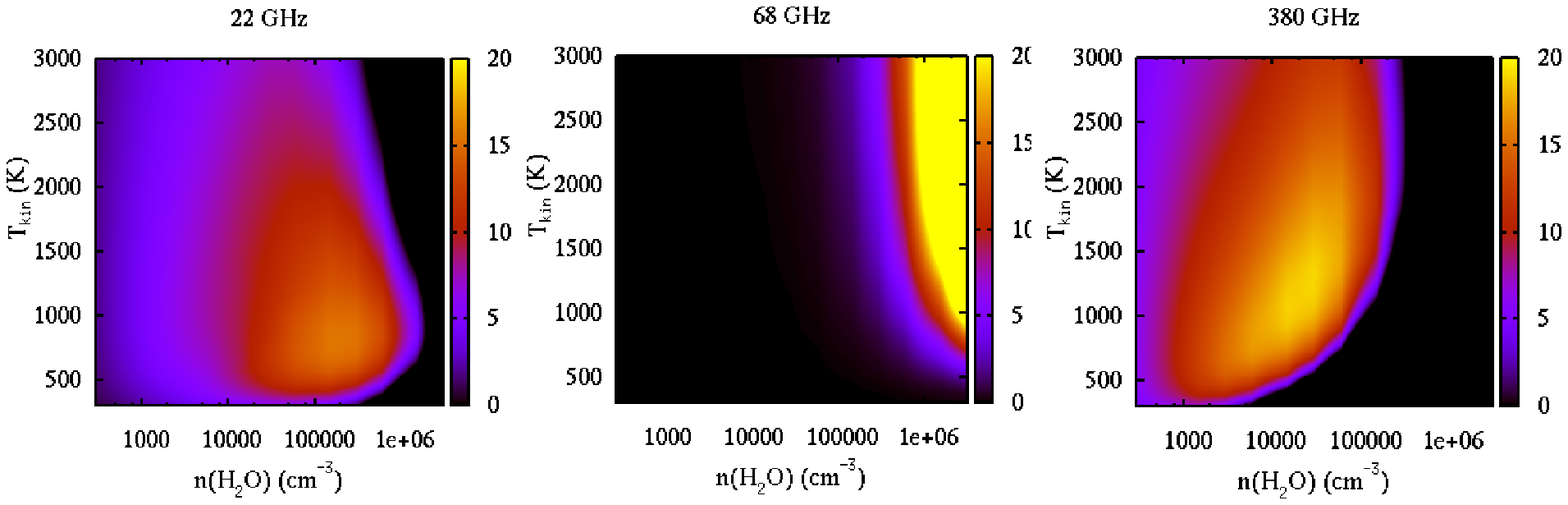}
  \caption{As for Fig.~\ref{oh2o_alma_v0_td50}, but for some
well-studied transitions not accessible to  \textit{SOFIA} or the
cycle-3 configuration of \textit{ALMA}.}
\label{oh2o_other_v0_td50}
\end{figure*}

Lastly in this section we plot, in Fig.~\ref{oh2o_other_v0_td50}, maser depths for three other maser lines from
Table~\ref{hugetab}: 22, 67 (or 67.8) and 380\,GHz. These lines are not detectable with
\textit{ALMA} in cycle 3, but 67\,GHz may eventually become accessible when
detectors for band 2 are constructed. The 380-GHz transition is in a position of strong atmospheric opacity,
and is consequently a poor target for ground-based observations. Depth in the 22-GHz transition is plotted, since
this is the most common water maser line; it will not be observable in future with \textit{ALMA} unless the
lower frequency of Band~1 is reduced below the current planned value of 31.3\,GHz.

\subsubsection{Effect of dust}
\label{sss:dust_effect}

The models discussed so far have used $T_d=50$\,K, so the pumping effect of radiation
emitted by the dust is very weak. Here we discuss models where the dust temperature
takes the values $T_d=1025$\,K and $T_d=1400$\,K (approximately the condensation temperature
of several refractory elements, \citep{2011piim.book.....D}), whilst the ranges of $T_K$ and
water number density are the same as in Section~\ref{ss_vycma_results}. The numbers
of transitions found to achieve a maser depth of at least $\tau=3.0$ at $T_d=1025$\,K were
60 for o-H$_2$O and 30 for p-H$_2$O. At the level of $\tau=10.0$, these numbers reduced,
respectively, to 31 and 20. Of the 60 o-H$_2$O lines recorded at $\tau=3.0$, 19 were
not already recorded at $T_d =50$\,K. The corresponding figure for p-H$_2$O was 18.
These lines, strongly inverted  only at the higher dust temperature, require a significant radiative
contribution in their pumping. They have been added to Table~\ref{hugetab}, but
may be easily identified by the frequencies printed
in an Italic font. The effect of increasing the dust temperature is to 
progressively destroy the inversions in most of the prominent maser transitions found
at $T_d=50$\,K.

\begin{figure*}
  \includegraphics[width=150mm,angle=0]{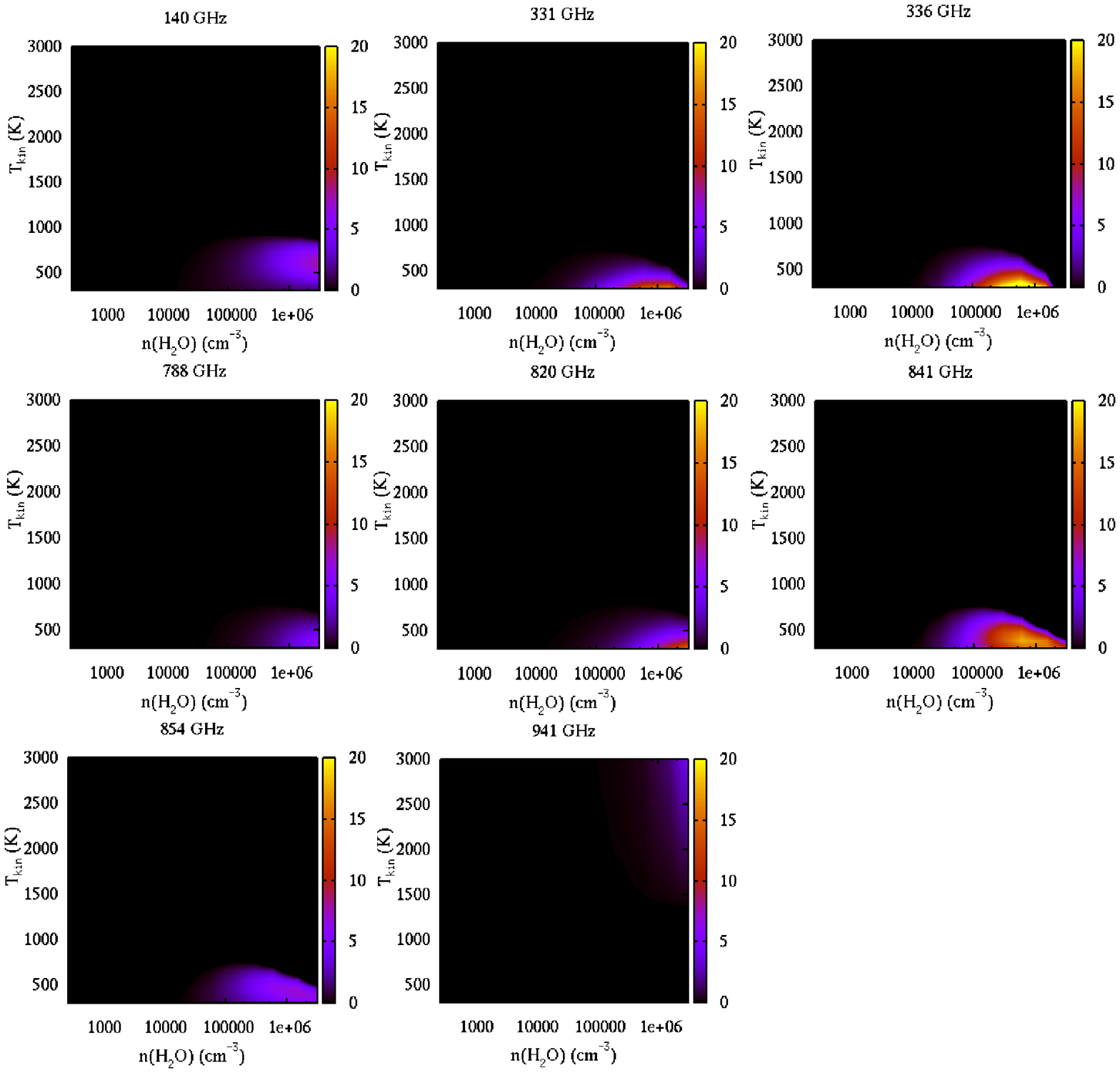}
  \caption{As for Fig.~\ref{oh2o_alma_v0_td50}, but for transitions visible to \textit{ALMA} that were
found to be significantly inverted (maser depth $\geq$3) at a dust temperature of
1025\,K, but not previously found at $T_d=50$\,K. These transitions require a significant
element of radiative pumping.}
\label{oh2o_alma_v0_td1025}
\end{figure*}

We plot the maser optical depths in the density/kinetic temperature plane for the additional
o-H$_2$O transitions visible to \textit{ALMA} in Fig.~\ref{oh2o_alma_v0_td1025}; a similar
plot for the \textit{ALMA} p-H$_2$O lines is presented in Fig.~\ref{ph2o_alma_v0_td1025}.
The locus of inversion for many of the transitions in these plots is notably different
from most transitions at $T_d=50$\,K. The sole exception is 941\,GHz, which resembles the
very high temperature and density family found at $T_d=50$\,K, for example 294\,GHz
in Fig.~\ref{oh2o_alma_v0_td50}. 
The remaining transitions have an inverted zone that is concentrated towards the bottom
right-hand corner of each plot, corresponding to $T_K < 800$\,K and number densities of
o-H$_2$O above 2$\times$10$^4$\,cm$^{-3}$. In four of the eight transitions the point of
peak maser depth is apparently outside the plotted region at a temperature $T_K < 300$\,K.
At this point, mention should also be made of the 268.15-GHz maser, discovered by
\citet{2010ApJ...720L.102T}, that appears in Table~\ref{almatab}: this transition narrowly missed
our classification as a strong maser at $T_d=1025$\,K, but would pass at higher dust
temperatures, for example 1250 and 1400\,K. If plotted in Figure~\ref{oh2o_alma_v0_td1025}, the
distribution of its maser depth would closely resemble that of 820\,GHz.
A similar story was found for the potential \textit{ALMA} p-H$_2$O transitions that are
plotted in Fig.~\ref{ph2o_alma_v0_td1025}:
\begin{figure*}
  \includegraphics[width=150mm,angle=0]{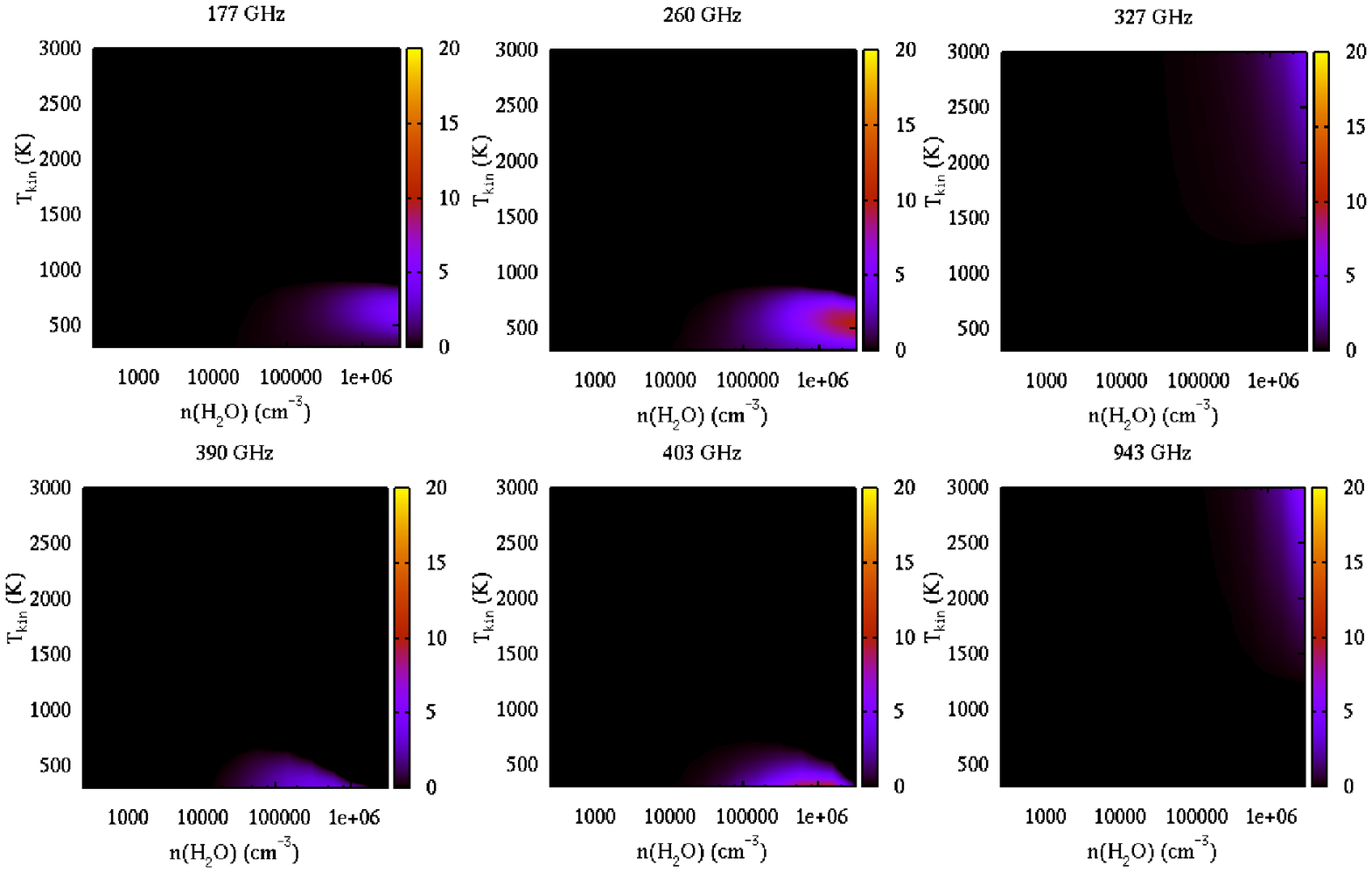}
  \caption{As for Fig.~\ref{ph2o_alma_v0_td50}, but for transitions visible to \textit{ALMA} that were
found to be significantly inverted (maser depth $\geq$3) at a dust temperature of
1025\,K, but not previously found at $T_d=50$\,K. These transitions require a significant
element of radiative pumping.}
\label{ph2o_alma_v0_td1025}
\end{figure*}
the transitions at 327 and 943\,GHz have an inverted region similar to the high-excitation
transitions at $T_d=50$\,K, whilst the remainder resemble the typical radiatively pumped
o-H$_2$O transitions in Fig.~\ref{oh2o_alma_v0_td1025}. The 943-GHz transition is the only
rovibrational example in Fig.~\ref{ph2o_alma_v0_td1025}, and, followed by
327\,GHz, has the highest upper state energy in the figure.

\begin{figure*}
  \includegraphics[width=150mm,angle=0]{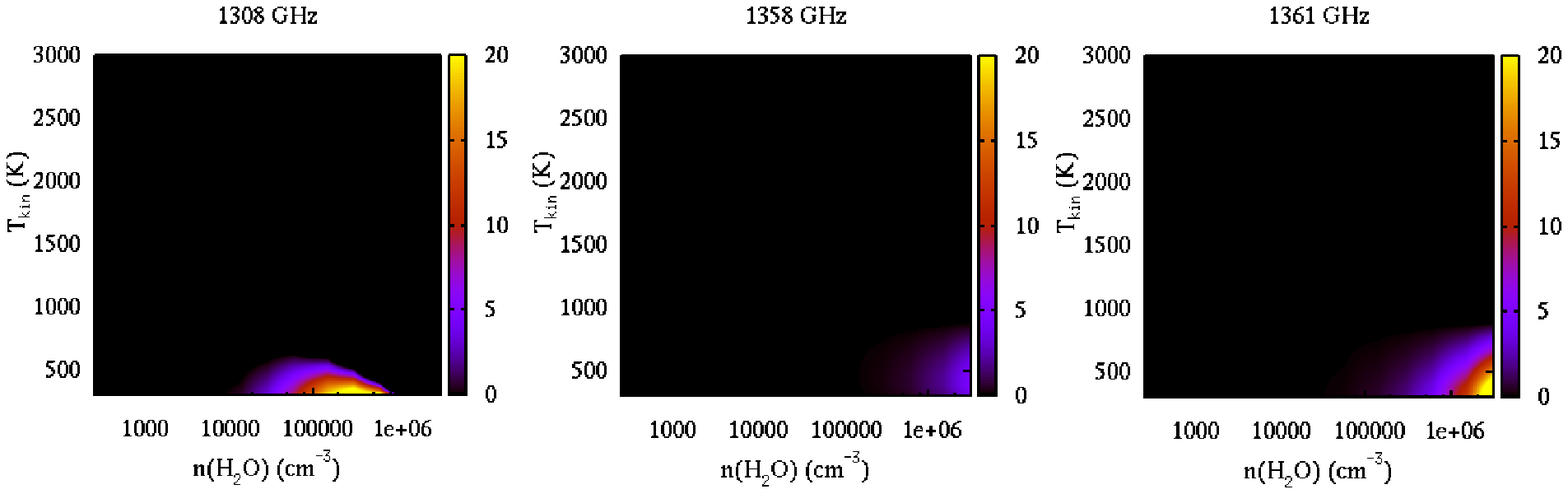}
  \caption{As for Fig.~\ref{oh2o_alma_v0_td1025}, but for transitions visible to \textit{SOFIA}.}
\label{oh2o_sofia_v0_td1025}
\end{figure*}

The new radiatively-pumped o-H$_2$O transitions visible to \textit{SOFIA} all occupy the 
low-kinetic temperature and high density locus of the plane (see Fig.~\ref{oh2o_sofia_v0_td1025}),
following the majority of the radiatively-pumped \textit{ALMA} transitions in this
respect. Although there is a sample of only three displayed transitions, the maser depth in the ground-state
transition at 1308\,GHz occupies a zone of the density/kinetic temperature plane at a
substantially lower density than the 1358 and 1361-GHz transitions, which are both
fully rovibrational: they have $(0,2,0)$ as a common lower vibrational state, with
$(0,0,1)$ (asymmetric stretch) as the upper state of 1358\,GHz and $(1,0,0)$ at
1361\,GHz.

\begin{figure*}
  \includegraphics[width=150mm,angle=0]{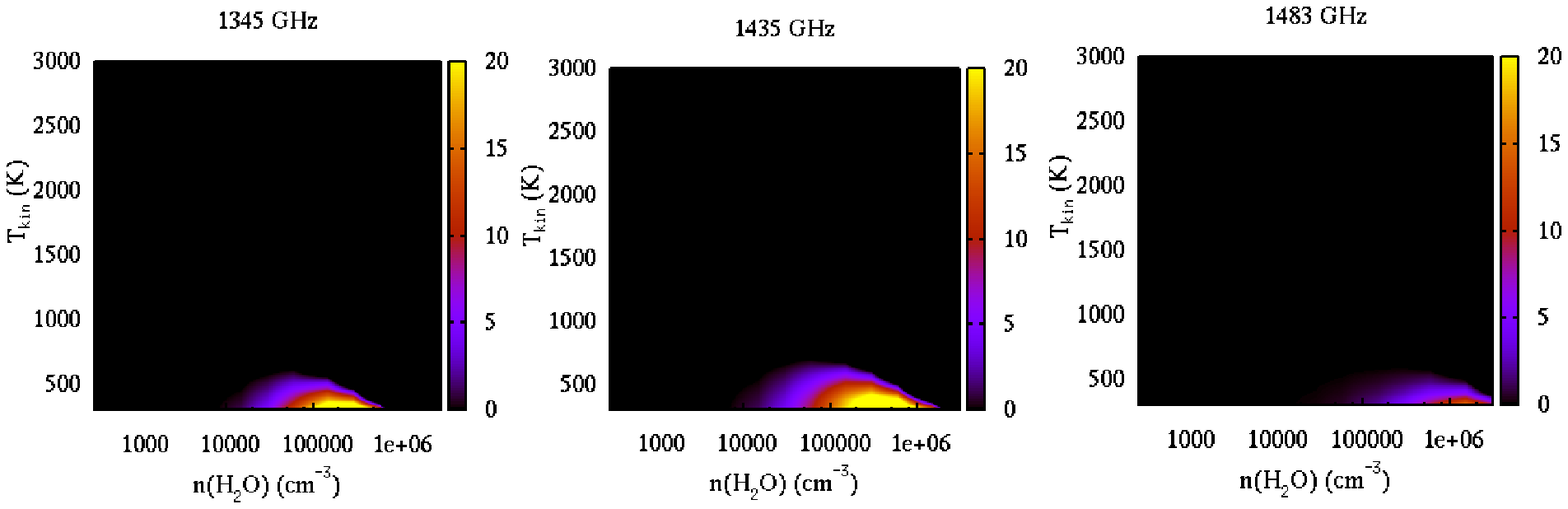}
  \caption{As for Fig.~\ref{ph2o_alma_v0_td1025}, but for transitions visible to \textit{SOFIA}.}
\label{ph2o_sofia_v0_td1025}
\end{figure*}

Maser depths for the radiatively pumped p-H$_2$O transitions visible to \textit{SOFIA} are shown in
Fig.~\ref{ph2o_sofia_v0_td1025}.
Both 1345- and 1435-GHz transitions are from the vibrational ground state, whilst
the 1483-GHz transition is rovibrational, with an upper level 5722\,K above
ground in the asymmetric stretching
mode $(0,0,1)$, transferring to a lower level in $(1,0,0)$. The locus of highest maser depth
for this transition does appear to lie to the higher density side of those for the
ground-state lines, but the effect is less pronounced than in the case of
the o-H$_2$O transitions.

The number of inverted transitions in both o-H$_2$O and p-H$_2$O (above a threshold maser depth
of 10) passes through a minimum at a dust temperature of approximately $T_d=650\,K$. When the
dust temperature reaches 1025\,K, as in the models discussed in detail above, 29 of the original
31 o-H$_2$O
maser transitions (from $T_d=$50\,K) no longer reach the threshold, but have been replaced by
30 radiatively pumped transitions. By the time the dust temperature reaches 1400\,K, all the
original lines have been lost, except 22\,GHz, which has faded and then reappeared at 1250\,K.
Very few maser transitions share this property of having both collisional and radiative pumping
systems (see below). The overall number of inverted transitions continues to rise with dust
temperature. At $T_d=1400$\,K, 25 new o-H$_2$O transitions have appeared above the threshold
that were not present at 1025\,K. 
The situation is similar for p-H$_2$O: at 1025\,K only 2 of the original 23 (at $T_d=50$\,K) transitions
still reach the threshold, and all have been lost at $T_d=1400$\,K. However, the transitions
at 96, 209 and 324\,GHz returned, having a radiative branch to their pumping mechanisms, as
for 22\,GHz (see above). At $T_d=1400$\,K, 39 transitions were found above the threshold maser
depth and, of these, 23 were not previously found at 1025\,K.

We now consider the effect of continuously varying the dust temperature over a
wide range for a restricted number of values of the kinetic temperature. For the moment,
the velocity shift is fixed at zero, and other
variables have their standard values. To represent as much information as possible
graphically, we use a particular symbol to represent each transition, as specified
in the key to each diagram. We then plot the maximum maser optical depth found in
the data as a function of the dust temperature, $T_d$. The extra information, provided
via the colour of each symbol, is the number density of H$_2$O at which the
maximum maser depth was found. 

\begin{figure}
  \includegraphics[bb=80 0 625 790,scale=0.55,angle=0]{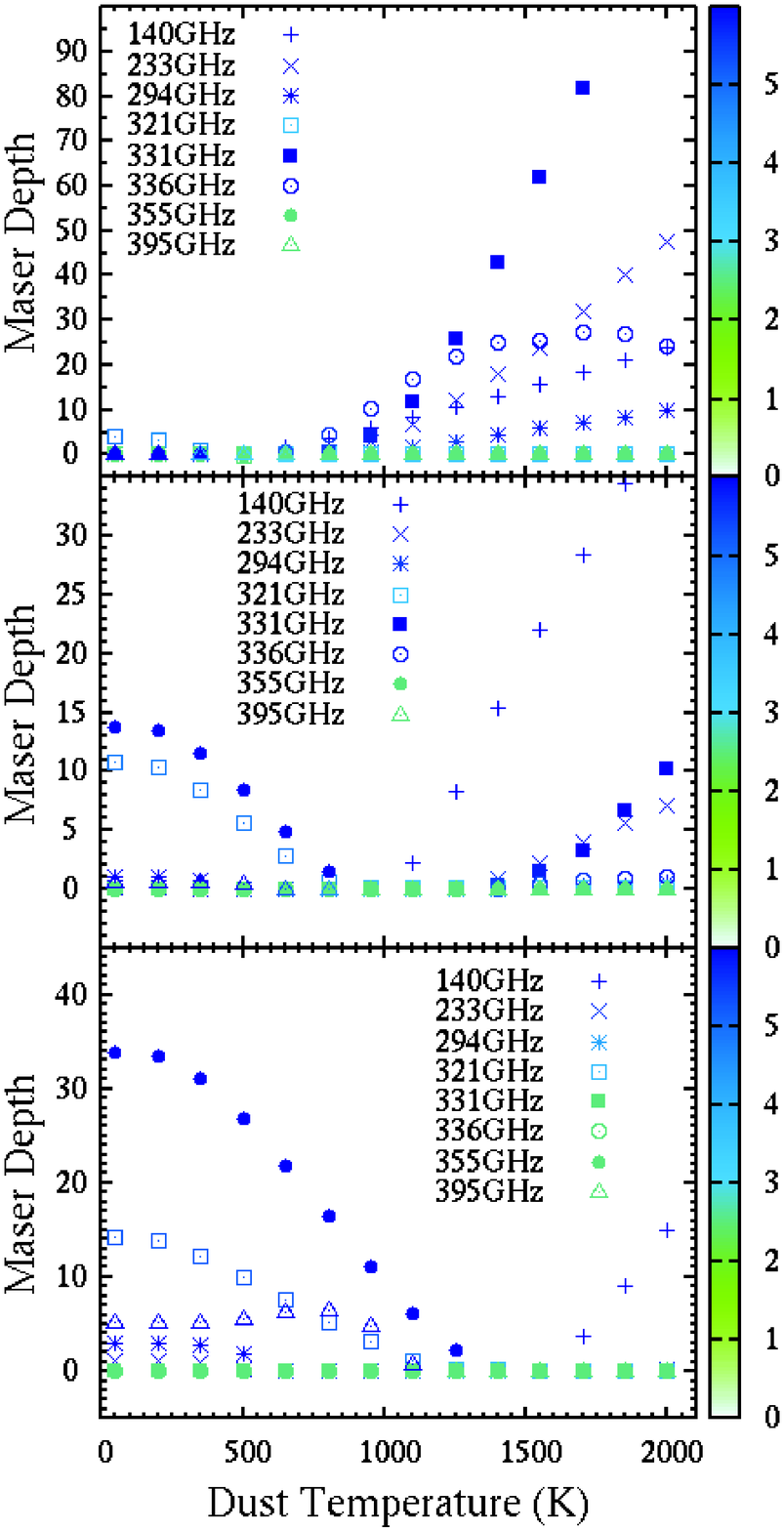}
  \caption{Maximum maser depths (negative optical depths) as a function of dust
temperature for o-H$_2$O transitions with frequencies below
400\,GHz visible to \textit{ALMA}. From top to bottom, the kinetic temperatures
for the three panels are 471.43, 985.71 and 1500.00\,K. Colours of the symbols 
represent the o-H$_2$O number density at which the maximum maser depth was found.
The colour scale is the log to the base 10 of the number density. }
\label{td_alma_lt400}
\end{figure}

In Fig.~\ref{td_alma_lt400} (top panel), we plot the effect of increasing dust temperature on
the inversions of o-H$_2$O maser lines at $T_K=471.43$\,K, with frequencies below 400\,GHz visible to \textit{ALMA}.
There are two families of lines: 321, 355 and 395\,GHz are rather weakly affected by
increasing dust temperature, and the effect is deleterious. These lines also have the
density of peak inversion somewhere below 10$^3$ o-H$_2$O molecules per cubic centimetre.
By contrast, the remaining transitions broadly follow the pattern of 140\,GHz, which
is not inverted at $T_d=50$\,K, but becomes inverted above $T_d=500$\,K, with increasing
maser depth thereafter. These transitions have a strong radiative component to their
pumping scheme. Moreover, as the inversions increase, this family of lines generally
has a density of peak maser depth that is at, or very close to, the maximum available
in the model, that is above 10$^6$\,o-H$_2$O\,cm$^{-3}$. Obviously the highest values
of $T_d$ plotted are physically unlikely, being above the sublimation temperature of most likely
minerals, but are shown to illustrate the effect of a very strong infra-red field.
The lower panels in Fig.~\ref{td_alma_lt400} represent slices through the parameter
space at higher kinetic tempeatures (985.71\,K and at the bottom, 1500\,K). The
main effect of increasing the kinetic temperature is to reduce the inverting effect
of the dust radiation. At 1500\,K, only the 140-GHz transition still shows a significant
increase in maser depth at the higher dust temperatures. Inversions are weakened
or destroyed in the other lines.

\begin{figure}
  \includegraphics[bb=80 0 625 790,scale=0.55,angle=0]{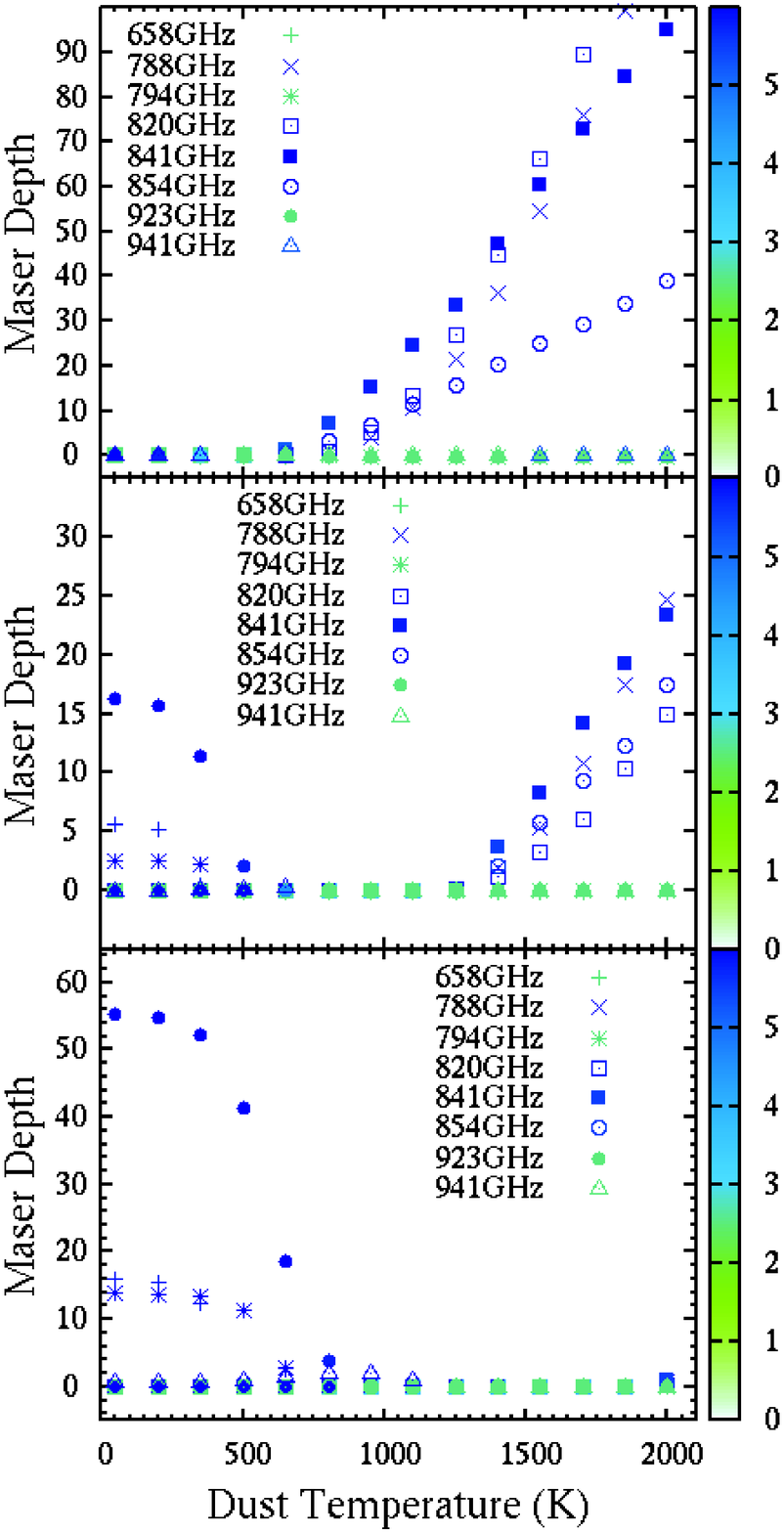}
  \caption{Maximum maser depths (negative optical depths) as a function of dust
temperature for o-H$_2$O transitions with frequencies above
600\,GHz visible to \textit{ALMA}. Kinetic temperatures and symbol colours
are as in Fig.~\ref{td_alma_lt400}. }
\label{td_alma_gt600}
\end{figure}

Figure~\ref{td_alma_gt600} has the same layout at Fig.~\ref{td_alma_lt400}, but is for
maser transitions with frequencies above 600\,GHz. In this case, there are four
strongly radiatively pumped lines at  $T_K=471.43$\,K: 788, 820, 841 and 854\,GHz, with
large maser depths in these lines appearing at $T_d > 700$\,K. The remaining transitions
are weakly inverted at this kinetic temperature, and little affected by dust radiation.
The lines that are radiatively pumped all achieve their maximum maser depths at
densities at, or close to, the largest studied in the model. The same radiatively pumped
group is evident at $T_K=985.71$\,K (middle panel) but with somewhat reduced maser
depths, that increase significantly beyond a dust temperature of $T_d > 1300$\,K. Large
inversions are evident at $T_d<500$\,K at 658, 794 and 923\,GHz, and it is clear that
inversions in these transitions are destroyed by radiation. The lower panel, for which
$T_K=1500.00$\,K shows no transitions being significantly pumped by the dust radiation.
As in the middle panel, large inversions at 658, 794 and 923\,GHz are destroyed by the
radiation, whilst 941\,GHz shows a peak maser depth at $T_d\sim800$\,K. 

\begin{figure}
  \includegraphics[bb=80 0 625 790,scale=0.55,angle=0]{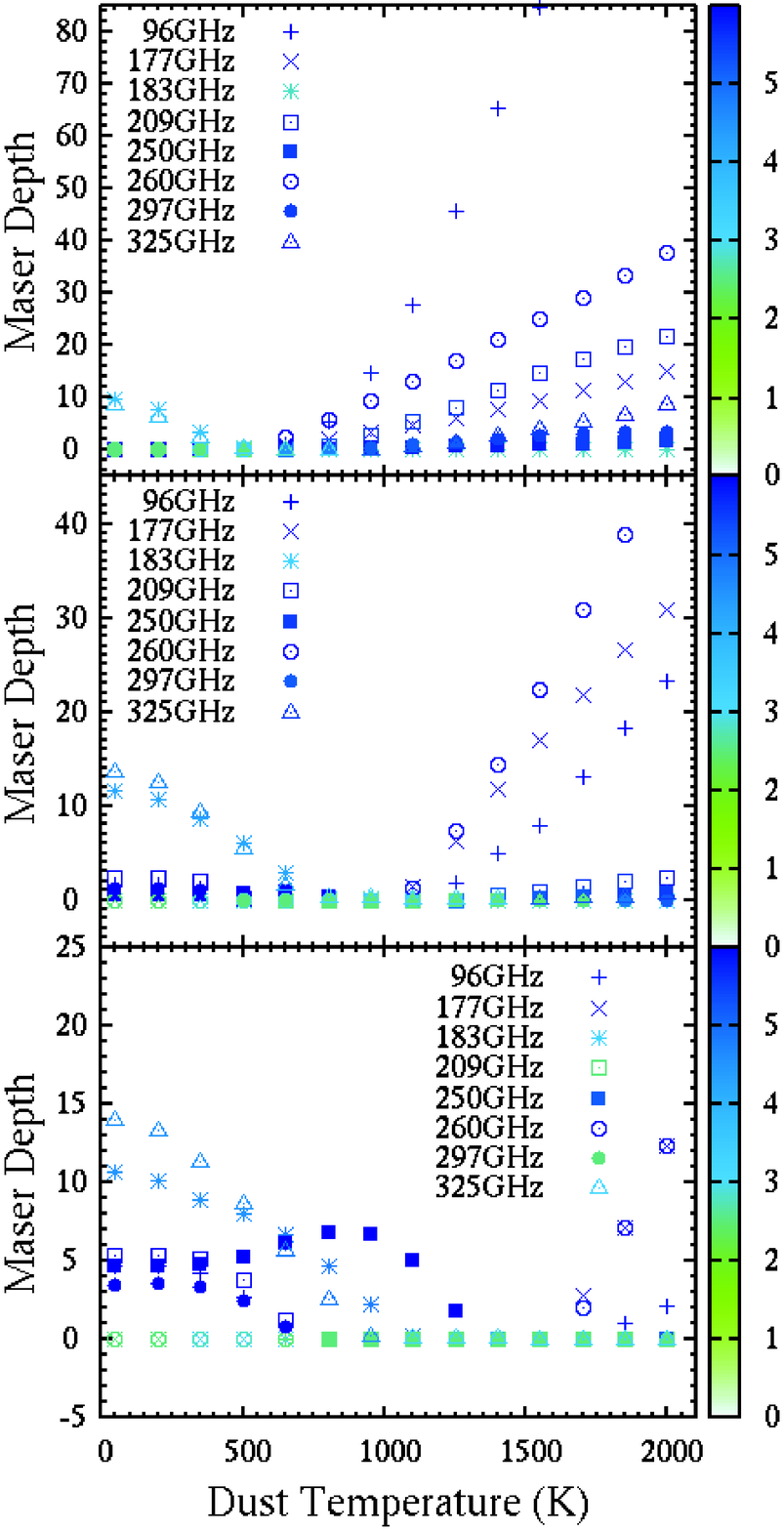}
  \caption{Maximum maser depths (negative optical depths) as a function of dust
temperature for p-H$_2$O transitions with frequencies up to, and including,
325\,GHz visible to \textit{ALMA}. Kinetic temperatures and symbol colours
are as in Fig.~\ref{td_alma_lt400}. }
\label{td_alma_lt350}
\end{figure}

\begin{figure}
  \includegraphics[bb=80 0 625 790,scale=0.55,angle=0]{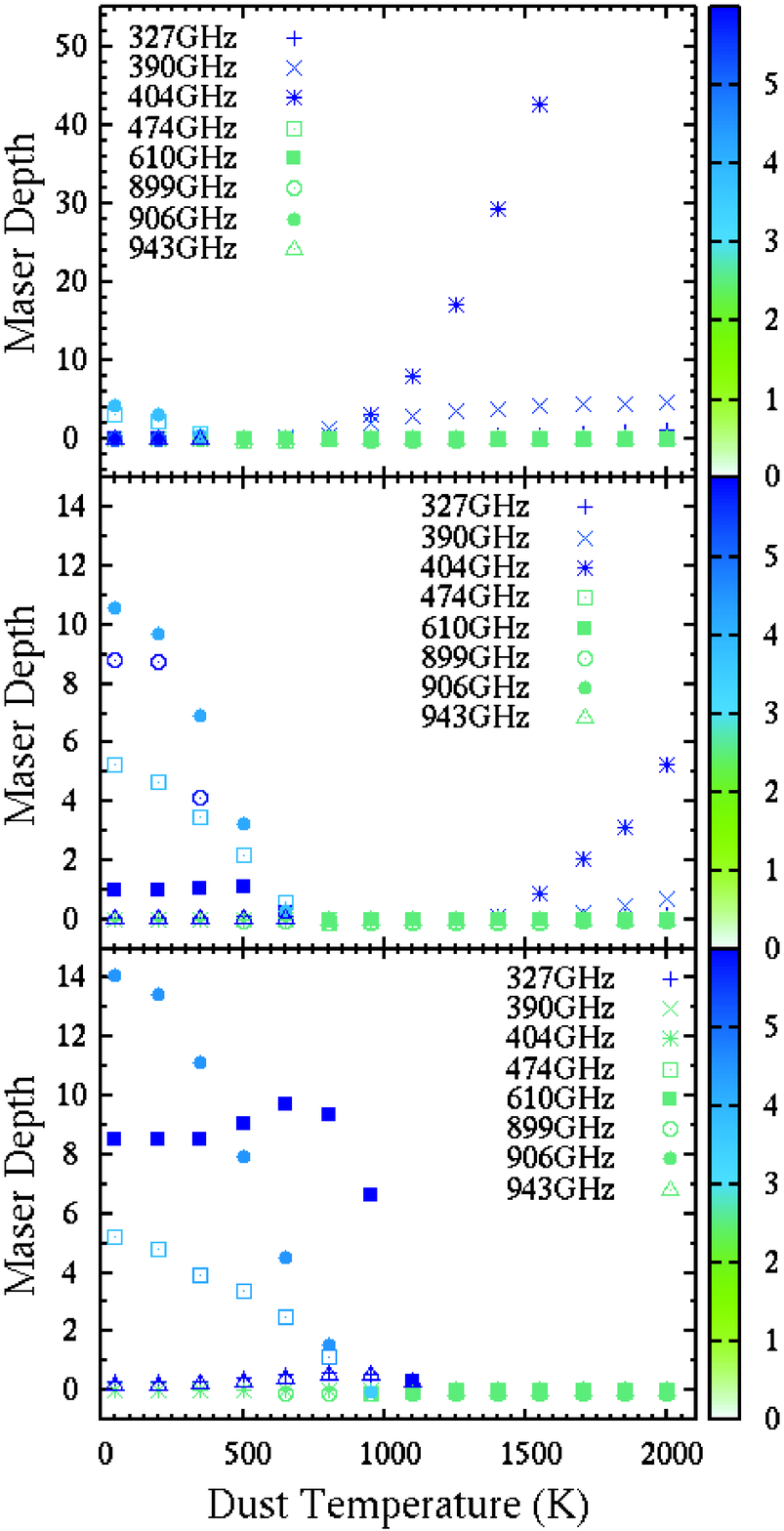}
  \caption{Maximum maser depths (negative optical depths) as a function of dust
temperature for p-H$_2$O transitions with frequencies above
325\,GHz visible to \textit{ALMA}. Kinetic temperatures and symbol colours
are as in Fig.~\ref{td_alma_lt400}. }
\label{td_alma_gt325}
\end{figure}

Maser depths for transitions in p-H$_2$O are shown in Fig.~\ref{td_alma_lt350} and
Fig.~\ref{td_alma_gt325}. The former figure is for transitions with frequencies up
to 325\,GHz, and the latter, for transitions at higher frequencies, beginning with
327\,GHz. At the lowest kinetic temperature of 471.43\,K (top panel, Fig.~\ref{td_alma_lt350}), the 183- and 325-GHz
masers initially decay with increasing dust temperature, reaching negligible inversion
at around $T_d=500$\,K. However, whilst the inversion at 183\,GHz remains close to zero
for higher values of $T_d$, the 325-GHz transition clearly has a radiatively-pumped
branch at high density, since the triangular symbols begin rising again from about 
$T_d=1250$\,K, reaching a maser depth of 8.9 at $T_d=2000$\,K. This behaviour is shared
with the well-known o-H$_2$O transition at 22\,GHz (see below). For 325\,GHz, the radiative
branch only appears at the lowest kinetic temperature: it is not repeated in the
two lower panels, where gain at 325\,GHz only decays with rising $T_d$.
At the lowest
kinetic temperature (top panel) four of the other transitions that appear are radiatively
pumped. Like the o-H$_2$O maser transitions in Fig.~\ref{td_alma_lt400} and Fig.~\ref{td_alma_gt600},
the radiative pumping becomes strong above $T_d \sim 700$\,K, and it is strongest for 96\,GHz,
with decreasing maser depth through 260, 209 and 177\,GHz. The radiatively pumped
lines also have their peak maser depth at, or near to, the
maximum available water density - a property that is also shared with the o-H$_2$O lines.
The remaining two transitions, at 250 and 297\,GHz, have small maser depths over the
full range of dust temperature.

As the kinetic temperature is increased to $T_K=985.71$\,K (Fig.~\ref{td_alma_lt350} middle panel), we see
a reasonably simple modification of the upper panel: the radiatively pumped transitions achieve lower
maser depths, and begin rising beyond a higher dust temperature of about 1000\,K. The 209-GHz
transition, though weakly masing throughout, now behaves somewhat like the 325-GHz transition in the
upper panel, with evidence of both collisional and radiative pumping. At $T_K=1500$\,K, 96, 177 and
260\,GHz remain the only radiatively pumped transitions; the remainder decay with increasing dust
temperature, although 250\,GHz shows a weak maximum in maser depth at $T_d \sim 800$\,K.

In Fig.~\ref{td_alma_gt325}, the higher frequency p-H$_2$O transitions break neatly into
two families: those at 403, 390 and 327\,GHz are radiatively pumped, though for 327\,GHz, the
effect is rather weak. As in previous graphs, the effect of radiative pumping falls with
increasing kinetic temperature, and there is no significant gain left in any of these
lines at $T_K=1500$\,K (bottom panel). The other seven transitions are predominantly collisionally
pumped, and mostly become stronger as $T_K$ rises. Maxima appear in the curves for the 610 and
943-GHz transitions at respective dust temperatures of 650 and 800\,K in the bottom panel.

\begin{figure}
  \includegraphics[bb=80 0 625 790,scale=0.55,angle=0]{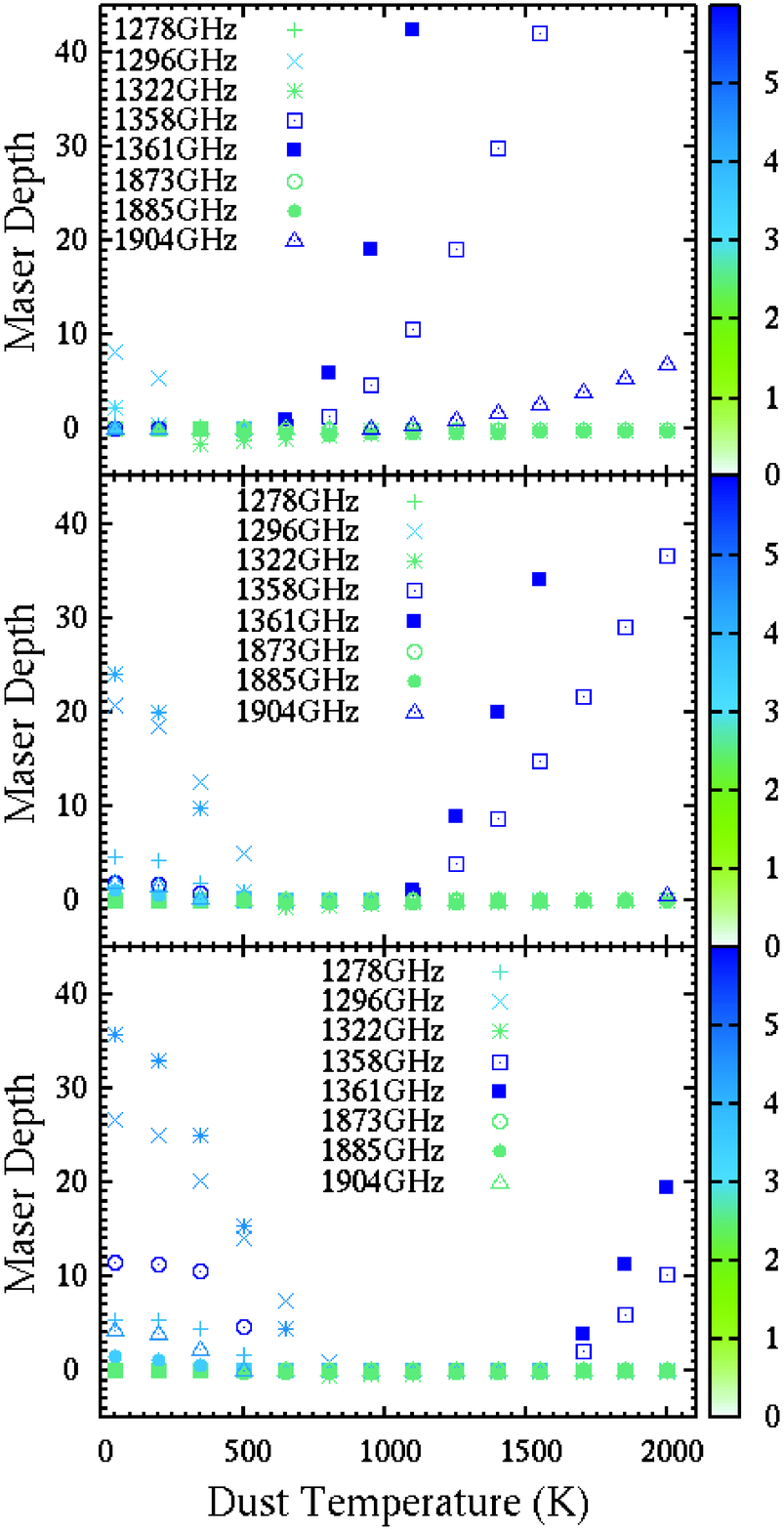}
  \caption{Maximum maser depths (negative optical depths) as a function of dust
temperature for o-H$_2$O transitions
visible to \textit{SOFIA}. Kinetic temperatures and symbol colours
are as in Fig.~\ref{td_alma_lt400}. }
\label{td_sofia_o}
\end{figure}

We now consider transitions visible to \textit{SOFIA}. The general separation into collisionally
and radiatively pumped families of lines is very similar to the behaviour observed at lower
frequencies, and the number of transitions is small enough that the most important can
be studied through one graph each for the o-H$_2$O and p-H$_2$O lines. The response of the
maser depth in eight transitions o-H$_2$O to variation of the dust temperature is plotted in
Fig.~\ref{td_sofia_o}. As for the \textit{ALMA} transitions, results are plotted for three
different kinetic temperatures, with $T_K$ increasing from the top to the bottom panel.

Five of the transitions shown, namely 1278, 1296, 1322, 1873 and 1885\,GHz, are clearly
collisionally pumped: their maser depths fall with increasing $T_d$, become more powerful
with increasing $T_K$, and their maximum depths are often found at a modest number density
of typically 10$^3$-10$^4$\,o-H$_2$O cm$^{-3}$. The two transitions at 1358 and 1361\,GHz
are radiatively pumped, showing maser depth rising with $T_d$, but generally falling with
$T_K$; the maximum maser depth is found at or near
the highest number density in the model. 
Previously noted effects for radiatively pumped lines also apply here: there is a
critical dust temperature for significant maser depth of typically $T_d=700$\,K in the
top panel (where T$_K$=471.43\,K), but increasingly delayed to higher $T_d$ as we progress
towards the bottom panel (higher T$_K$). The 1904-GHz transition displays a more peculiar
behaviour, appearing to be radiatively pumped in the top panel, but collisionally pumped
at $T_K=985.71$ and 1500\,K.

\begin{figure}
  \includegraphics[bb=80 0 625 790,scale=0.55,angle=0]{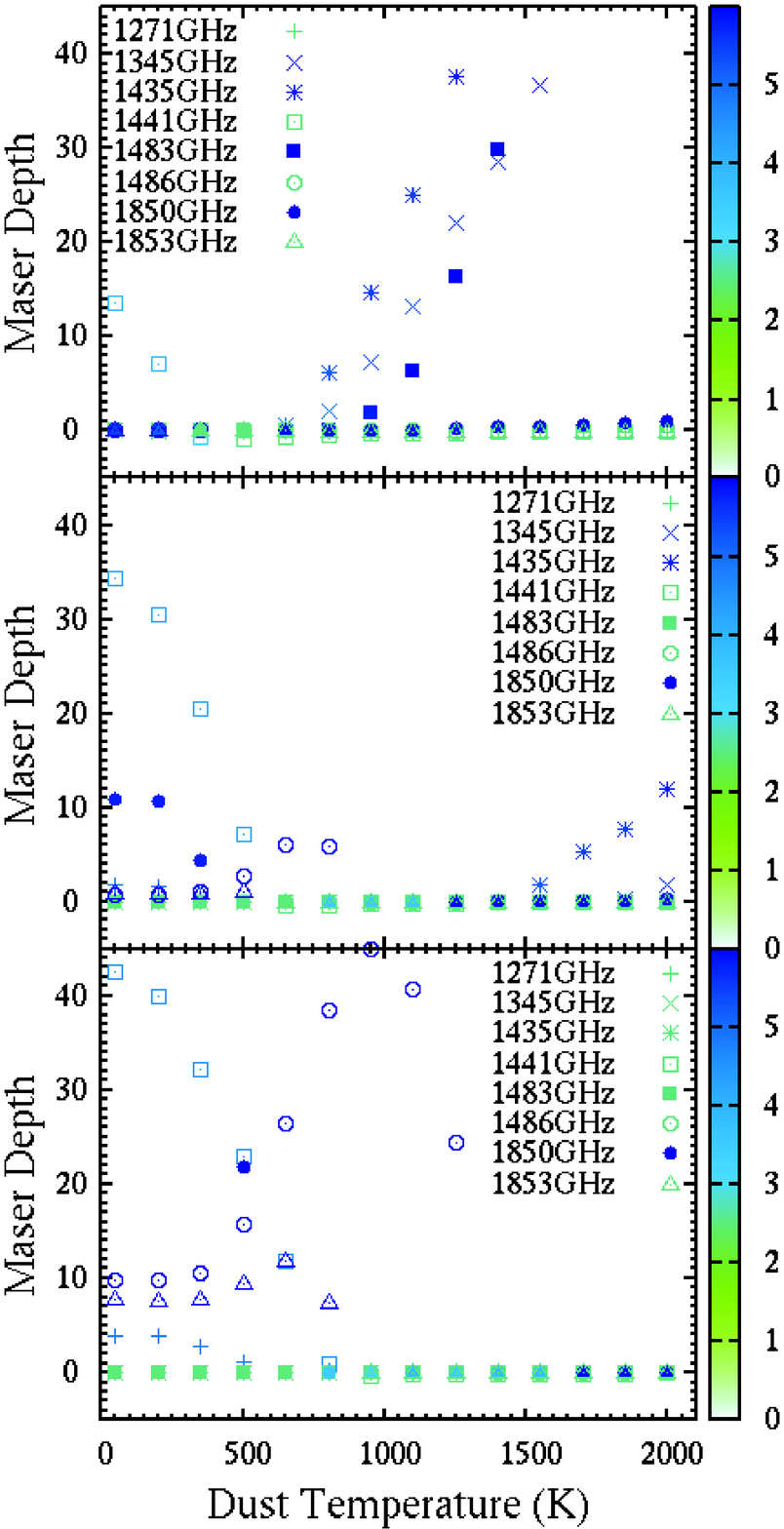}
  \caption{Maximum maser depths (negative optical depths) as a function of dust
temperature for p-H$_2$O transitions
visible to \textit{SOFIA}. Kinetic temperatures and symbol colours
are as in Fig.~\ref{td_alma_lt400}. }
\label{td_sofia_p}
\end{figure}

The variation of maser gain with $T_d$ for eight p-H$_2$O transitions visible to \textit{SOFIA}
is shown in Fig.~\ref{td_sofia_p}.
The major oddity in this case is the 1486-GHz transition. Although not significantly inverted
at $T_K=471.43$\,K, the maser depth in this line shows a maximum as a function of $T_d$ at both
the higher kinetic temperatures. The effect is particularly strong at $T_K=1500$\,K (bottom
panel) where the peak maser depth occurs at $T_d=950$\,K. Higher dust temperatures result in
a very rapid decline in the inversion. The transition shares with radiatively pumped lines the
property of having maximum maser depths associated with very high number densities.
A weak maximum also appears for 1853\,GHz, but it is
perhaps best to consider this as a collisionally pumped transition (along with 1271, 1441 and
1850\,GHz). The transitions at 1345, 1435 and 1435\,GHz exhibit typical radiatively pumped
behaviour, but none remain in the bottom panel.

\subsubsection{Effect of velocity gradient}
\label{sss:velgrads}

The discussion so far has considered results from slabs without a velocity gradient.
We now consider the effect of introducing positive and negative velocity gradients
through the slab. In a slab with a velocity gradient that is defined as positive in this
work, an observer situated on the optically thin (observer's) side of the slab would see
material approaching faster at each successive depth point towards the optically thick
boundary. Radiation from the most remote layers therefore appears more blue shifted.
In a slab with a negative velocity gradient, it is the outermost layers that approach
the observer fastest, and radiation from more distant slabs is progressively redshifted.
A typical AGB star envelope, for example, would have a negative velocity gradient in
this scheme, except for its innermost shock-dominated zone. The velocity gradients are
constant over the model, so that the velocity shift varies linearly with depth; velocity
variation is therefore concentrated in the geometrically thicker slabs. In most of the
results presented in this section, only the total velocity shift through the model
is considered.

When considering variation of inversion, or maser depth, with velocity shift, we would
ideally hold all other parameters at some set of standard values. However, from the
variation with dust temperature, studied in Section~\ref{sss:dust_effect} above, we
can see that there is no good single choice of $T_d$: we need at least two, one to
represent typical conditions for collisionally pumped transitions, and a second, higher, value for
those transitions that are predominantly radiatively pumped. The value of $T_d=1025$\,K, specified
as standard in Table~\ref{phys_var} is suitable for the radiatively pumped transitions, but we
also use $T_d = 350$\,K as representative of collisionally-pumped transitions.

Radiatively pumped transitions also tend to show their largest maser depths at low
kinetic temperatures compared to the collisionally pumped subset. It is therefore
also impossible to choose a single representative value of $T_K$, and we use the
same three values as in Section~\ref{sss:dust_effect}. The overall result is that we
display variation of maser gain for groups of transitions in blocks of 6 graphs,
corresponding to 2 values of $T_d$, and 3 of $T_K$.

\begin{figure}
  \includegraphics[bb=50 0 595 810,scale=0.5,angle=0]{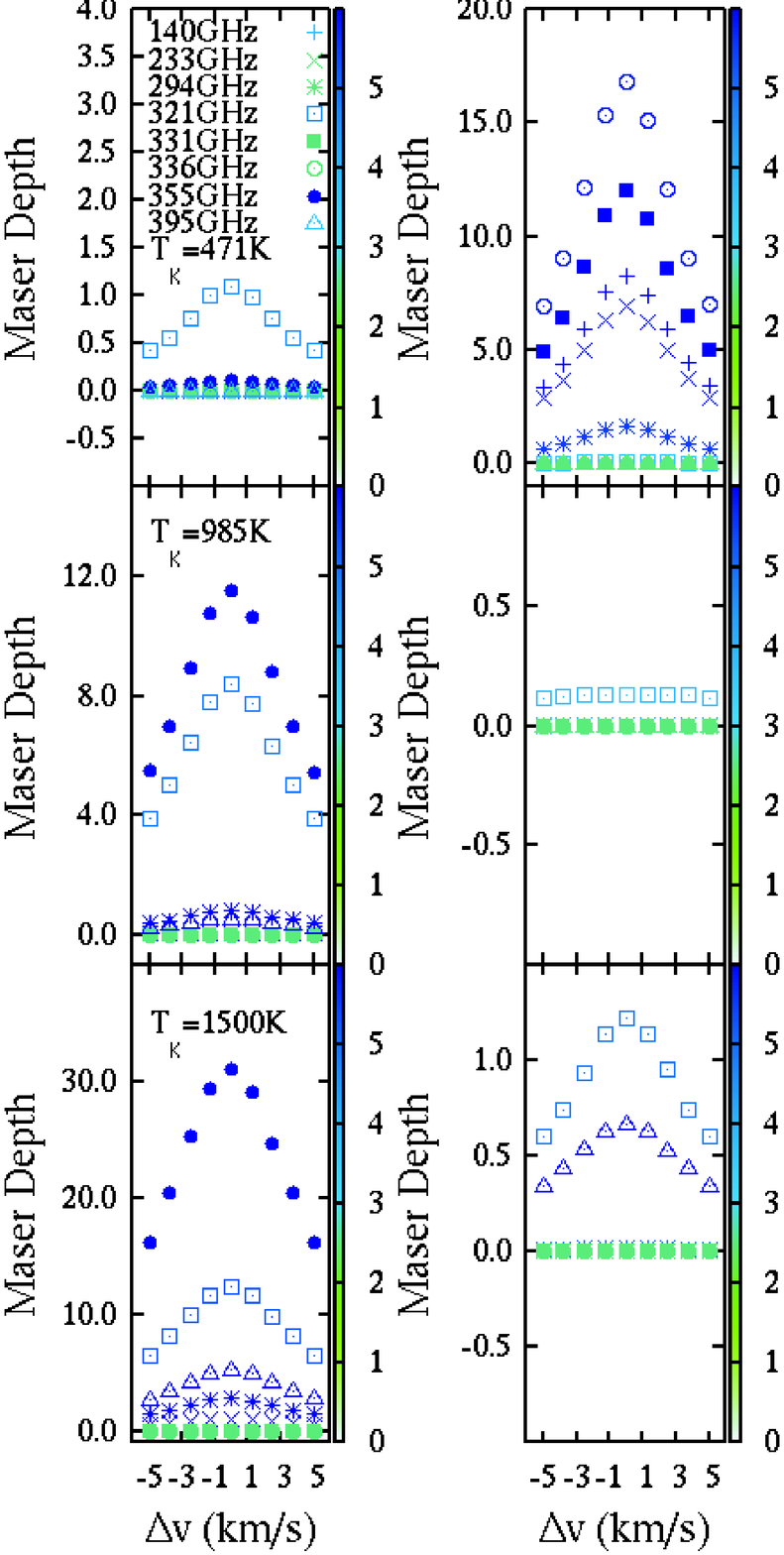}
  \caption{Maximum maser depths (negative optical depths) as a function of velocity
shift at 3 kinetic and 2 dust temperatures. Kinetic temperature increases downwards, and
is the same for any pair of panels horizontally; the value is marked in the left-hand
panel of each pair. Panels in a given column share the same dust temperature, with the value marked at
the top of the column. Maser transitions are identified by the symbols listed in the
key that appears in the top-left-hand panel only. The \textit{colour} of a symbol, in
any panel denotes the base-10 logarithm of the o-H$_2$O number density at which the
maximum maser depth was found, as in Section~\ref{sss:dust_effect}. The colour palette
appears to the right of each panel.}
\label{veloc_alma_ortho_lo}
\end{figure}

\begin{figure}
  \includegraphics[bb=50 0 595 810,scale=0.5,angle=0]{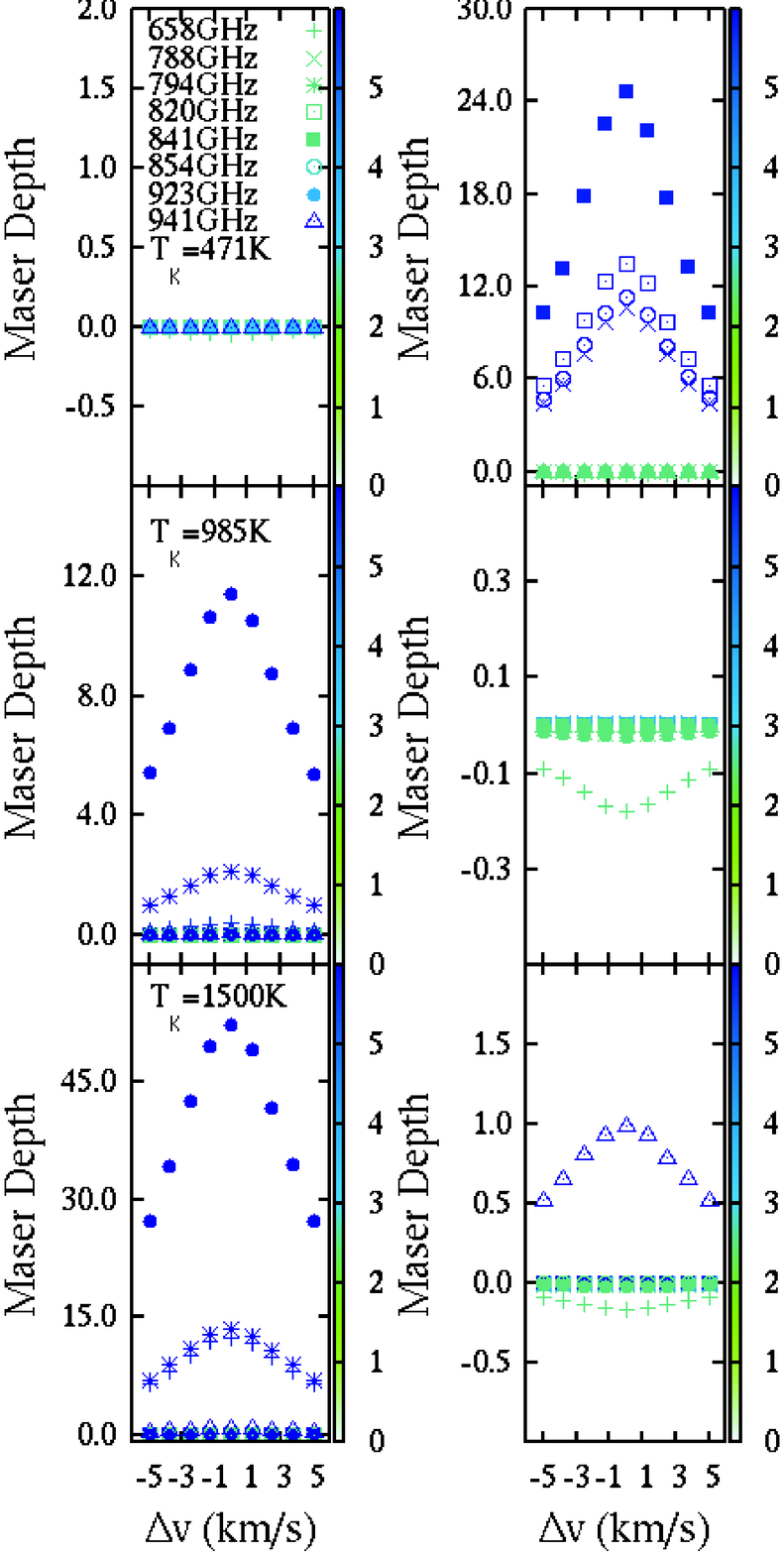}
  \caption{As for Fig.~\ref{veloc_alma_ortho_lo}, but for a group of o-H$_2$O transitions
with rest frequencies $>$400\,GHz.}
\label{veloc_alma_ortho_hi}
\end{figure}

The first plot of this type is Fig.~\ref{veloc_alma_ortho_lo}
for the group of o-H$_2$O masers visible to ALMA, with
frequencies below 400\,GHz, as considered for the effect of $T_d$ in Figure~\ref{td_alma_lt400}.
Perhaps the main result is that, at least for this subset of powerful maser transitions, the
effect of velocity shifts are close to  symmetric about $\Delta v =0$. For large maser depths,
the effect of increasing $|\Delta v|$ is always deleterious. We also see that the contrast,
$\tau_0/\tau_5$, where $\tau_0,\tau_5$ are the respective maser depths at $\Delta v=0$ and
$\Delta v = 5$ generally increases with $\tau_0$. We consider some of these statistics more
quantitatively below. It is also apparent that, for all velocity shifts, high maser depth
correlates with high number density of water (darkest blue colours), whilst green symbols
(low density) are associated with weak gain or absorption across all velocity shifts. These
conclusions are unchanged in general terms for the set of o-H$_2$O transitions with frequencies
above 400\,GHz, displayed in Fig.~\ref{veloc_alma_ortho_hi}.
The higher frequency transitions are almost completely inactive, at all velocity shifts, in
the top-left panel (T$_K=350$\,K and T$_d=1025$\,K. The appearance of modest inversions at
321,395 and 941\,GHz in bottom right-hand panels suggests that these transitions may be
pumped both collisionally and radiatively.

\begin{figure}
  \includegraphics[bb=50 0 595 810,scale=0.5,angle=0]{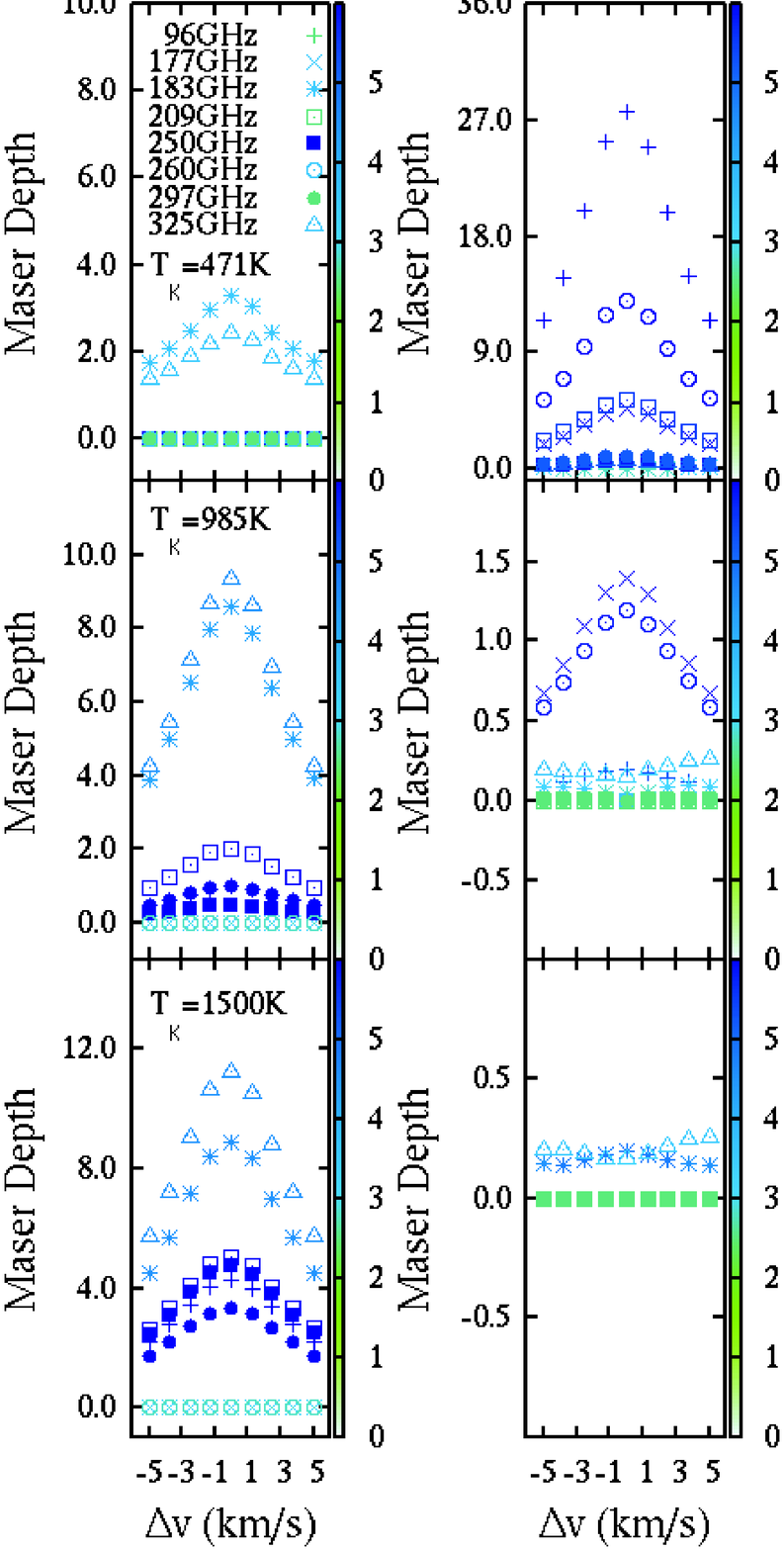}
  \caption{As for Fig.~\ref{veloc_alma_ortho_lo}, but for a group of p-H$_2$O transitions
with rest frequencies $\leq$325\,GHz.}
\label{veloc_alma_para_lo}
\end{figure}

\begin{figure}
  \includegraphics[bb=50 0 595 810,scale=0.5,angle=0]{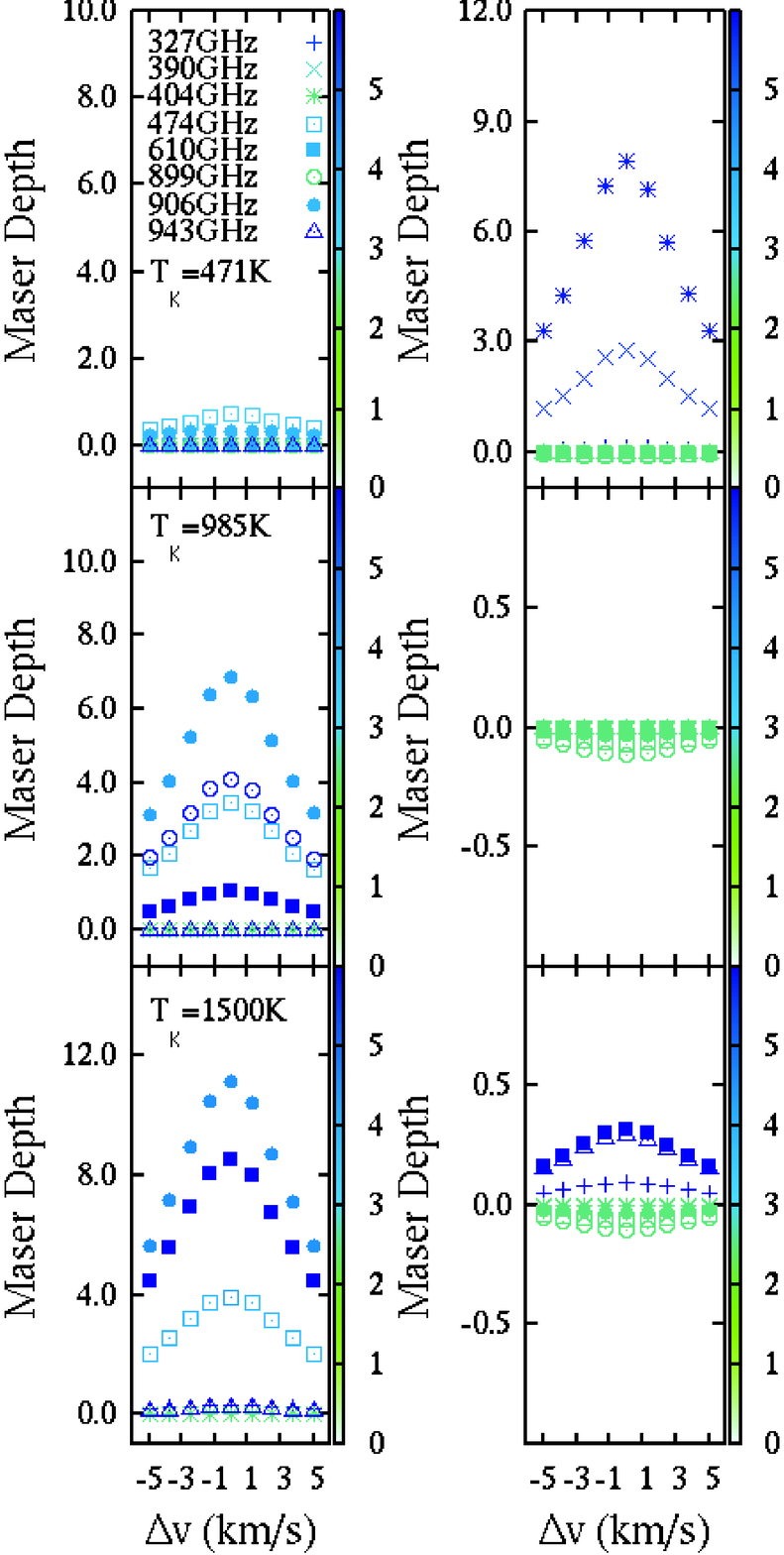}
  \caption{As for Fig.~\ref{veloc_alma_ortho_lo}, but for a group of p-H$_2$O transitions
with rest frequencies $>$325\,GHz.}
\label{veloc_alma_para_hi}
\end{figure}

The effect of velocity shifts on the strongest 
\textit{ALMA} maser transitions of p-H$_2$O is shown
in Fig.~\ref{veloc_alma_para_lo} and Fig.~\ref{veloc_alma_para_hi}.
In Fig.~\ref{veloc_alma_para_lo}, we see for the first time evidence of strongly
asymmetric behaviour of maser depth with velocity shift. This occurs in the right-hand
($T_d=1025$\,K) column, and the two lower panels at $T_K=$985 and 1500\,K. In both
these panels, the 325-GHz maser depth is largest at $\Delta v = 5$\,km\,s$^{-1}$, and
weakest for zero velocity shift. However, in both cases the maser depth in this transition
is small, reaching only 0.25. At 985\,K, the even smaller maser depth at 183\,GHz is
close to symmetric, but smallest at zero shift.
There is no similar behaviour in Fig.~\ref{veloc_alma_para_hi}, which covers the strongest
p-H$_2$O maser transitions with frequencies above 325\,GHz. There is evidence for weak
radiative pumping at 610\,GHz and a strong correlation between maser activity and high
number density in the right-hand column.

\begin{figure}
  \includegraphics[bb=50 0 595 810,scale=0.5,angle=0]{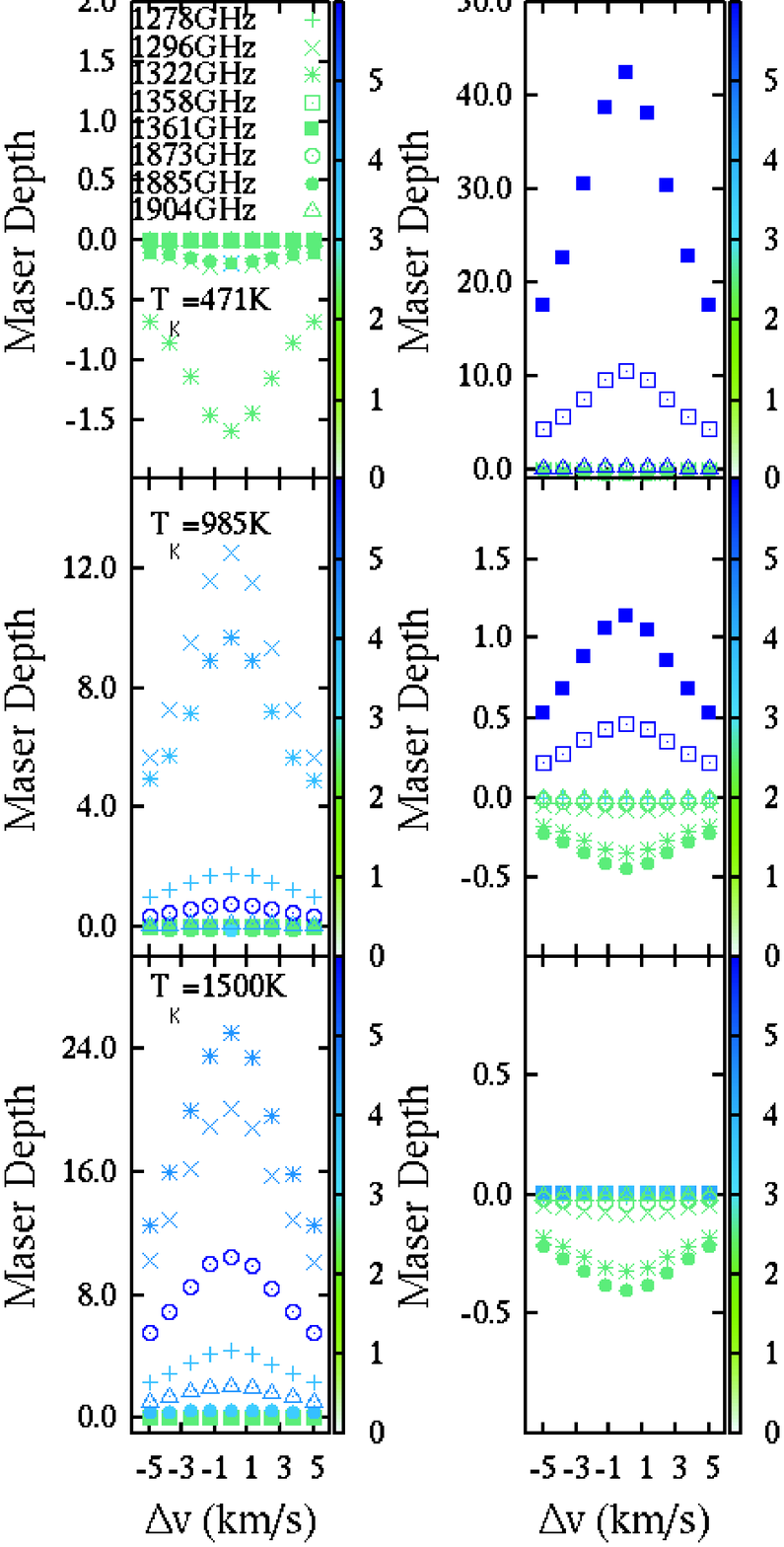}
  \caption{As for Fig.~\ref{veloc_alma_ortho_lo}, but for a group of o-H$_2$O transitions
suitable for observations with \textit{SOFIA}.}
\label{veloc_sofia_ortho}
\end{figure}

\begin{figure}
  \includegraphics[bb=50 0 595 810,scale=0.5,angle=0]{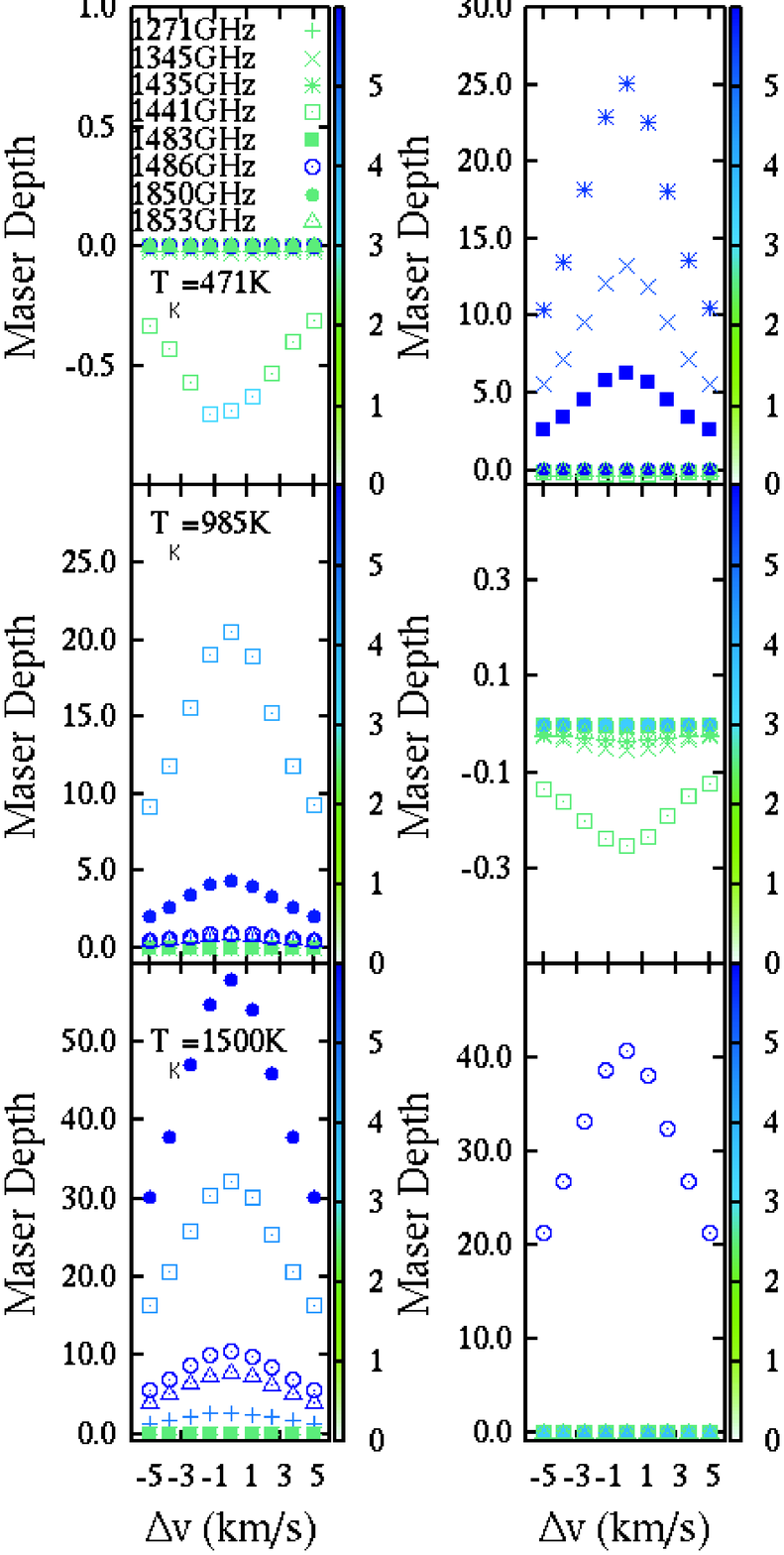}
  \caption{As for Fig.~\ref{veloc_alma_ortho_lo}, but for a group of p-H$_2$O transitions
suitable for observations with \textit{SOFIA}.}
\label{veloc_sofia_para}
\end{figure}

The effect of velocity shift on the strongest o-H$_2$O maser transitions visible to
\textit{SOFIA} is shown in Fig.~\ref{veloc_sofia_ortho}.
The top left-hand panel of Fig.~\ref{veloc_sofia_ortho} shows the strongest absorption
found for any of the transitions plotted in Figures~\ref{veloc_alma_ortho_lo}-\ref{veloc_sofia_para}: 
it is in the transition at
1322\,GHz. The effect of velocity shift on this absorption is, however, close to symmetric as
for most inverted transitions. An increase of kinetic temperature to 985\,K is enough to transform
the 1322-GHz absorption to a large positive maser depth (middle left panel), and this has become
the strongest maser transition at $T_K=1500$\,K (bottom left panel). At high kinetic and dust
temperature (bottom right-hand panel), there is some absorption, but little or no maser activity,
suggesting that none of the transitions plotted in Fig.~\ref{veloc_sofia_ortho} are both
collisionally and radiatively pumped. As noted previously, at the higher dust temperature particularly,
strong maser action correlates with high density.

The final set of plots of this type, Fig.~\ref{veloc_sofia_para}, for selected strong p-H$_2$O transitions
visible to \textit{SOFIA}, shows a significant absorption asymmetry at 1441\,GHz in the top-right panel,
where the most negative maser depth is at a shift of -1\,km\,s$^{-1}$, rather than at zero. There is also
a noticeable variation in the density of maximum absorption as a function of velocity shift in this
transition. The 1441-GHz transition is inverted at the higher temperatures in the left-hand panels, but
is also quite strongly absorbing at $T_d=1025$\,K and $T_K=985$\,K (middle right-hand panel), where the
strongest absorption is at zero shift. The 1486-GHz transition is strongly inverted in both bottom
panels, and is therefore likely to have both collisional and radiative pumping schemes.

For a more quantitative discussion of asymmetry and contrast, we define the following pair
of statistics: the asymmetry parameter is
\begin{equation}
q_a = \sum_{i=1}^4 ( \tau(\Delta v_i) - \tau(-\Delta v_i) ) / \tau (0) ,
\label{eq_asympar}
\end{equation}
where the sum is over the positive non-zero velocity shifts, and $\tau$ represents
the maser depth at the specified shift or $\tau(0)$ for zero shift. A systematic trend
of larger maser depths at positive shifts yields a positive value of $q_a$; a perfectly
symmetric distribution gives $q_a=0$. The contrast parameter is
\begin{equation}
q_c = 2 \tau (0) / (\tau(5) + \tau(-5)),
\label{eq_contpar}
\end{equation}
which is the ratio of the maser depth at zero shift to the mean of the depths at
the largest positive and negative shifts. Results are discussed below 
for the 48 transitions plotted in Fig.\ref{veloc_alma_ortho_lo}-Fig\ref{veloc_sofia_para}.
In some cases, a statistic is given in terms of the six panels in which the transition
appears, whilst in other cases, the average over the six panels only is considered.

Of the 48 transitions, 31 have a negative mean asymmetry and 17 positive. The line with
the strongest negative asymmetry on average is 233\,GHz with $q_a=-0.5525$; the strongest
positive asymmetry is in 325\,GHz with $q_a=0.3417$. Both results are dominated by panels
with rather weak maser depths. The transitions with the smallest (largest) mean contrast is
331\,GHz with $q_c=0.8385$ (1486\,GHz with $q_c=2.1463$). It may be more instructive to consider
maximum and minimum values that relate to individual panels in  Fig.\ref{veloc_alma_ortho_lo}-Fig\ref{veloc_sofia_para},
rather than to averages over the 6 panels in which data is plotted for a particular frequency.
At this level of detail, the greatest positive asymmetry is $q_a = 1.9391$ for 294\,GHz, middle
right-hand panel, Fig.~\ref{veloc_alma_ortho_lo}, and the greatest negative asymmetry is
$q_a = -3.5148$ at 233\,GHz, again in the middle right-hand panel of  Fig.~\ref{veloc_alma_ortho_lo}.
Both of these extremes correspond to distributions where the maser depths are $\ll$1.
The greatest contrast is $q_c=2.5348$ at 321\,GHz (Fig.~\ref{veloc_alma_ortho_lo}, top left-hand
panel) and the weakest for an inverted transition is $q_c=0.0435$ (bottom right-hand panel of
the same figure, but for 140\,GHz). Whilst the weakest contrast is again found under conditions
where the maser depth is $\ll$1, the strongest is associated with a modest maser depth
of $\tau \simeq 1$ at zero shift.

The limited effect of velocity fields that is evident in the discussion above also extends
to transitions that do not appear in Fig.\ref{veloc_alma_ortho_lo}-Fig\ref{veloc_sofia_para}, but
is not negligible. 
Lists of o- and p-H$_2$O transitions with maximum maser depth $>$3 were prepared for $T_d=1025$\,K and
$T_d=350$\,K and 
velocity shifts of zero and both $\pm$5\,km\,s$^{-1}$. It is important to note that these
maximum maser depths were found with respect to number density and kinetic temperature, rather
than number density and velocity shift (as represented by the symbol colours 
in Fig.\ref{veloc_alma_ortho_lo}-Fig\ref{veloc_sofia_para}). Lists at $\pm$5\,km\,s$^{-1}$ never
contained new maser transitions that were not already present at zero velocity shift. In
fact, the lists at higher shifts were generally substatially shorter than those at zero
shift. However, the effects of introducing a velocity shift were not found to be negligible,
since, in every case examined, a small subset of transitions were found to have a larger
maximum maser depth at one or both of the extreme shifts than the maximum found for
zero shift. For o-H$_2$O at $T_d=1025$\,K, the 140,1308 and 1359-GHz transitions
(3 of 20 strong masers at $\Delta v=-5$\,km\,s$^{-1}$) had larger maxima than at zero shift.
In this case, the same set of transitions also had higher maxima at
$\Delta v=+5$\,km\,s$^{-1}$. At $T_d=350$\,K, 11 of 21 strong masers at $\Delta v=-5$\,km\,s$^{-1}$
had higher maxima than at zero shift (321,658,556,902,793,1739,1321,1295,1694,1643 and 1538\,GHz).
The 1295-GHz transition was lost from this list at $\Delta v=+5$\,km\,s$^{-1}$.
A broadly similar set of results was found for p-H$_2$O: 6 of 16 strong masers had larger
maximum depths at $\Delta v=-5$\,km\,s$^{-1}$ than at zero shift for $T_d=1100$\,K
(96,1109,1486,1478,1435 and 1407\,GHz). The 1109 and 1478-GHz lines were not included
at $\Delta v=+5$\,km\,s$^{-1}$. At $T_d=35$\,K, 7 of 15 transitions had an enhanced maximum
at $\Delta v=-5$\,km\,s$^{-1}$: 533,905,899,592,1077,1440 and 1849\,GHz). This list was the
same, bar the loss of 899\,GHz at $\Delta v=+5$\,km\,s$^{-1}$. Plots were made of the
contrast as a function of maser depth at zero shift for several lines individually and
for all lines together, but no simple relationship was apparent.

\subsubsection{Effect of microturbulence and dust modelling}
\label{sss:testing}

In this section we briefly discuss some test models where input parameters have
been varied significantly away from their standard values or formulas. Specifically,
we consider the effect of varying the microturbulent speed, changing the size
distribution of the dust, and artificially limiting the effects of line
overlap.

The microturbulent speed is one of the standard cloud variables, and is used
as a temperature-independent line-broadening parameter only. We consider its effect
on a small subset of maser transitions for the standard, and four additional
values. The results are shown in Table~\ref{test_vt}. For the standard values of the
other cloud variables, see Table~\ref{phys_var}.
\begin{table}
\caption{Effect of microturbulent speed on maser depth}
\begin{tabular}{@{}lrrrrr}
\hline
           &\multicolumn{5}{c}{$v_T$(km\,s$^{-1}$)} \\
           & 0.0 & 1.0 & 2.0 & 3.0 & 4.0 \\
\hline
$\nu$(GHz) &     &      &     &     &     \\
\hline
22                          & 1.5533& 1.3019 & 0.9768 & 0.7620 & 0.6232 \\
68                          & 0.8378& 0.6309 & 0.4032 & 0.2781 & 0.2071 \\
321                         & 2.3217& 1.8155 & 1.2112 & 0.8534 & 0.6402 \\
380                         & 1.2271& 1.0040 & 0.7242 & 0.5517 & 0.4434 \\
448                         & 1.6629& 1.3013 & 0.8327 & 0.5493 & 0.3857 \\
595                         & 0.2536& 0.1554 & 0.0567 & 0.0117 &-0.0075 \\
658                         &-13.113&-10.546 &-7.4581 &-5.5634 &-4.4021 \\
1361                        &-3.8623&-3.0824 &-2.1547 &-1.6060 &-1.2720 \\ 
\hline
96                          &-0.6384&-0.4899 &-0.3195 &-0.2226 &-0.1672 \\
183                         & 2.0100& 1.6156 & 1.1314 & 0.8459 & 0.6760 \\
325                         & 0.4698& 0.4186 & 0.3581 & 0.3100 & 0.2762 \\
474                         & 0.4509& 0.1923 &-0.0356 &-0.0973 &-0.1019 \\
547                         & 0.6922& 0.4810 & 0.2569 & 0.1460 & 0.0907 \\
609                         & 0.1189& 0.0810 & 0.0426 & 0.0246 & 0.0158 \\
906                         &-0.1061&-0.0572 &-0.0446 &-0.0554 &-0.0655 \\
1435                        &-5.7881&-4.5243 &-3.0364 &-2.1662 &-1.6518 \\
\hline
\end{tabular}
\label{test_vt}
\end{table}

The principal effect of the microturbulent speed is to lower the maser
depth of inverted transitions as $v_T$ is increased; this is in accord
with expectations, since increasing $v_T$ reduces the optical depth in
pumping transitions. Even the collisional scheme suffers from this effect,
since radiative processes are required to couple the backbone levels
(see Section~\ref{ss:theo}). Transitions that are in absorption (negative
depths in Table~\ref{test_vt}) also lose absorbing power with increasing
$v_T$, as shown by the entries for 658 and 1361\,GHz. The rows for 595\,GHz
and 474\,GHz
show that weakly inverted transitions under standard conditions can be
forced into weak absorption by increasing $v_T$ to 4\,km\,s$^{-1}$. A
differential effect is demonstrated (amongst other
transitions) by the 380- and 448-GHz rows: 448\,GHz
has the larger maser depth at $v_T=0$\,km\,s$^{-1}$, but the 380\,GHz depth
is the larger for $v_T>3$\,km\,s$^{-1}$. At 906\,GHz, a weakly absorbing
transition at $v_T=0$ initially becomes still more weakly absorbing
as $v_T$ increases. However, it does not pass into inversion, but
the absorbing strength passes through a minimum between $v_T=2$ and
3\,km\,s$^{-1}$, beyond which the absorbing power again increases.

\begin{figure}
  \includegraphics[angle=0,scale=0.7]{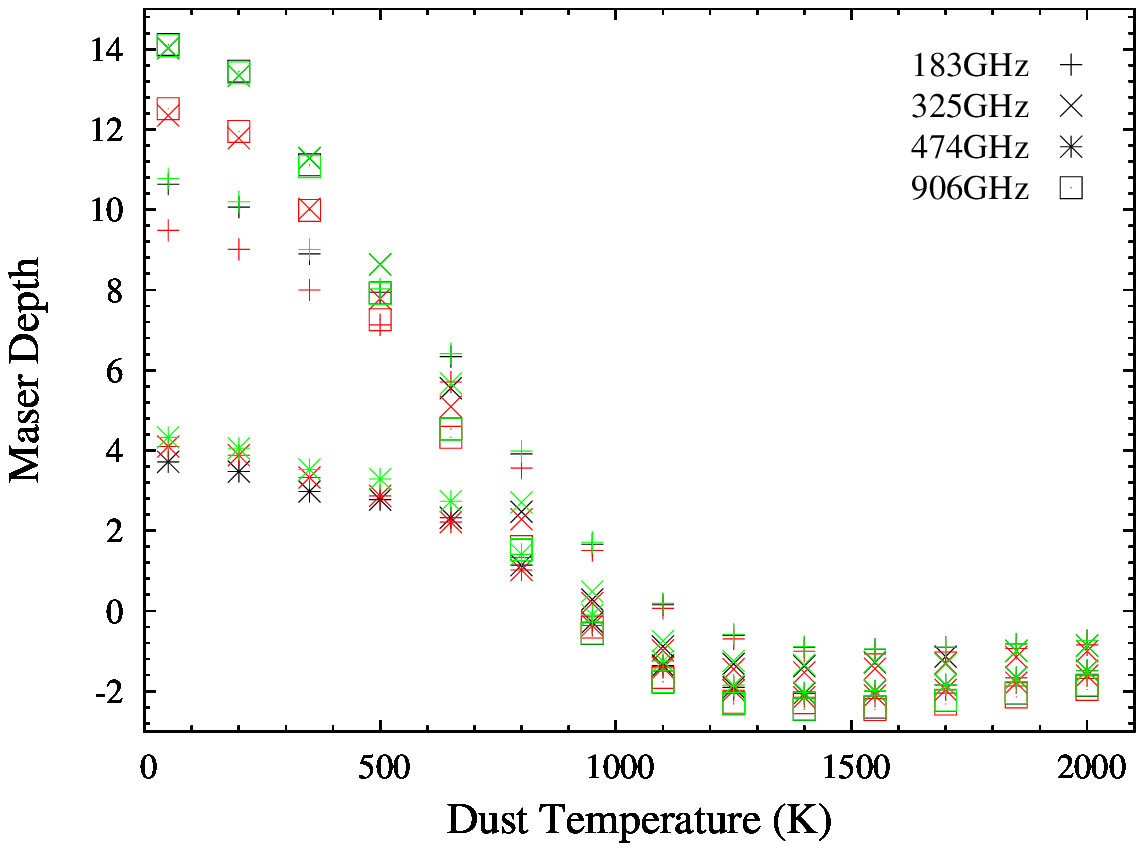}
  \caption{Maser depths in 4 maser transitions of p-H$_2$O as a function
of dust temperature. Other physical conditions are standard, except that
$T_K=1500$\,K. Colours denote the usual model (black), monodispersed dust (red)
and no line overlap (green).}
\label{figtests}
\end{figure}

We test the robustness of the dust model by moving from the spectrum of
sizes with upper and lower limits, as introduced in Section~\ref{s:model}, to
monodispersed dust, where the single grain radius is equal to the mean radius
of the original distribution ($\bar{a}=1.667$\,nm). 
The dust mass fraction and opacity behaviour
with wavelength were as used in the main model. We present the results for
four maser transitions of p-H$_2$O in Fig.~\ref{figtests}, plotting maser
depths as a function of dust temperature for the standard (black symbols) and monodispersed
models (red symbols). Other input parameters were as in Table~\ref{phys_var}, except that
$T_K=1500$\,K was used instead of $1750$\,K, because more data were available
at the lower temperature.
It can be seen that changing to a monodispersed dust model typically changes
the maser depths by a few per cent, but does not change the general trend
of strong masers falling into absorption as $T_d$ is increased. For three
of the transitions shown, the standard model has generally higher maser
gains for a given $T_d$, but maser depths are more positive in the monodispersed
model at 474\,GHz.

We also plot in Fig.~\ref{figtests} (green symbols) results for a model that uses the standard
dust, but has no line overlap. This restriction was imposed by limiting the size
of overlapping groups to a maximum of 1, so that all transitions have (artifically)
independent line shapes. For the transitions in Fig.~\ref{figtests}, results are also
generally modest differences from the standard model. At 905\,GHz, omission of line
overlap slightly depresses the maser depths and reduces absorption at higher values
of $T_d$. However, at the other three frequencies, maser depths are somewhat larger
without overlap. A more quantitative effect appears in the 471-GHz transition, which
is weaker than the plotted lines under these conditions: in the standard model this line has a maser depth
of 2.19 at $T_d=50$\,K (2.53 with monodispersed dust), but without line overlap the
transition is in absorption (maser depth -0.094). In the overlap-free model, the 471-GHz
transition is only very weakly inverted at $T_d=350$ and $500$\,K, rather than being
continuously inverted over the range $T_d=50-800$\,K.

\subsection{AGB Stars}
\label{sss:agbchanges}

We have seen in Section~\ref{ss:lumps} that the scale size of maser clouds is
expected to be approximately proportional to the radius of the host star.
The results discussed up to this point are all based on slabs of thickness
2.25$\times$10$^{14}$\,cm, considered appropriate for a red supergiant star,
such as VY~CMa (see Section~\ref{s:model}). We now briefly discuss the effect
of reducing the cloud thickness to something more appropriate for the extended
atmosphere of an AGB star, namely 4.5$\times$10$^{13}$\,cm.

Only two slab thicknesses have been considered, so we tabulate the effect of
changing scale for a number of transitions of both spin species of H$_2$O.
We choose conditions where each transition is a reasonably strong maser for
the thicker slab, so the selected
water number density, $T_K$, $T_D$ and corresponding maser depths
are listed for each transition. Velocity
shift is zero in all cases and the constant microturbulent speed is 1\,km\,s$^{-1}$.
Results are tabulated in Table~\ref{agb_tab}. Note that $T_K$ is limited to a maximum
of 1500K, since this was the maximum value used for the thinner slab.
\begin{table}
\caption{The effect of changing slab thickness on maser depths achieved
for selected strong maser transitions; $R$ is the ratio of the maser depth
in the thick (supergiant) model to the depth in the thin (AGB star) model.}
\begin{tabular}{@{}lrrrrrr}
\hline
$\nu$    & $n_{H{_2}O}$       & $T_K$     & $T_d$    & $\tau_{SG}$    &  $\tau_{AGB}$ & $R$      \\
(GHz)    & cm$^{-3}$         &  K       &   K      &               &              &        \\
\hline
22       & 1.5(5)          & 814       & 50       & 15.45         & 2.624        & 5.888  \\
321      & 6.0(4)          & 1500      & 50       & 13.35         & 2.676        & 4.989  \\
336      & 3.0(5)          & 471       & 1025     & 12.82         & 1.869        & 6.859  \\
380      & 3.0(6)          & 1500      & 50       & 33.85         & 6.424        & 5.269  \\
447      & 6.0(3)          & 900       & 50       & 10.79         & 2.843        & 3.795  \\
658      & 6.0(5)          & 1500      & 50       & 15.90         & 2.971        & 5.352  \\
841      & 6.0(5)          & 471       & 1025     & 24.62         & 3.774        & 6.524  \\
902      & 1.5(6)          & 1500      & 50       & 11.47         & 2.267        & 5.060  \\ 
922      & 3.0(6)          & 1500      & 50       & 55.26         & 7.000        & 7.894  \\
1296     & 3.0(4)          & 1500      & 50       & 26.59         & 5.785        & 4.596  \\
1308     & 1.5(5)          & 471       & 1025     & 12.61         & 1.682        & 7.497  \\
1322     & 1.5(4)          & 1500      & 50       & 35.67         & 8.047        & 4.433  \\
1361     & 3.0(6)          & 471       & 1025     & 42.42         & 5.743        & 7.386  \\
\hline
183      & 3.0(4)          & 1071      & 50       & 11.56         & 3.159        & 3.659  \\
260      & 3.0(6)          & 557       & 1025     & 15.69         & 2.587        & 6.065  \\
325      & 3.0(4)          & 1329      & 50       & 14.17         & 3.676        & 3.855  \\
403      & 1.5(6)          & 300       & 1025     & 25.92         & 3.395        & 7.635  \\
899      & 1.5(6)          & 1500      & 50       & 45.70         & 9.131        & 5.005  \\
906      & 3.0(4)          & 1500      & 50       & 14.12         & 3.657        & 3.861  \\
1441     & 3.0(4)          & 1500      & 50       & 42.61         & 10.06        & 4.236  \\
1850     & 1.5(6)          & 1500      & 50       & 63.96         & 0.000        & $\infty$ \\
\hline
\end{tabular}
\label{agb_tab}
\end{table}

The supergiant slab is geometrically five times thicker than the AGB slab. For a constant
and equal gain coefficient throughout both slabs, we would naively expect that the ratio
of maser depths would be close to 5.0 for any line, with deviations due to more subtle radiative
transfer effects. With the exception of the 1850-GHz transition of p-H$_2$O, which is not
inverted in the thinner model, all examples of the ratio $R$ are indeed within a factor of
2 of the naively expected value. For the listed transitions of o-H$_2$O, the mean value of
$R$ is 5.811 with a standard deviation of 1.306. For the p-H$_2$O transition, excluding  
1850\,GHz, the corresponding figures are 4.902 and 1.472. The grand average is 5.493 with
a standard deviation of 1.400.

\section{Predictions for Observations}
\label{s:predict}

Our list of predicted maser lines from zero frequency to 1910\,GHz
is presented in Table~\ref{hugetab}. The typeface
in column~1 indicates whether the transition is predominantly pumped by collisions
or radiation. Column~9 indicates whether the maser is potentially observable in one
of the \textit{ALMA} or \textit{SOFIA} bands or, for lower frequency transitions, the
radio band in which the maser might be observed with a different ground-based instrument.
The final column gives the expected ratio of the maximum maser depth in the transition to
the maximum found at 22\,GHz. A couple of caveats should be observed regarding this ratio: firstly, the
figure takes no account of saturation, so numbers significantly greater than 1 would not
be reflected in brightness ratios of observed masers; secondly, under the conditions where
the radiative transitions were added ($T_d=1025$\,K), 22-GHz is not a strong maser, so
its depth at $T_d=50$\,K was used, and the maser depths are therefore not compared under 
the same radiation field. Kinetic temperatures for the two maser depths could be
different for all transitions.

It is arguably more informative to attempt to predict which maser lines are closely
associated with each other, and under what conditions. There are various reasonable
ways to attempt this:
one idea is a points-based pair-wise correlation
function. This has the advantage of relatively simple visual presentation, and more
complex groupings can reasonably easily be constructed from it. However, the exact
number of points to award for, for example, finding two maser transitions with
potentially saturating maser depths under the same physical conditions, is open to
much debate. We therefore defer the construction of such a function to future work. 

A more qualitative picture can be deduced from examining the groupings that we
have already found. Grouping of transitions by pumping scheme has been discussed in
detail in Section~\ref{sss:dust_effect}, and collisionally and radiatively pumped transitions
may be identified in Table~\ref{hugetab}. The transitions that have both radiative and
collisional branches in their pumping schemes (22,96,209,321,325,395,941 and 1486\,GHz) 
may be problematic in the analysis of observational data, since they may have unusual,
or previously unexpected spatial association with other transitions.

Based on the loci of inverted regions in the number-density/kinetic temperature plane,
we can perhaps group the  masers displayed in Fig.~\ref{oh2o_alma_v0_td50}-Fig.~\ref{oh2o_other_v0_td50} into four families:
firstly the `classic' set, comprising 321,906,1271,1278,1296,1322,1441,1885 and 1904\,GHz.
These masers have a collisional pump, with a peak maser depth reasonably close to
$T_K=1500\,K$ and $n_{H_2}=10^9$\,cm$^{-3}$. The 22-GHz transition is shared between this
group and the `low-temperature' set (183,325,380,447,474,620 and 916\,GHz), which form
at a number density close to, or rather below, that for the classic set, and at $T_K\sim1000$\,K.
A third group, with 658\,GHz as the typical member, comprising 658,902,899 and 1850\,GHz, have a peak maser depth associated with
$T_K\sim2500$\,K or higher, but with a number density that is substantially lower than the
maximum used in the model, perhaps of order 10$^{10}$\,cm$^{-3}$. The remaining transitions
form a fourth, hot, dense, group that has a peak maser depth at a density and kinetic temperature
beyond the top right-hand corner of the plotted plane. 

Masers with a predominantly radiative pump add three transitions, 327,941 and 943\,GHz to the
hot, dense group above. The remainder can perhaps be split into a `warm' group (140,177,260,788 and
1358\,GHz) that have $1000>T_K>300$\,K for peak maser depth. The remainder (`cold') all appear to peak
at a kinetic temperature below the minimum plotted, 300\,K.


\begin{table*}
 \centering
 \begin{minipage}{140mm}
  \caption{Water maser lines predicted to be significantly inverted in the radiative transfer
models from Section~\ref{s:comput}}
\label{hugetab}
  \begin{tabular}{@{}llrrrrlrrlr@{}}
  \hline
   $\nu$  & $\lambda$ & o/p & $ v_{up}$           & $v_{lo}$             & Rot$_{up}$ & Rot$_{lo}$ & T$_{up}$ & Band     & Det. & $\tau/\tau_{22}$ \\
   GHz    &   $\umu m$&     &($v_1,v_2,v_3$)&($v_1,v_2,v_3$) &$J_{K_a,K_c}$&$J_{K_a,K_c}$ &   K     &   &      &                 \\
 \hline
 \textit{2.159980}$^\dagger$ & 138794  & p & (0,1,0) & (0,1,0) & 4$_{2,2}$ & 5$_{1,5}$ & 2766.54 & S     & N & 0.48  \\ 
 \textit{4.581939}$^\dagger$ & 65429.2 & o & (0,0,1) & (0,0,1) & 5$_{2,3}$ & 6$_{1,6}$ & 6036.89 & C     & N & 0.30  \\
 \textit{6.328131}$^\dagger$ & 47374.6 & p & (0,2,0) & (0,2,0) & 4$_{3,1}$ & 5$_{2,4}$ & 5177.49 & C     & N & 3.25  \\
 \textit{7.257424}$^\dagger$ & 41308.4 & o & (0,1,0) & (0,1,0) & 9$_{3,6}$ &10$_{2,9}$ & 4178.85 & C     & N & 0.75  \\
 12.00880          & 24964.4  & o   & (0,1,0) & (0,1,0) & 4$_{2,3}$  & 3$_{3,0}$  & 2745.23 & X        & N & 0.26  \\
 \textit{12.47825}$^{\dagger \#}$ & 24025.2 & p & (0,0,0) & (0,0,0) & 15$_{7,9}$ & 16$_{4,12}$ & 5213.8 & K$_a$ & N & 0.27 \\
 22.23508          & 13482.8  & o   & (0,0,0) & (0,0,0) & 6$_{1,6}$  & 5$_{2,3}$  & 643.501 & K        & Y & 1.00  \\
 67.80396          & 4421.46  & o   & (0,1,0) & (0,1,0) & 4$_{1,4}$  & 3$_{2,1}$  & 2620.89 & A2       & N & 3.5   \\
 96.26116          & 3114.366 & p   & (0,1,0) & (0,1,0) & 4$_{4,0}$  & 5$_{3,3}$  & 3064.07 & A3       & Y & 0.78  \\
\textit{117.32655}$^{\dagger \#}$ & 2555.197 & o & (1,0,0) & (0,0,1) & 6$_{2,5}$ & 6$_{0,6}$ & 6036.40 & - & N & 0.35 \\
\textit{119.99594} & 2498.36  & p   & (0,1,0) & (0,1,0) & 2$_{2,0}$  & 3$_{1,3}$  & 2502.74 & -        & N & 7.12  \\
 137.04852         & 2187.491 & p   & (0,2,0) & (0,2,0) & 3$_{3,1}$  & 4$_{2,2}$  & 5036.49 & A4       & N & 0.26  \\
\textit{139.61429} & 2147.290 & o   & (0,0,0) & (0,0,0) & 14$_{6,9}$ & 15$_{3,12}$ & 4438.42 & A4      & N & 0.50 \\
\textit{177.31707} & 1690.710 & p   & (0,0,0) & (0,0,0) & 15$_{6,10}$ & 16$_{3,13}$ & 4954.02 & A5     & N & 0.34 \\
 183.31012         & 1635.439 & p   & (0,0,0) & (0,0,0) & 3$_{1,3}$  & 2$_{2,0}$  & 204.708 & A5       & Y & 0.76  \\
 209.11837         & 1433.602 & p   & (0,1,0) & (0,1,0) & 5$_{5,1}$  & 6$_{4,2}$  & 3461.92 & A5       & N & 0.80  \\
 232.68670         & 1288.395 & o   & (0,1,0) & (0,1,0) & 5$_{5,0}$  & 6$_{4,3}$  & 3461.93 & A6       & Y & 0.30  \\
 250.75179$^\dagger$ & 1195.557 & p   & (0,2,0) & (0,2,0) & 9$_{2,8}$  & 8$_{3,5}$  & 6141.09 & A6       & N & 3.19  \\
\textit{259.95218} & 1153.260 & p   & (0,0,0) & (0,0,0) & 13$_{6,8}$ & 14$_{3,11}$ & 3953.93 & A6      & N & 0.66 \\
 262.89775         & 1140.339 & p   & (0,1,0) & (0,1,0) & 7$_{7,1}$  & 8$_{6,2}$  & 4474.50 & A6       & N & 0.26  \\
\textit{268.14912} & 1118.006 & o   & (0,2,0) & (0,2,0) & 5$_{5,2}$  & 7$_{4,3}$  & 6026.20 & A6       & Y & 0.20  \\
 293.66449         & 1020.867 & o   & (0,1,0) & (0,1,0) & 6$_{6,1}$  & 7$_{5,2}$  & 3933.59 & A7       & Y & 0.41  \\
 297.43928         & 1007.911 & p   & (0,1,0) & (0,1,0) & 6$_{6,0}$  & 7$_{5,3}$  & 3933.59 & A7       & Y & 0.61  \\
 321.22564         & 937.6215 & o   & (0,0,0) & (0,0,0) & 10$_{2,9}$ & 9$_{3,6}$  & 1861.26 & A7       & Y & 0.96  \\
 325.15292         & 922.0045 & p   & (0,0,0) & (0,0,0) & 5$_{1,5}$  & 4$_{2,2}$  & 454.339 & A7       & Y & 0.92  \\
\textit{326.68672} & 917.675  & p   & (1,0,0) & (1,0,0) & 5$_{1,5}$  & 4$_{2,2}$  & 5722.71 & A7       & N & 0.29 \\
\textit{331.12373} & 905.378  & o   & (0,2,0) & (0,2,0) & 3$_{2,1}$  & 4$_{1,4}$  & 4881.44 & A7       & Y & 1.14 \\
\textit{336.22794} & 891.634  & o   & (0,1,0) & (0,1,0) & 5$_{2,3}$  & 6$_{1,6}$   & 2955.22& A7       & Y & 1.78 \\
 354.80858         & 844.9410 & o   & (0,0,0) & (0,0,0) & 17$_{4,13}$& 16$_{7,10}$ & 5780.91 & A7      & Y & 3.91  \\
 A380.19737        & 788.517  & o   & (0,0,0) & (0,0,0) & 4$_{1,4}$  & 3$_{2,1}$  & 323.494 & -        & Y & 1.23  \\
\textit{390.13451} & 768.433  & p   & (0,0,0) & (0,0,0) & 10$_{3,7}$ & 11$_{2,10}$ & 2213.07 & A8      & N & 0.26 \\
 395.17191$^\#$    & 758.638  & o   & (0,1,0) & (0,0,0) & 14$_{7,8}$  & 14$_{14,1}$ & 7165.67 & A8      & N & 2.73 \\
\textit{403.49242} & 742.994 & p    & (0,2,0) & (0,2,0) & 4$_{2,2}$  & 5$_{1,5}$   & 5029.91 & A8      & N & 0.69 \\
 439.15081         & 682.664  & o   & (0,0,0) & (0,0,0) & 6$_{4,3}$  & 5$_{5,0}$  & 1088.77  & A8      & Y & 1.56 \\
 A448.00108        & 669.177  & o   & (0,0,0) & (0,0,0) & 4$_{2,3}$  & 3$_{3,0}$  & 432.157  & A8      & N & 0.70 \\
 474.68913         & 631.5553 & p   & (0,0,0) & (0,0,0) & 5$_{3,3}$  & 4$_{4,0}$  & 725.089  & A8      & Y & 0.35 \\
 488.49113         & 613.711  & p   & (0,0,0) & (0,0,0) & 6$_{2,4}$  & 7$_{1,7}$  & 867.262  & A8      & N & 0.39 \\
\textit{516.23018} & 580.734  & p   & (0,2,0) & (0,2,0) & 2$_{2,0}$  & 3$_{1,3}$  & 4747.98  &  -      & N & 0.49 \\
 530.34286         & 565.280  & o   & (0,0,0) & (0,0,0) & 14$_{3,12}$ & 13$_{4,9}$ & 3671.04  & -      & N & 0.42  \\
 534.24045$^{\dagger \#}$ & 561.156  & p& (0,0,0) & (0,0,0) & 18$_{4,14}$ & 17$_{7,11}$ & 6369.75 & -      & N & 3.6   \\
 540.75406         & 555.397  & o   & (1,0,0) & (1,0,0) & 1$_{1,0}$  & 1$_{0,1}$   & 5321.34  & -      & N & 0.91  \\
 546.69052         & 548.376  & p   & (0,1,0) & (0,1,0) & 5$_{2,4}$  & 4$_{3,1}$  & 2912.32   & -      & N & 2.5   \\
 554.05526$^\dagger$ & 541.087  & o   & (0,2,0) & (0,2,0) & 3$_{1,2}$ & 2$_{2,1}$   & 4797.82   & -      & N & 0.37  \\
\textit{556.84175}$^\dagger$ & 538.380 & o & (0,2,0) & (0,2,0) & 3$_{3,0}$ & 4$_{2,3}$ & 5009.95 & -     & N & 0.29  \\
 557.98548$^{\dagger \#}$ & 537.276 & o & (0,0,0) & (0,0,0) & 19$_{4,15}$ & 18$_{7,12}$& 6980.73  & -      & N & 2.82  \\
\textit{563.94516}$^{\dagger \#}$ & 531.599 & p & (1,0,0) & (0,2,0) & 8$_{0,8}$ & 7$_{5,3}$ & 6285.41 & - & N & 1.82  \\
 566.49747$^\dagger$ & 529.203  & o   & (0,2,0) & (0,2,0) & 8$_{2,7}$  & 7$_{3,4}$  & 5858.35   & -      & N & 4.80  \\
\textit{571.91369} & 524.192  & o   & (0,0,0) & (0,0,0) &12$_{6,7}$ & 13$_{3,10}$ & 3474.27   & -      & N & 0.52  \\
 593.70815         & 504.949  & p   & (0,1,0) & (0,1,0) & 9$_{2,8}$  & 8$_{3,5}$  & 3871.20   & -      & N & 3.95  \\
\textit{598.49487}$^{\dagger \#}$ & 500.911 & o & (1,0,0) & (0,0,1) & 9$_{2,7}$ & 9$_{2,8}$ & 6931.69 & -  & N & 0.31 \\
 610.79543$^\dagger$ & 490.823  & p   & (0,2,0) & (0,2,0) & 7$_{2,6}$  & 6$_{3,3}$  & 5602.88   & A9     & N & 4.57 \\
 A620.70095        & 482.990  & o   & (0,0,0) & (0,0,0) & 5$_{3,2}$  & 4$_{4,1}$  & 732.072   & A9     & Y & 0.49 \\
 658.00655         & 455.6070 & o   & (0,1,0) & (0,1,0) & 1$_{1,0}$  & 1$_{0,1}$  & 2360.34   & A9     & Y & 4.67 \\
 753.74777$^{\dagger \#}$ & 397.737 & o & (0,2,0) & (1,0,0) & 6$_{6,1}$  & 5$_{5,0}$  & 6848.41   & -     & N & 0.57  \\
\textit{766.79360} & 390.968 & p    & (0,0,0) & (0,0,0) & 11$_{5,7}$ & 12$_{2,10}$ & 2820.32  & -      & N & 0.91 \\
 775.49588         & 386.582  & p   & (0,2,0) & (0,2,0) & 2$_{0,2}$  & 1$_{1,1}$  & 4635.71   & -      & N & 1.66  \\
\textit{788.04615} & 380.425 & o    & (0,0,1) & (1,0,0) & 7$_{2,6}$ & 7$_{2,5}$   & 6405.98   & A10    & N & 0.41 \\
 793.61632         & 377.754  & o   & (0,2,0) & (0,2,0) & 1$_{1,0}$  & 1$_{0,1}$  & 4606.87   & A10    & N & 6.49 \\
\textit{820.22402} & 365.500 & o    & (0,0,1) & (1,0,0) & 5$_{1,5}$ & 5$_{1,4}$   & 5865.78   & A10    & N & 1.04 \\
 823.67060         & 363.971  & o   & (1,0,0) & (0,0,1) & 5$_{3,2}$  & 4$_{3,1}$  & 5976.63   & A10    & N & 0.49 \\
 832.69841         & 360.025  & p   & (1,0,0) & (1,0,0) & 9$_{2,8}$  & 8$_{3,5}$  & 6786.90   & A10    & N & 0.41 \\
\textit{841.05071} & 356.449 & o    & (0,0,0) & (0,0,0) & 10$_{5,6}$ & 11$_{2,9}$ & 2472.87   & A10    & N & 1.15 \\
 847.45203$^{\dagger \#}$ & 353.757 & o & (0,2,0) & (1,0,0) & 9$_{5,4}$  & 8$_{4,5}$  & 6884.11  & A10    & N & 0.34 \\
\textit{854.04981} & 351.024 & o    & (0,0,0) & (0,0,0) & 12$_{5,8}$ & 13$_{2,11}$ & 3273.78  & A10    & N & 0.46 \\
\hline
\end{tabular}
\end{minipage}
\end{table*}
\begin{table*}
 \centering
 \begin{minipage}{140mm}
  \contcaption{}
  \begin{tabular}{@{}llrrrrlrrlr@{}}
 899.30212         & 333.361  & p   & (0,1,0) & (0,1,0) & 2$_{0,2}$  & 1$_{1,1}$  & 2395.53   & A10    & N & 6.03 \\
 902.60941         & 332.139  & o   & (0,1,0) & (0,1,0) & 3$_{1,2}$  & 2$_{2,1}$  & 2550.12   & A10    & N & 1.02 \\
 906.20590         & 330.821  & p   & (0,0,0) & (0,0,0) & 9$_{2,8}$  & 8$_{3,5}$  & 1554.44   & A10    & N & 1.00 \\
 A916.17158        & 327.223  & p   & (0,0,0) & (0,0,0) & 4$_{2,2}$  & 3$_{3,1}$  & 454.339   & A10    & N & 0.42 \\
 923.11335         & 324.762  & o   & (0,1,0) & (0,1,0) & 6$_{2,5}$  & 5$_{3,2}$  & 3109.63   & A10    & N & 5.48 \\
 939.30214$^{\dagger \#}$ & 319.165 & o & (1,0,0) & (0,0,1) & 8$_{4,5}$  & 8$_{2,6}$  & 6843.43  & A10    & N & 0.75 \\
\textit{941.06194}$^\dagger$ & 318.568 & o & (1,0,0) & (0,0,1) & 5$_{4,1}$ & 4$_{4,0}$ & 6123.82 & A10   & N & 0.27 \\
\textit{943.33754}$^{\dagger \#}$ & 317.799 & p & (1,0,0) & (0,0,1) & 9$_{4,6}$ & 9$_{2,7}$ & 7152.58 & A10 & N & 0.39 \\
 968.04698         & 309.697  & o   & (0,1,0) & (0,1,0) & 8$_{2,7}$  & 7$_{3,4}$  & 3590.01   & -      & N & 6.5   \\
 970.31505         & 308.964  & p   & (0,0,0) & (0,0,0) & 5$_{2,4}$  & 4$_{3,1}$  & 598.835   & -      & Y & 2.10  \\
 978.77430         & 306.293  & p   & (1,0,0) & (1,0,0) & 2$_{0,2}$  & 1$_{1,1}$  & 5360.83   & -      & N & 0.29  \\
 988.41446         & 303.306  & p   & (1,0,0) & (0,0,1) & 5$_{4,2}$  & 4$_{4,1}$  & 6126.05   & -      & N & 0.28  \\
 997.65562         & 300.496  & p   & (1,0,0) & (1,0,0) & 5$_{2,4}$  & 4$_{3,1}$  & 5848.86   & -      & N & 0.48  \\
1000.85357         & 299.536  & p   & (0,2,0) & (0,2,0) & 2$_{1,1}$  & 2$_{0,2}$  & 4683.75   & -     & N & 2.42  \\
1016.81003$^{\#}$   & 294.836  & o   & (1,0,0) & (0,0,1) & 6$_{2,5}$  & 5$_{2,4}$  & 6042.03   & -     & N & 0.49  \\
1077.76304         & 278.161  & p   & (0,1,0) & (0,1,0) & 7$_{2,6}$  & 6$_{3,3}$  & 3335.88   & -     & N & 7.64  \\
1080.23873         & 277.524  & p   & (1,0,0) & (0,0,1) & 7$_{4,4}$  & 7$_{2,5}$  & 6566.61   & -     & N & 2.23  \\
1085.01419         & 276.302  & o   & (0,2,0) & (0,2,0) & 8$_{3,6}$  & 7$_{4,3}$  & 6078.28   & -     & N & 3.2   \\
1086.46274         & 275.934  & p   & (1,0,0) & (1,0,0) & 1$_{1,1}$  & 0$_{0,0}$  & 5313.86   & -     & N & 0.62  \\
1099.29823         & 272.712  & o   & (1,0,0) & (1,0,0) & 6$_{3,4}$  & 5$_{4,1}$  & 6176.58   & -     & N & 0.49  \\
\textit{1101.13030}& 272.258  & p   & (0,0,0) & (0,0,0) & 11$_{6,6}$ & 12$_{3,9}$ & 3029.89   & -     & N & 0.39  \\
\textit{1109.59787}& 270.181  & p   & (0,0,0) & (0,0,0) & 9$_{5,5}$  & 10$_{2,8}$ & 2068.93   & -     & N & 1.52  \\
1153.12682         & 259.982  & o   & (0,0,0) & (0,0,0) & 3$_{1,2}$  & 2$_{2,1}$  & 249.436   & -     & N & 0.41  \\
1158.32385         & 258.815  & o   & (0,0,0) & (0,0,0) & 6$_{3,4}$  & 5$_{4,1}$  & 933.742   & -     & N & 0.85  \\
1172.52583         & 255.680  & p   & (0,0,0) & (0,0,0) & 7$_{4,4}$  & 6$_{5,1}$  & 1334.83   & -     & N & 0.27  \\
1205.78910         & 248.627  & p   & (0,1,0) & (0,1,0) & 1$_{1,1}$  & 0$_{0,0}$  & 2352.37   & -     & N & 2.10  \\
1228.30364$^{\dagger \#}$ & 244.069 & o & (1,0,0) & (1,0,0) & 8$_{2,7}$ & 7$_{3,4}$  & 6511.90    & -    & N & 0.67  \\
1271.47318$^\dagger$ & 235.783 & p    & (0,0,0) & (0,0,0) & 13$_{3,11}$& 12$_{4,8}$ & 3234.48    & L1$_{lo}$ & N & 0.46 \\
1278.26592         & 234.530 & o    & (0,0,0) & (0,0,0) &  7$_{4,3}$ & 6$_{5,2}$  & 1339.85    & L1$_{lo}$ & N & 0.34 \\
1296.41106         & 231.247 & o    & (0,0,0) & (0,0,0) &  8$_{2,7}$ & 7$_{3,4}$  & 1274.19    & L1$_{lo}$ & N & 1.9  \\
\textit{1307.96330} & 229.205 & o   & (0,0,0) & (0,0,0) & 8$_{4,5}$ & 9$_{1,8}$   & 1615.34    & L1$_{lo}$ & N & 1.7 \\
1312.04322$^\dagger$ & 228.492 & o    & (1,0,0) & (0,0,1) &  6$_{4,3}$ & 6$_{2,4}$  & 6322.70    & L1$_{lo}$ & N & 5.6  \\
1322.06480         & 226.760 & o    & (0,0,0) & (0,0,0) &  6$_{2,5}$ & 5$_{3,2}$  & 795.521    & L1$_{lo}$ & N & 2.5  \\
\textit{1344.67616} & 222.947 & p   & (0,0,0) & (0,0,0) & 7$_{4,4}$ & 8$_{1,7}$   & 1334.83    & L1$_{lo}$ & N & 2.1 \\
1349.66583$^{\dagger \#}$ & 222.123 & p & (1,0,0) & (1,0,0) &  7$_{2,6}$ & 6$_{3,3}$ & 6263.37    & L1$_{lo}$  & N & 1.08 \\
\textit{1358.25249}$^{\dagger \#}$ & 220.719 & o & (0,0,1) & (0,2,0) & 7$_{2,6}$ & 6$_{6,1}$ & 6405.98 & L1$_{lo}$ & N & 0.31 \\
\textit{1361.28260} & 220.227 & o   & (1,0,0) & (0,2,0) & 5$_{2,3}$ & 5$_{5,0}$ & 5893.15      & L1$_{lo}$ & N & 1.80 \\ 
1362.64126$^\dagger$ & 220.008  & o   & (1,0,0) & (1,0,0) &  6$_{2,5}$ & 5$_{3,2}$  & 6042.03    & L1$_{lo}$  & N & 0.80 \\
\textit{1410.19282}$^{\dagger \#}$ & 212.589 & p & (0,0,1) & (1,0,0) & 6$_{2,5}$ & 6$_{2,4}$ & 6114.16 & -        & N & 1.42 \\
\textit{1435.00863} & 208.913 & p   & (0,0,0) & (0,0,0) & 9$_{4,6}$ & 10$_{1,9}$  & 1929.25    & L1$_{hi}$ & N & 2.00 \\
1440.78167         & 208.076 & p    & (0,0,0) & (0,0,0) &  7$_{2,6}$ & 6$_{3,3}$  & 1020.97    & L1$_{hi}$  & N & 2.9  \\
1457.14608$^\dagger$ & 205.739 & o    & (1,0,0) & (0,0,1) &  6$_{3,4}$ & 5$_{3,3}$  & 6176.58   & L1$_{hi}$  & N & 0.48 \\
1474.75342$^{\dagger \#}$ & 203.283 & o & (1,0,0) & (0,0,1) &  6$_{4,3}$ & 5$_{4,2}$  & 6322.70   & L1$_{hi}$  & N & 0.40 \\
\textit{1482.73603}$^\dagger$ & 202.188 & p & (0,0,1) & (1,0,0) & 4$_{1,4}$ & 4$_{1,3}$ & 5722.43 & L1$_{hi}$ & N & 1.04 \\
1486.47075$^{\dagger \#}$ & 201.680 & p & (0,2,0) & (1,0,0) &  9$_{3,7}$ & 8$_{0,8}$  & 6383.81   & L1$_{hi}$  & N & 9.6  \\
1491.24429$^{\dagger \#}$ & 201.035 & o & (0,2,0) & (0,0,1) &  8$_{6,3}$ & 8$_{2,6}$  & 6869.92   & L1$_{hi}$  & N & 0.27 \\
1494.05754         & 200.656 & p   & (0,1,0) & (0,1,0) &  2$_{2,0}$ & 2$_{1,1}$   & 2508.50    & L1$_{hi}$  & N & 0.58 \\
1511.69242$^\dagger$ & 198.315 & p   & (1,0,0) & (1,0,0) &  6$_{3,3}$ & 5$_{4,2}$   & 6198.60    & L1$_{hi}$  & N & 0.40 \\
1517.77556$^{\dagger \#}$ & 197.520 & p & (0,2,0) & (0,2,0) & 9$_{3,7}$ & 8$_{4,4}$   & 6383.81   & L1$_{hi}$  & N & 1.9  \\
1525.77486$^\dagger$ & 196.485 & p    & (0,2,0) & (0,2,0) & 4$_{1,3}$  & 3$_{2,2}$  & 4947.37    & -    & N & 5.4   \\
1526.93003$^{\dagger \#}$ & 196.336 & p & (0,2,0) & (0,0,1) & 7$_{6,2}$  & 7$_{2,5}$  & 6973.47   & -    & N & 0.89  \\
1538.69161$^\dagger$ & 194.835 & o     & (0,2,0) & (0,2,0) & 3$_{0,3}$  & 2$_{1,2}$  &  4732.52  & -    & N & 9.4   \\
1541.96702         & 194.422 & p     & (0,0,0) & (0,0,0) & 6$_{3,3}$  & 5$_{4,2}$  & 951.828   & -     & N & 0.29  \\
1564.36368$^\dagger$ & 191.638 & p   & (0,0,1) & (0,0,1) & 5$_{3,2}$  & 5$_{2,3}$   &  6112.18   & -     & N & 1.21  \\
\textit{1574.23203}& 190.437 & o   & (0,0,0) & (0,0,0) & 6$_{4,3}$  & 7$_{1,6}$   &  1013.21   & -     & N & 0.69  \\
\textit{1596.25248}& 187.810 & o   & (0,0,0) & (0,0,0) & 8$_{5,4}$  & 9$_{2,7}$   & 1729.31    & -     & N & 1.28  \\
1632.16726         & 183.677 & o   & (1,0,0) & (1,0,0) & 2$_{1,2}$  & 1$_{0,1}$   & 5373.72    & -     & N & 1.33  \\
1643.91917         & 182.364  & o   & (0,1,0) & (0,1,0) & 3$_{0,3}$  & 2$_{1,2}$  & 2491.83    & -     & N & 16.0  \\
1675.14173$^\dagger$ & 178.965  & p   & (1,0,0) & (1,0,0) & 7$_{3,5}$  & 6$_{4,2}$  & 6413.85    & -     & N & 0.91  \\
1690.31359$^\dagger$ & 177.359  & p   & (1,0,0) & (0,0,1) & 6$_{4,2}$  & 5$_{4,1}$  & 6333.45    & -     & N & 0.31  \\
1689.24140$^{\dagger \#}$ & 177.471 & o & (0,2,0) & (0,0,1) & 6$_{6,1}$  & 6$_{2,4}$  & 6730.46    & -    & N & 25.3  \\
1698.86164$^{\dagger \#}$ & 176.466 & p & (1,0,0) & (0,0,1) & 7$_{2,6}$  & 6$_{2,5}$  & 6263.37   & -     & N & 0.42  \\
1718.69484$^\dagger$ & 174.430  & o   & (0,0,1) & (0,0,1) & 6$_{3,3}$  & 6$_{2,4}$  & 6342.21    & -    & N & 4.51  \\
1722.88565$^{\dagger \#}$ & 174.006 & o & (0,2,0) & (0,2,0) & 10$_{4,7}$ & 9$_{5,4}$  & 6966.79   & -     & N & 0.74  \\
1730.29897$^{\dagger \#}$ & 173.260 & p & (0,2,0) & (0,2,0) & 11$_{3,9}$ &10$_{4,6}$  & 7081.84   & -     & N & 0.84  \\
1737.03274$^{\dagger \#}$ & 172.588   & o   & (0,2,0) & (0,2,0) & 10$_{3,8}$ & 9$_{4,5}$  & 6718.75 & -   & N & 4.35  \\
1740.39813            & 172.255 & o & (0,1,0) & (0,1,0) & 8$_{3,6}$  & 7$_{4,3}$   & 3784.29   & -    & N & 4.68  \\
\hline
\end{tabular}
\end{minipage}
\end{table*}
\begin{table*}
 \centering
 \begin{minipage}{140mm}
  \contcaption{}
  \begin{tabular}{@{}llrrrrlrrlr@{}}
1753.91550            & 170.927  & o  & (0,1,0) & (0,1,0) & 2$_{1,2}$  & 1$_{0,1}$  & 2412.93  & -    & N & 3.01  \\
1766.19875            & 169.738  & p  & (0,0,0) & (0,0,0) & 7$_{3,5}$  & 6$_{4,2}$  & 1175.05  & -    & N & 1.57  \\
1820.50238$^{\dagger \#}$ & 164.675 & o   & (1,0,0) & (1,0,0) &  6$_{3,4}$ & 7$_{0,7}$  & 6176.58  & L2   & N & 0.35 \\
1849.74076$^{\dagger \#}$ & 162.072 & p   & (1,0,0) & (0,2,0) &  6$_{0,6}$ & 7$_{3,5}$  & 5892.28  & L2   & N & 9.1 \\
1851.95160$^\dagger$ & 161.879 & o   & (0,2,0) & (0,0,1) &  6$_{6,1}$ & 5$_{4,2}$  & 6340.80 & L2        & N & 1.55 \\
1853.18949$^\dagger$ & 161.771 & p   & (1,0,0) & (0,0,1) &  5$_{4,2}$ & 5$_{2,3}$  & 6126.05 & L2        & N & 7.0 \\
1872.98540$^\dagger$ & 160.061 & o   & (0,2,0) & (0,2,0) &  2$_{1,2}$ & 1$_{0,1}$  & 4658.67 & L2        & N & 6.57 \\
1881.40504$^\dagger$ & 159.344 & o   & (0,0,1) & (0,0,1) &  6$_{3,3}$ & 5$_{4,2}$  & 6342.21 & L2        & N & 0.28 \\
1884.88789         & 159.050 & o   & (0,0,0) & (0,0,0) &  8$_{4,5}$ & 7$_{5,2}$  & 1615.34 & L2        & N & 0.33 \\
1903.64426         & 157.483 & o   & (0,0,0) & (0,0,0) & 12$_{3,10}$ & 11$_{4,7}$ & 2823.63 & L2       & N & 0.51 \\
\hline
\end{tabular}\\
Frequencies are taken from the JPL catalogue, and are derived from laboratory
measurements unless marked with superscript $\dagger$. A superscript $\#$ indicates
an uncertainty larger than 200\,kHz. Transitions preceded by the letter A in column~1 are unobservable due
to deep atmospheric absorption. A frequency in italic script indicates a transition found
as a maser at $T_d=1025$\,K, but not in the set found with $T_d=50$\,K. In column~9, the relevant \textit{ALMA}
or \textit{SOFIA} instrumental band is given, or the radio band for low frequency transitions that
fall below \textit{ALMA} Band~2.
\end{minipage}
\end{table*}

\section{Discussion}
\label{discuss}

The results presented in this paper span a considerable parameter space in several physical
variables, giving some predictive power with regard to the typical conditions where certain
maser transitions will be strongly inverted, and to the strength of association between
masers at different frequencies. The very large number of water maser lines even offers some
prospect of attempting the inverse problem - the recovery of physical conditions from the
observed brightnesses of maser features at VLBI resolution - since the number of available
lines considerably exceeds the number of formal variables. There are, however, a number of
issues relating to other parameters that have been considered constant in our models, but
that could, in practice, vary significantly from the values adopted.

\subsection{Effect of collisional coefficients}
\label{ss:coll_effect}

The importance of collisions can be roughly estimated by comparing radiative and collisional
downward rates in a particular transition. The ratio of the Einstein A-value to the second-order
collisional rate coefficient is one estimate of $n_c$, the critical density, above which the
transition becomes thermalized. There is, however, no single prescription for calculating
$n_c$ exactly, and much depends upon the network of other transitions that is considered to be
coupled to the upper level of the transition of interest. For example, estimates of $n_c$ at 2000\,K for
the 658-GHz transition in $(0,1,0)$ differ by more than an order of magnitude, depending on
whether we consider just the downward transitions from the upper state ($n_c\sim $1.3$\times$10$^{11}$\,cm$^{-3}$)
or include all upward collisional rates from the upper state ($n_c\sim $2.5$\times$10$^{10}$\,cm$^{-3}$).

More generally, theoretical estimates of $n_c$ for many H$_2$O maser transition for
kinetic temperatures typical of the current work ($T_K\sim 500-3000$\,K) lie in the approximate
range 10$^{10}$-10$^{11}$\,cm$^{-3}$, below which we expect maser action to be possible.
In general terms, this analysis agrees with the results 
presented in Fig.~\ref{oh2o_alma_v0_td50}-Fig.~\ref{oh2o_other_v0_td50}.
In many cases, we can see a sharp (compared to the low-density side) transition between
strong inversion and no inversion as we cross $n_c$ to higher density. However, $n_c$ is quite a strong function
of temperature for some transitions. The right-hand edge of all these figures corresponds to
a number density of 10$^{11}$\,cm$^{-3}$, so it is clear that many transitions have a significantly
larger $n_c$, particularly at high $T_K$.

Computed inversions and maser depths depend of course on the collisional rate coefficients
adopted (for details of the set we have used, see Section~\ref{ss:molecdat}). In that section
we also noted that use of a newer set of rate coefficients by \citet{2013A&A...553A..70D} did
not introduce large qualitative changes to molecular level populations, so
our model predictions should change only in detail on changing to any reasonable set of
rate coefficients. Somewhat larger uncertainties might appear in higher vibrational states,
where the data quality deteriorates. However, vibrationally inelastic collisions are unlikely
to play a major role in maser pumping: they are generally too small. As we
mentioned earlier in Section~\ref{ss:molecdat}, a typical number density
of H$_2$, say 10$^9$\,cm$^{-3}$ produces collisional de-excitation rates at 300\,K, for the vibrational
transitions $(0,1,0)-(0,0,0)$, $(0,2,0)-(0,1,0)$ and $(0,2,0)-(0,0,0)$, of respectively
1.3$\times$10$^{-3}$, 2.3$\times$10$^{-3}$ and 6$\times$10$^{-4}$\,s$^{-1}$. These are small relative
to corresponding radiative rates, even allowing for an order of magnitude increase in the
collisional rates as $T_K$ rises to 3000\,K.

Many of the predicted maser transitions in Table~\ref{hugetab} have upper energy levels
with T$_{up}$ above 5000\,K. If these transitions are predominantly radiatively pumped, this
is not a great difficulty, but if we are relying on collisions to populate levels this
far above the ground state it is necessary to consider both the survival of water molecules and
the existence of such high kinetic temperatures in stellar envelopes. Zones with
kinetic temperatures above 5000\,K do exist in the AGB star models by \citet{2011MNRAS.418..114I},
but are geometrically thin (of order 0.005\,AU) and lie very close to the star. However,
they might be thicker in RSGs, and possibly accessible to water transitions that share
the SiO maser zone, rather than the 22-GHz maser shell. Analysis of the formation and
destruction of water by \citet{1979ApJ...229..560E} yielded an equilibrium temperature
of 7660\,K that is, however, dependent on other reactions in a chemical network. Assuming
that water is not significantly depleted at 2000\,K, on this basis it would be 10 times
less abundant at 5000\,K. It should also be pointed out that masers predicted in the current
work are all found in conditions where $T_K\leq 3000$\,K.

\subsection{Families of transitions}
\label{ss:families}

We have found three families of maser transitions on the basis of their response
to increasing energy density of infrared radiation. The first, and largest, family
suffer reduced maser depths with rising dust temperature. We call these the collisionally-pumped
transitions and they are probably inverted by a process similar to the DeJong scheme
(see Section~\ref{ss:theo}), though it cannot be this precise mechanism in the many
transitions of this type that do not include a backbone level. The collisionally-pumped
transitions may be identified in Table~\ref{hugetab} by the frequency in column~1 given
in standard type. Collisionally pumped masers have mostly ceased to be inverted for
dust temperatures above 500-700\,K. In the density-kinetic temperature plane, their
region of inversion typically shows a clear lower bound in temperature and a distinct,
though temperature-dependent critical density. The upper limit of the inverted zone
generally extends beyond the limit of 3000\,K used in the present model.

The second family of transitions have maser depths that generally increase with rising
dust temperature. A minimum number of inverted transitions is typically found
at dust temperatures of 500-600\,K, and on the higher side of this region, collisionally
pumped transitions are swiftly replaced by a smaller, but still large, family of radiatively
pumped masers. The radiatively pumped family may be identified by the frequency in
italic type in Table~\ref{hugetab}. Radiatively pumped transitions become dominant
when $T_d>700-1000$\,K, depending on the transition. The highest dust temperatures used
in the model are probably not realistic; nonetheless, most of the radiatively pumped
transitions are substantially inverted by $T_d\sim 1500$\,K. The radiatively pumped masers
mostly have an inverted region in density/kinetic temperature space that looks very
different to the pattern of collisionally pumped transitions. The upper kinetic temperature
limit is visible, and inverted zones are concentrated towards the highest number 
densities. In many cases, the inverted zone extends to temperatures well below the
minimum 300\,K used in the models. A minority, for example 327 and 943\,GHz are instead
inverted at high density and high kinetic temperature.

A third, rather select family of transitions can be pumped, under rather different
conditions, both radiative and collisional processes. The regions of the density/kinetic temperature
plane under which the collisional and radiative branches of the pumping scheme resemble
the typical patterns discussed above, but these transitions can exist as strong masers
in both of them. Membership of this family includes at least the following transitions:
22, 96, 209, 321, 325, 395, 941 and 1486\,GHz. Note that these transitions appear as
collisionally pumped in Table~\ref{hugetab}.

\subsection{Line overlap and velocity shifts}
\label{ss:oversplat}

The dominant effect of velocity shifts appears to be to simply reduce the optical depth
of active pumping transitions. This conclusion still holds for collisionally pumped lines,
since even these still require a radiative transition to drain the non-backbone energy
level in the DeJong scheme. However, the fact that some transitions have peak maser depths
that are not at zero velocity shift implies that velocity-driven line overlap is
important in some transitions. In our plotted data, however, maser depths are remarkably
symmetrically distributed over the modelled velocity shifts. A possible effect, worthy of
future study, is that an effect of line overlap is to shift the position of maximum
maser gain about in the $n,T_K,T_d$-space, rather than to change the magnitude significantly.
For some transitions, for example 471\,GHz, qualitatively different behaviour can appear
in models where line ovelap is artifically suppressed (see Section~\ref{sss:testing}) though the
effect of overlap is more usually a modest change in the maser depth.

\subsection{Comparison with other recent work}
\label{ss:nest}

Another recent modelling paper that is closely related to the present work, but dealing
more specifically with the 658-GHz transition is \citet{2015MNRAS.449.2875N}, which
uses the same radiative transfer method, very similar spectroscopic and molecular collisional
data and a subset of the physical conditions used in the present work. Minor differences
are that we ignore the contribution of He to collisions, \citet{2015MNRAS.449.2875N} uses
a microturbulent speed of 3\,km\,s$^{-1}$ (compared to our 1), and \citet{2015MNRAS.449.2875N}
uses a stratification that is based on the star IK~Tau, rather than our more general slab.
The dust models are very different: \citet{2015MNRAS.449.2875N} use monodispersed dust with
a size of 300\,nm that is heated by external starlight. This should be compared to our
model in Section~\ref{s:model}.
 
Results in the two papers are presented in somewhat different ways. However, the qualitative
results regarding the 658-GHz transition agree very well: this transition is in the collisional
family, and its gain is suppressed by large quantities of warm dust, or by radiation from the
host star. Changing the velocity shift over a few km\,s$^{-1}$ has a modest (few per cent) effect
on the maser depth or gain coefficient. The additional maser lines predicted by \citet{2015MNRAS.449.2875N}
are also found in our Table~\ref{hugetab}. The only transition for which \citet{2015MNRAS.449.2875N}
had a negative result was 336\,GHz, and this is probably because it is in our radiatively
pumped family, and is found only at kinetic temperatures far lower than the 1100\,K used
by \citet{2015MNRAS.449.2875N} (and see our Figure~\ref{oh2o_alma_v0_td1025}).

For a quantitative comparison, we consider Figure~3 of \citet{2015MNRAS.449.2875N}, and
choose a number density of 10$^{10}$\,cm$^{-3}$ of H$_2$, corresponding roughly to
their maximum gain coefficient. Using our standard fraction of 3$\times$10$^{-5}$ of
o-H$_2$O, we obtain a average gain coefficient of just under 10$^{-13}$\,cm$^{-1}$, for
$T_K=1100$\,K and (presumably) cold dust. Our closest model comparison has the same
number density of H$_2$, $T_K=1157$\,K, and $T_d=50$\,K. Our maser depth of 6.08 under
these conditions yields and average gain coefficient of 2.7$\times$10$^{-14}$\,cm$^{-1}$.

Observing prospects for the frequency range covered in this work look set to improve
in future. Participation of \textit{ALMA} as a phased array in VLBI
is under development and observations at Bands 3, 6 and 7 will
eventually be possible at resolutions of tens of micro-arcsec
\citep{2014arXiv1406.4650T}. Sub-microarcsecond resolution observations in the
0.3-17-mm range should become possible following the launch of \textit{Millimetron}.
For details of this space telescope see\footnote{www.millimetron.ru/index.php/en/}, but
one mode of operation is to be a space-based VLBI antenna that can partner an
Earth-based array.

\section{Conclusions}
\label{conclusion}
We have summarized current observational and theoretical knowledge of astrophysical
water masers, with particular emphasis on the masers in red supergiant and AGB stars.
We have considered in detail the opportunities for water maser observations with \textit{ALMA} and
\textit{SOFIA}, with a brief overview of some other telescopes.
For comparison with existing observations, and to make predictions
of yet undetected maser transitions, we ran tens of thousands of ALI computational models
of both spin species of H$_2$O,
spanning a considerable parameter space in density, kinetic and dust temperatures and
in velocity shift. The frequency range covered by these models is from
zero to 1910\,GHz, spanning the entire coverage of \textit{ALMA} and
\textit{SOFIA}, but also including frequency ranges that fall outside the specific
instrumental bands of these telescopes.

Results of the computational models have been presented as maser depths (negative
optical depths). The maser depth has been plotted
as a function of kinetic temperature and water number density
in Section~\ref{ss_vycma_results} and Section~\ref{sss:dust_effect}, where we have concentrated on the transitions observable  
with \textit{ALMA} and \textit{SOFIA}. However, model data was also produced 
in equivalent form for all transitions, including
the well-known 22-GHz transition, at frequencies between the radio S-band and
1.91\,THz. Parameters of all these maser transitions have been presented in a master
table, Table~\ref{hugetab}.

The locus of significantly inverted regions in the kinetic temperature versus number density
plane fell into two distinct patterns, corresponding to collisional and radiative
pumping schemes. The division of maser transitions into pumping families based on sensitivity
to a dust radiation field was investigated in more detail in Section~\ref{sss:dust_effect}.
We found that we could separate the maser transitions found by the model into three groups
dependent on their main pumping mode (collisional, radiative or both), with the sensitivity
of peak maser depth to increasing dust temperature supporting the provisional assignment,
based on the loci of
inverted regions in the number density/kinetic temperature plane. A dust temperature of
approximately 650\,K produces a minimum in the number of observable maser lines with
increasing numbers of collisionally (radiatively) pumped transitions observable
below (above) this temperature. While collisionally pumped maser transitions may
exhibit peak maser depths over a broad range of number density, radiatively
pumped transitions almost invariably have their peak maser depth at a number density
close to the maximum modelled value.
A much weaker conclusion is that full rovibrational
maser transitions become inverted under conditions of higher kinetic temperature and
density than those confined to a particular vibrational state.

In contrast to the profound effects of varying the dust temperature in the model,
the effects of applying a velocity shift, see Section~\ref{sss:velgrads}, were mostly modest.
The predominant effect of velocity shifts of both signs is to reduce the maser depth
with increasing magnitude of shift. Line overlap produces some asymmetry in this
effect, but truly asymmetric profiles (where the peak maser depth is offset to a
shift of magnitude at least 1\,km\,s$^{-1}$) are rare and, at least for the conditions
plotted in Section~\ref{sss:velgrads}, concentrated to weak masers and transitions in absorption.

A small number of test jobs were run to study the effects of variation in microturbulent
speed (between 0 and 5\,km\,s$^{-1}$) and parameters of both the dust model and line overlap in Section~\ref{sss:testing}.
The effect of varying the microturbulent speed is predominantly a weakening in optical
depth or maser depth as expected; changing the dust model from a power-law size spectrum
to monodispersed dust typically produced changes of a few per cent in maser depths, as did
artifically suppressing line overlap. Moving from a model thickness appropriate to
a supergiant to a model more representative of an AGB star, thinner by a factor of
5, produced mostly commensurate changes in maser depth (see Section~\ref{sss:agbchanges}).

We have briefly considered likely spatial association
between certain groups of transitions, and it is hoped that one or more groups of
these transitions may be used in future to attempt the inverse problem from
multi-frequency interferometric observations.
Multiple transitions can be detected within the variability timescales 
by \textit{ALMA} and \textit{SOFIA}, and imaged at high resolution by \textit{ALMA} 
(complemented by 22-GHz data if needed).  Combinations of maser lines which co-incide 
or avoid each other on various scales will provide tight constraints on physical conditions 
such as number density and temperature, as well as the more traditional use in 
kinematic modelling. The high brightness temperature of masers means that their observational 
properties can be measured with very high precision, on angular scales corresponding to roughly 1-10\,au 
within the Galaxy, comparable to the sizes of individual masing clouds.

\section*{Acknowledgments}

Computations were carried out on the Legion supercomputer at the
HiPerSPACE Computing Centre, University College London, which
is funded by the UK Science and Technology Facilities Council (STFC).
AMS acknowledges support from UrFU
Competitiveness Enhancement Program.

\bibliographystyle{mn2e}
\bibliography{h2o_parms}

\appendix

\section[]{RADEX data files}
\label{app_filehack}

The RADEX data files used in the present work, one for each spin-species
of water, each have three principal blocks. Block~1 lists energy levels
and their associated quantum numbers; Block~2 lists allowed transitions with
Einstein A-value, frequency in GHz and level numbers that tie the transition
to levels from Block~1; Block~3 contains the collisional data, but needs
no further discussion here. The peculiarities of these files, as addressed below,
relate to a version of o-h2o.dat downloaded on 2012-Jul-25 and of p-h2o.dat
downloaded on 2013-Nov-08.

The first problem is that the transition frequencies listed in Block~2 are
not the same as frequencies calculated from the difference of the relevant energy
levels in Block~1. In the worst cases, corresponding to fully rovibrational
transitions involving levels with large values of $J$, the difference may amount to several GHz. It is the latter value
of the frequency (derived from energy differences) that corresponds better (typically to an accuracy of
order 100\,kHz) to frequencies in the JPL database. Since the JPL frequencies
mostly have laboratory measured accuracy of 200\,kHz or better, we have
generally used these, and employed the quantum numbers from Block~1 to double-check the
correct identification of transitions.

The second problem is that the vibrational quantum numbers in Block~1 are
not given in the conventional order. If we label $v_1$,$v_2$ and $v_3$ as
the symmetric stretch, bending and asymmetric stretch, respectively, then the
conventional bracket of vibrational quantum numbers is $(v_1,v_2,v_3)$, and
this form is used throughout the present work. In Block~1 of the data files, the
order \textit{appears to be} $(v_3,v_1,v_2)$, and this is correctly implemented
for $v_1$ and $v_2$. However, some levels that must, on the basis of the JPL
catalogue, involve levels within $v_3=1$, are actually represented as ground vibrational state levels.
This made it somewhat awkward to resolve rovibrational transitions that
involved excited stretching modes.

\end{document}